# Finding the UV-Visible Path Forward: Proceedings of the Community Workshop to Plan the Future of UV/Visible Space Astrophysics

A COPAG SIG Workshop held at NASA Goddard Space Flight Center, June 25-26, 2015

Paul A. Scowen (ASU)[1], Todd Tripp (U. Mass)[2], Matt Beasley (Planetary Resources, Inc.)[3], David Ardila (Aerospace Corp.)[4], B-G Andersson (SOFIA Science Center)[5], Jesús Maíz Apellániz (CSIC)[6], Martin Barstow (U. Leicester)[7], Luciana Bianchi (JHU)[20], Daniela Calzetti (U. Mass)[2], Mark Clampin (NASA-GSFC)[8], Christopher J. Evans (ROE)[9], Kevin France (U. Colorado)[10], Miriam García García (IAC)[11], Ana Gomez de Castro (Universidad Complutense de Madrid)[12], Walt Harris (LPL)[13], Patrick Hartigan (Rice U.)[14], J. Christopher Howk (U. Notre Dame)[15], John Hutchings (NRC-CA)[16], Juan Larruquert (CSIC)[17], Charles F. Lillie (Lillie Consulting LLC)[18], Gary Matthews (Harris)[19], Stephan McCandliss (JHU)[20], Ron Polidan (Northrup Grumman)[21], Mario R. Perez (NASA HQ)[22], Marc Rafelski (STScI)[23], Ian U. Roederer (U. Michigan)[24], Hugues Sana (AIAP)[25], Wilton T. Sanders (NASA GSFC)[8], David Schiminovich (Columbia U.)[26], Harley Thronson (NASA-GSFC)[8], Jason Tumlinson (STScI)[23], John Vallerga (SSL)[27], Aida Wofford (UNAM)[28]

[1] Arizona State University, School of Earth & Space Exploration, PO Box 876004, Tempe, AZ 85287-6004
[2] University of Massachusetts, Department of Astronomy, 710 North Pleasant Street, Amherst, MA 01003-9305
[3] Planetary Resources, Inc., 6742 185th Ave NE, Redmond, WA 98052
[4] The Aerospace Corporation, 2310 E El Segundo Blvd., El Segundo, CA 90245
[5] SOFIA Science Center/USRA, NASA Ames Research Center , M.S. N232-12, Moffett Field, CA 94035
[6] Centro de Astrobiología, CSIC-INTA, ESAC campus, Camino bajo del castillo s/n, 28 692 Villanueva de la Cañada, Madrid, Spain
[7] University of Leicester, Leicester Institute of Space & Earth Observation, Physics Building, University Road, Leicester LE1 7RH, UK
[8] NASA Goddard Space Flight Center, Code 660, 8800 Greenbelt Road, Greenbelt MD 20771
[9] UK Astronomy Technology Centre, Royal Observatory, Blackford Hill, Edinburgh, EH9 3HJ, UK
[10] University of Colorado, Laboratory for Atmospheric and Space Physics, 600 UCB, Boulder, CO 80309
[11] Centro de Astrobiología, CSIC-INTA. Ctra. Torrejón a Ajalvir km.4, E-28850 Torrejón de Ardoz, Madrid, Spain
[12] Dept de Fisica de la Tierra, Astronomia y Astrofisica I, (S.D. Astronomia y Geodesia), Fac. CC Matematicas Plaza de Ciencias 3, 28040 Madrid, Spain
[13] University of Arizona, Lunar and Planetary Lab, 1629 University Blvd., Tucson, AZ 85721-0001
[14] Rice University, Dept. of Physics and Astronomy, Mail Stop 108, Houston TX 77005-1892
[15] University of Notre Dame, Department of Physics, Notre Dame, IN 46556
[16] 5071 West Saanich Rd, Victoria, B.C. V9E 2E7, Canada
[17] Instituto de Optica. CSIC, C/ Serrano 144, 28006 Madrid (Spain)
[18] 6202 Vista del Mar #364, Playa del Rey, CA 90293
[19] 221 Thomas White Blvd, Chester, MD 21619
[20] The Johns Hopkins University, Dept. of Physics and Astronomy, 3400 N. Charles Street, Baltimore, Maryland 21218
[21] Polidan Science Systems & Technologies, LLC, PO Box 457, Redmond, OR 97756
[22] NASA Headquarters, Mail Stop 3U23, 300 E Street SW, Washington, DC 20546
[23] Space Telescope Science Institute (STScI), 3700 San Martin Drive, Baltimore, MD 21218
[24] University of Michigan, Department of Astronomy, 1085 S. University Ave., Ann Arbor, MI 48109
[25] Institute of Astrophysics, KU Leuven, Celestijnlaan 200D, Leuven, Belgium
[26] Columbia University, Dept. of Astronomy, 550 W. 120th St., New York, NY 10027-6623
[27] Space Sciences Laboratory, Univ. of California, Berkeley, 7 Gauss Way, Berkeley, CA 94720-7450
[28] Instituto de Astronomía, UNAM, Ensenada, CP 22860, Baja California, Mexico

# Abstract

We present the science cases and technological discussions that came from the workshop entitled "*Finding the UV-Visible Path Forward*" held at NASA GSFC June 25-26, 2015. The material presented outlines the compelling science that can be enabled by a next generation space-based observatory dedicated for UV-visible science, the technologies that are available to include in that observatory design, and the range of possible alternative launch approaches that could also enable some of the science. The recommendations to the Cosmic Origins Program Analysis Group from the workshop attendees on possible future development directions are outlined.

# 1 Introduction

The Cosmic Origins Program Analysis Group (COPAG) Science Interest Group (SIG) #2 convened a workshop[29] at NASA Goddard Space Flight Center in late June 2015 to discuss the future evolution and development of UV-Visible Astronomy within the context of what is possible technologically today and how that might enable the evolution of compelling science within the next decade. The intent of this open workshop was to bring the UV/Visible Space Astrophysics community together, and to provide input to NASA on the question of flagship missions that NASA should consider for the 2020 Decadal Survey of Astronomy and Astrophysics.

The workshop was also intended to enable the various SIG2 subcommittees to work together towards their respective goals. While the meeting was intended to be face- to-face, call-in facilities were made available and collaborative tools such as Webex and others were used to provide as creative and productive an environment as possible.

This paper is an accumulated set of reports from the speakers at the event, and the session chairs, to give a published overview of what was discussed and the conclusions that were reached during the workshop. It is intended to serve as a reference and snapshot of where the field stood as of that date, with the understanding that even today as the report is being published that many of those boundaries are being advanced as part of the Large Mission concept studies referred to in the Recommendations section at the end of the paper. This was the intended result of this workshop – to deliver a considered set of recommendations to the COPAG Executive Committee concerning strategies to advance development of space-based UV-visible astronomy for them to pass on to Dr. Paul Hertz, the Director of Astrophysics in the NASA Science Mission Directorate.

# 2 The NASA Astrophysics Program – Mario R. Perez, Wilton T. Sanders

The national investments on NASA astrophysics research, science and associated technology developments continue to be stable and significant. Available funding continues to remain either flat or slightly growing year after year in actual dollars.

The NASA astrophysics program is divided into five large programmatic themes or projects:

[29] http://asd.gsfc.nasa.gov/conferences/uvvis/

- Physics of the Cosmos
- Cosmic Origins
- Exoplanet Exploration
- Explorer Program
- Research and Analysis

In this presentation, we report on technologies activities both related to inception and maturation aspects, and provide an update on the Astrophysics Explorers Program.

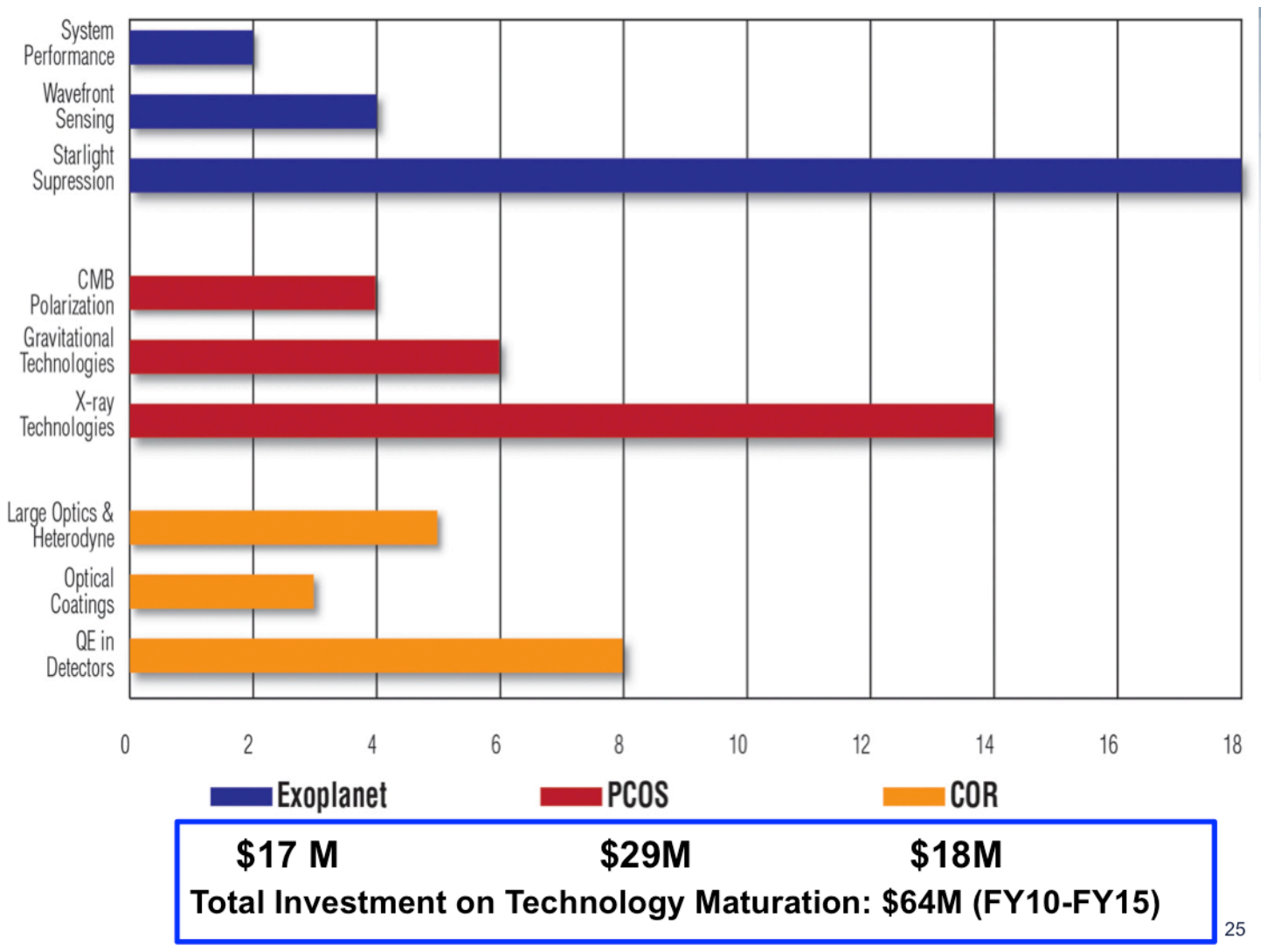

**Figure 2.1** - The selected SAT investigations up to FY2015 for each of the prioritized areas within the themes are described in figure on the left. The total investments are also described for each theme. These activities have greatly advanced the state-of-the-art in some of the areas related to these applied technologies.

## 2.1 Strategic Astrophysics Technologies (SAT)

In 2009, the Astrophysics Division pioneered a program for technology maturation known as the Strategic Astrophysics Technology (SAT). This technology solicitation has three components associated with each of the themes, in which the Astrophysics Division is divided, namely, Physics of the Cosmos (PCOS), Cosmic Origins (COR) and Exoplanet Exploration (EXEP). This program supports the maturation of technologies of mid-range Technology Readiness Level (TRL) already developed and tested in the laboratory (TRL ≤ 3) to a point where they can be incorporated into a flight mission with an acceptable level of risk (TRL 5 or 6). The SAT program is intended to fill the so-called "Mid-TRL Gap" of technologies that have potential but are not sufficiently mature, making them ill-suited to be part of flight programs or to be funded under basic research programs.

The funded SAT proposals are not required or expected to complete the entire development process during the period of their grants. It is expected that the proposers are able to identify verifiable milestones and provide a realistic schedule to achieve these milestones. The technologies emphasized in the SAT program are basically enabling the achievement of science drivers, as opposed to enhancing aspects of further scientific interest. The community can collaborate by identifying technology gaps, which are annually prioritized by a Technology Management Board, which is published in the Program Annual Technology Reports (PATRs). These technology gaps inform technology development needs

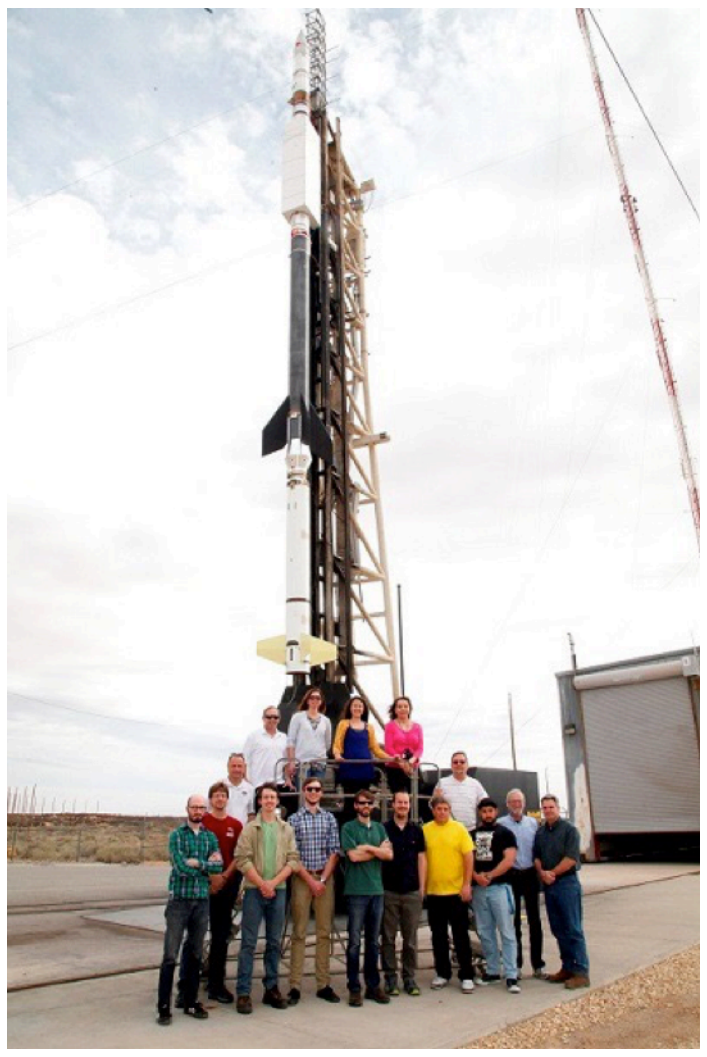

**Figure 2.2** - The Colorado High-Resolution Echelle Stellar Spectrograph 2 (CHESS-2) was successfully launched on 21 February 2016. The team is depicted on the left. CHESS studied translucent clouds in the interstellar medium by analyzing the UV absorption spectra of the two most abundant molecules, molecular hydrogen $H_2$ and CO. These data will be used to study where planets like Earth came from the by determining the raw materials available as building blocks.

## 2.2 Sub-Orbital Program

The Astrophysics Division continues to fund several payloads, via the APRA annual solicitation, for both balloon and sounding rocket experiments. The annual number of astrophysics flights in the last five years, for each activity, has been about 3-5, throughout different seasonal campaigns. These payload flights can accomplish technology maturation (TRL by testing components in relevant environments.

## 2.3 Astrophysics Explorers Program

The goal of NASA's Explorers Program is to provide frequent flight opportunities for high quality, high value, focused astrophysics science investigations that can be accomplished under a not-to-exceed cost cap and that can be developed relatively quickly, generally in 36 months or less, and executed on-orbit in less than three years.

The Explorers Program accomplishes these world-class space science investigations utilizing efficient management approaches to contain mission cost through commitment to, and control of, design, development, and operations costs. The Program also seeks to enhance public awareness of and appreciation for space science by incorporating educational and public outreach activities as integral parts of the investigations.

The Explorers Program strives to:

- Advance scientific knowledge of astrophysics processes and systems;
- Add scientific data and other knowledge-based products to data archives for all scientists to access;
- Lead to scientific progress and the publishing of results in the peer-reviewed literature to encourage, to the maximum extent possible, the fullest commercial use of the knowledge gained;
- Implement technology advancements prepared in related programs; and
- Announce scientific progress and results in popular media, scholastic curricula, and materials that can be used to inspire and motivate students to pursue careers in science, technology, engineering, and mathematics.

Though historically not always this way, the program currently administers only Principal Investigator (PI)-led science investigations for the Heliophysics and Astrophysics Divisions of NASA's Science Mission Directorate (SMD). Competitive selection ensures that the most current and best science that can be done within the cost cap will be accomplished.

Since the early 1990s, the Explorers Program has provided several classes of flight opportunities for addressing astrophysics science objectives. These mission classes are defined by their AO Cost Caps and are designed to increase the number of flight opportunities in response to recommendations from the scientific community. The Medium Explorer (MIDEX) missions solicited in 2016 have an AO Cost Cap of $250M, roughly double the AO Cost Cap of the Small Explorer (SMEX) missions class.

Explorer Missions of Opportunity (MO) are investigations generally characterized by being part of a host space mission other than a strategic SMD mission. These partner investigations are conducted on a no-exchange-of-funds basis with the organization sponsoring the mission. Missions of Opportunity also include small complete missions and new science investigations utilizing existing spacecraft. NASA generally solicits proposals for MOs with each Astrophysics Explorers Program AO issued.

Current ongoing activities in the Astrophysics Explorers Program are:

- Downselection of the 2014 Astrophysics Explorers Program Concept Study Reports for both a Small Explorer (SMEX) and a Mission of Opportunity (MO) flight investigation.
- Preparing the draft (and later the final) AOs for the 2016 Astrophysics Explorers Program Medium Explorer (MIDEX) and for a Mission of Opportunity via the Second Stand Alone Missions of Opportunity Notice (SALMON-2). A Community Announcement regarding these solicitations was issued on December 16, 2015.

# 3 From Cosmic Birth to Living Earth: The Future of UVOIR Space Astronomy - Jason Tumlinson

In 2015 the Associated Universities for Research in Astronomy (AURA) published a new vision for the 2030s entitled "From Cosmic Birth to Living Earths: The Future of UVOIR Space Astronomy". The report caps a two-year AURA-chartered study of how the ambitious goals of exoplanet and cosmic origins science can be combined into a single flagship general observatory dubbed the "High-Definition Space Telescope" (HDST). Led by Co-Chairs Julianne Dalcanton (University of Washington) and Sara Seager (MIT), the study panel defines HDST as a flagship 12-m general observatory with broadband (0.1 - 3 micron) sensitivity, several novel modes of instrumentation such as high-performance optical/NIR coronography and far-UV multi-object spectroscopy, and unprecedented stability in support of high-precision astrometry, spectroscopy, and exoplanet characterization. HDST's headline science goals are to (1) detect and characterize dozens of Earth-like planets in the habitable zones of nearby stars, looking for biosignature gases in their optical/NIR spectra, and (2) to revolutionize studies of stars, galaxies, and the ultimate ingredients of life in the cosmos with high-resolution imaging and spectroscopy across the full UVOIR bandpass.

HDST builds on NASA and ESA's foundation of missions for exoplanet discovery and characterization: Kepler, JWST, TESS, and Plato have or are expected to make fundamental advances in finding and studying planets around other stars. But even if all these missions exhaust their full potential, they will not be able to find and examine the atmospheres of dozens of Earth-like planets in the habitable zones of nearby stars. JWST in particular will excel at detecting the atmospheres of planets larger than Earth transiting stars smaller than the Sun (i.e. SuperEarths orbiting M-dwarfs) where conditions are more favorable for an infrared telescope with coronagraphic contrast at only the $10^6$ level. Reaching the atmospheres of dozens of Earth-twin planets requires contrast ratios of order $10^{10}$, and so is beyond even the future generations of giant ground-based telescopes, which might reach a handful of Earth-size planets by surveying up to a few dozen M stars. NASA's WFIRST will advance key starlight suppression technologies and search for giant planets and SuperEarths at the $\sim 10^9$ level, but again will not reach dozens of Earth-twin planets. A large stride toward characterizing dozens of Earth-twin planets and searching them for biosignatures requires 10-million-fold ($10^{10}$) starlight suppression and collecting area large enough to obtain revealing exoplanet spectra, which in turn require a large telescope with an extremely stable wavefront and advanced starlight suppression technology (a coronagraph and/or starshade). The AURA study advocates that NASA and its community and industrial partners take steps to meet these technological challenges while the array of current and near-future missions lays the groundwork for HDST's search for "Living Earths".

HDST also promises revolutionary gains in capability that will make tremendous advances in our understanding of astrophysical objects and processes from the smallest stars to the most massive black holes. As a generally available flagship in the mold of the Great Observatories, HDST will unleash the widespread creativity of the astronomical community with transformative capabilities: it will have 25x the pixel density per area as HST in the optical, 4x better resolution than JWST in the NIR, up to 100x the point-source UV spectroscopic sensitivity, multiobject UV spectroscopy for up to 100 sources in a ~3 arcmin FOV, and extremely stable wavefronts to provide precise PSFs over long timelines. HDST's optical-band spatial resolution corresponds to ~100 pc or less physical scales *at all redshifts*, which will reveal the internal structure of high-z galaxies that even JWST will not resolve (Fig 1). HDST will also have the UV sensitivity to map the weak emission from metals in the circumgalactic medium of galaxies at $z < 2$, where gas flows drive galaxy fueling and transformation. This same UV MOS capability will enable detailed dissection of star-formation and AGN-driven feedback source-by-source in nearby galaxies. These new UVOIR capabilities promise to transform our understanding galaxies, stars, and the origins of the chemical elements.

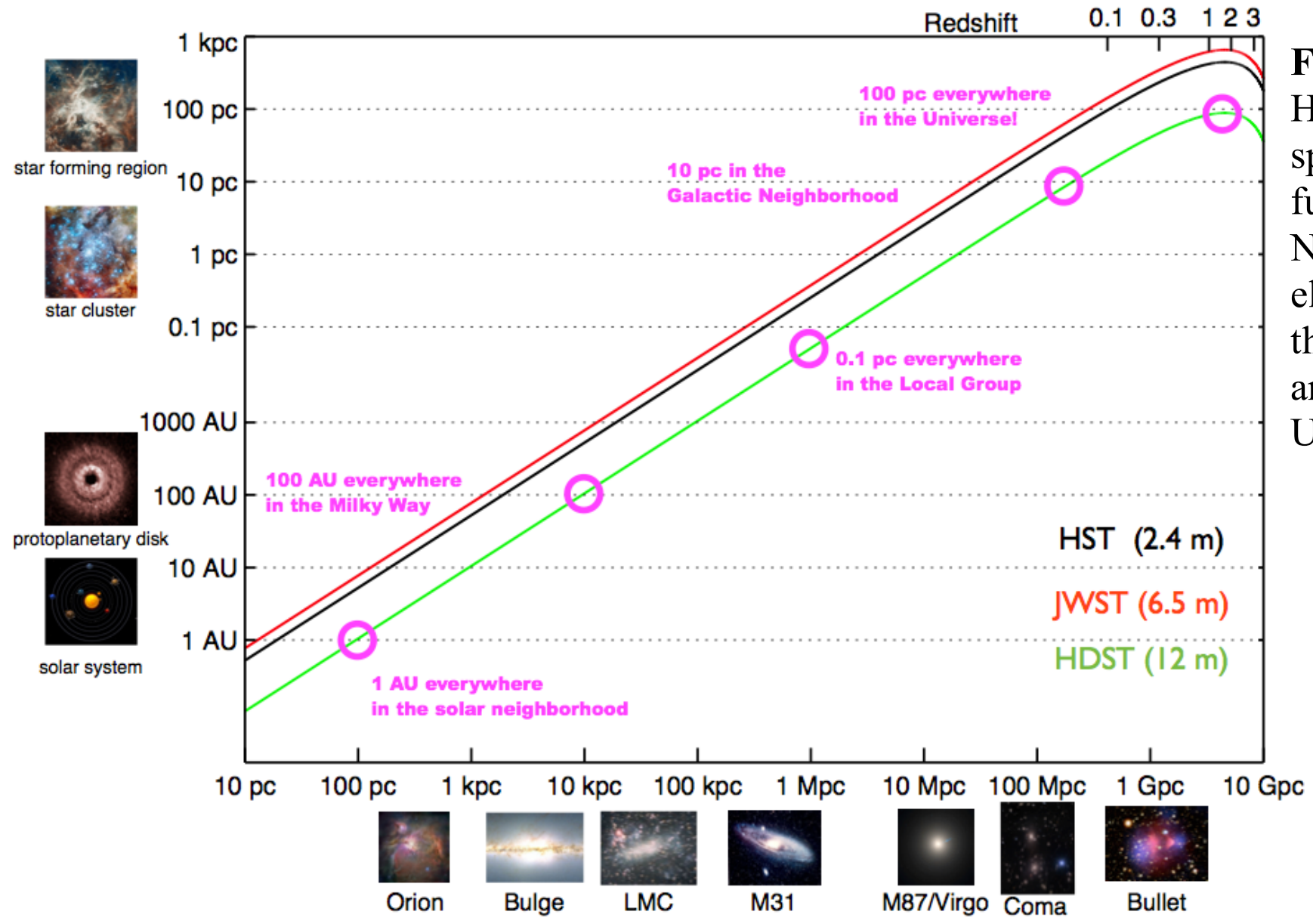

**Figure 3.1 -** The physical size of HDST's diffraction limited spatial resolution element, as a function of distance / redshift. Note that the HDST resolution element is < 100 AU anywhere in the Milky Way and < 100 pc anywhere in the observable Universe.

The AURA study also carefully examined the technology requirements for HDST and concluded that, with the proper investments this decade, HDST will be feasible and affordable for a new start in the next decade and launch in the 2030s. The high degree of starlight suppression required for HDST's exoplanet science is the greatest technology challenge, but new designs for coronagaphy are advancing rapidly. For cosmic origins science, the greatest technology needs are for large-format and high-QE UV and optical detectors and UV coatings that support sensitivity down to 0.1 micron while minimally affecting the optical wavefront. The report makes specific recommendations on these points that the committee hopes will feed into priorities for tech development funding in the next few years and into NASA's Science and Technology Definition Teams for flagship missions (where the HDST-like concept is known as LUVOIR). More details about HDST and contact information are available at www.hdstvision.org.

# 4 The Advanced Technology Large-Aperture Telescope (ATLAST): 2007 – 2015 – Mark Clampin / Harley Thronson

In mid-2007, NASA issued a call for proposals for "Astrophysics Strategic Mission Concept Studies," seeking to identify concepts for scientifically compelling missions and to help identify technology developments such missions might require in advance of the 2010 NRC Decadal Survey. The Advanced Technology Large-Aperture Space Telescope (ATLAST) was one of the concepts selected to develop a technology plan to enable a large UVOIR space telescope to be considered by the NRC for development in the 2020s. The priority scientific goals of ATLAST, which would remain for its decade-long series of studies, incorporated both general astrophysics and exoplanet science on equal footing.

The ATLAST study consisted of three UVOIR telescope concepts: an 8 m monolithic mirror telescope and two segmented telescopes, one with a 9.2 m primary that could fit into an existing Evolved Expendable Launch Vehicle, and one with a 16.8 m primary. The 8 m and 16 m designs required a heavy lift launcher similar to the current Space Launch System. At the time, the ATLAST team considered an aperture about 8 m to be the minimum size needed to characterize the atmospheres of a significant number of terrestrial-mass planets in the habitable zones of their host stars, as well as providing the required spatial resolution and collecting area for other UV/optical/near-IR science goals.

In 2009, the ATLAST concept and science goals were submitted for consideration by the 2010 Decadal Survey, which responded with its highest-priority "medium" activity, investment in technologies to "[Prepare] for a planet-imaging mission beyond 2020" with mission-specific funding of ~$200 M over the decade of the 2010s. This technology development program was recommended to "[Prepare] for a planet-imaging mission beyond 2020."

In response to the NRC's recommendations, in spring 2013 NASA Goddard Space Flight Center initiated an internally funded assessment of a large-aperture UVOIR space observatory specifically intended to be sufficiently well-characterized to be considered by the 2020 Decadal Survey. This design retained the ATLAST acronym and the study was joined by JPL, the Space Telescope Science Institute, and NASA MSFC, along with regular discussions with the AURA HDST team. The ATLAST study was concluded in late 2015 with study results passed to the Large UV/Optical/IR (LUVOIR) Surveyor team.

The ATLAST scenario planning process recognized uncertainties in major science goals for this wavelength range in the 2020s, characteristics of plausible future launch vehicles, as well as capabilities for astronauts and robotic systems at notional operations site for an ATLAST-type mission. Moreover, at the time of the ATLAST assessment, NASA HQ had yet to prepare plans for the 2020 Decadal Survey. The final reference design was, therefore, intended to be flexible enough to be a successful candidate for future development, regardless of the scientific, programmatic, and engineering environment of the 2020s. In addition, as mandated by Congress, the ATLAST designs were all intended to be serviceable, if not actually serviced, when eventually operational.

The segmented design in one concept for ATLAST would be adaptable to different launch vehicles by increasing or decreasing the size of the segments, their number, or the adopted deployment scheme. Unfortunately, in the event with limited study resources, only a single segmented design, a 9.2 m configuration, was studied in depth. This aperture was validated to fit within a five-meter launch vehicle fairing, an industry standard, which was adopted to build confidence that the design was feasible and costs are controllable. This 9.2 m aperture is made up of 36 hexagonal segments and reflects the design heritage of JWST.

# 5 Science

## 5.1 Workshop discussion of compelling ultraviolet-visible science drivers and facility requirements – Todd Tripp

Context: This workshop did not occur in a vacuum; several previous (and extensive) studies had already considered the science drivers, telescope design, instrument suite, etc. for a future UV-optical telescope, and shortly before this workshop occurred, a committee commissioned by the Association of Universities for Research in Astronomy (AURA) published a detailed concept study of a 12m UV-visible telescope dubbed *The High Definition Space Telescope (HDST).* In addition, a team at Goddard, JPL, STScI, and Marshall had spent several years studying various concepts for an 8m to 16m UV-visible telescope known as *The Advanced Technology Large Aperure Space Telescope* (*ATLAST)*, and a variety of materials were available from the *ATLAST* group. The Astro2010 Decadal Survey, while not explicitly recommending construction of a UV-optical telescope for development in the 2010-2020 period, did recognize the scientific importance of a successor to the *Hubble Space Telescope* and recommended funding for technology development to enable a UV-visible telescope in the 2020-2030 decade. Building on the Astro2010 report, the NASA 2013 Astrophysics Roadmap sketched a notional UV-visible mission called *The Large UV-Optical-IR Surveror (LUVOIR)* and provided a detailed discussion of science drivers for such a facility. Finally, a few months before this workshop, the NASA Cosmic Origins Program Analysis Group solicited white papers on large astrophysics missions to be studied by NASA in preparation for the 2020 Decadal Survey.

Science meeting goals: Given this context of rich information from very detailed existing studies, the science subcommittee identified several goals for this workshop with the philosophy that we should complement and build on previous work: (1) Are there recurring science themes in the various materials above (as well as the current research interests of the astronomical community) that point to "killer apps" for a future UV-visible-IR facility? (2) Are there important science drivers that were missed or underappreciated in previous studies? (3) What are the most important technical requirements for a future telescope? With these goals in mind, eleven speakers were invited to present science cases for a future UV-visible telescope during the science breakout session. Some speakers were selected to discuss topics where little information was available from the sources above (e.g., science drivers for ultraviolet polarimetry) while others were chosen to review hot topics that could be killer apps to motivate a future facility. One concept in the studies summarized above stands out: at this time, there is broad support for a future telescope that can search for and characterize habitable exoplanets but can also serve as a general astrophysical observatory that would be useful to a broad scientific community. Previous studies have placed significantly more emphasis on the science and technical requirements for habitable exoplanet studies, and NASA has appointed a program analysis group devoted exclusively to exoplanets, so for this workshop, we elected to focus on general astrophysics science drivers (i.e., not including exoplanets) and their technical requirements. Brief summaries of the presentations in the science breakout group are included in this paper.

Emerging themes: After the science breakout session was completed, a townhall-style discussion occurred with all workshop attendees. This discussion briefly summarized the science breakout talks, but the emphasis was primarily on emerging themes and telescope requirements for a general astrophysics observatory. Many excellent talks were presented, and we refer the reader to the science summaries elsewhere in this paper. Speakers were asked to provide remarks about the bandpass, spectral resolution, angular resolution, field of view, multiplexing capability, etc. required by their science, and the most prominent technical requirement expressed by speakers was to have access to the far-ultraviolet band, with good sensitivity to wavelengths as short of 1000 Å at least, and several speakers pointed out benefits of sensitivity all of the way to 912 Å. This may not be surprising – as ground-based optical telescopes grow larger and increasingly capable, the ultraviolet band naturally becomes the dominant reason to deploy a UV-visible telescope in space. The low backgrounds of space are also highly advantageous for IR astronomy as well, of course, and while some speakers commented on near-IR science, most of the talks focused on the science enabled by access to the ultraviolet. This requirement for good sensitivity in the far-UV was emphasized in the townhall discussion because good sensitivity in the far ultraviolet has been achieved in the past with various optical coatings that could be problematic for the observatory's dual purpose of studying habitable exoplanets, which can impose stringent requirements on optics in entirely different ways. Coatings that have been preferred in previous exoplanet observatory proposals/concepts (e.g., silver) could preclude ultraviolet astrophysics entirely. Therefore, it is crucial during the next few years to invest in and develop optical coatings that are suitable for both exoplanet and ultraviolet instruments. Further discussion of coatings for ultraviolet astrophysics is provided elsewhere in this paper. Another recurring theme in the science breakout talks is that future science will often require spectroscopy with good to excellent spectral resolution. Design of high-resolution spectrographs is a well-developed field, so this initially seems like a straightforward requirement, but some high-resolution spectrograph designs benefit from large-format detectors, and therefore the development of large-format ultraviolet detectors with good quantum efficiency is another area of potential investment to prepare for a future UV-visible observatory. Large-format detectors would also be highly advantageous for imaging instruments. It has been frequently noted that the UV detectors in instruments on current and past space-based telescopes have poor quantum efficiency, so development of more sensitive UV detectors is also a promising avenue for expanding the science capability of a future space telescope. Finally, there was broad consensus that a future UV-optical-IR flagship telescope must be an observatory with a suite of instruments that provides very broad capability. Given the amount of time that is required to build a flagship-class observatory, it is very difficult to predict the science questions that will be most pressing when the facility is finally deployed. For this reason, the members of this science subcommittee (and most scientists that have provided feedback in this process) broadly agree that flexibility in capabilities is crucial to enable astrophysicists of the future to fully exploit the observatory for science that we might not even imagine currently.

## 5.2 UV-Optical Flagship and Massive Stars - A. Wofford, M. García García, J. Maíz Apellániz, L. Bianchi, C. Evans, H. Sana et al.

**What are massive stars?** Classically, these are stars that ignite carbon non-degenerately and thus have initial masses of greater than ~8 $M_\odot$. They are the progenitors of type II and Ib/c supernovae (SNe), produce black holes, and are possible progenitors of gamma-ray bursts. Although they are rare (~0.26% by number and ~23% by mass assuming a Kroupa 2001 stellar initial mass function over the mass range 0.1-120 $M_\odot$), and have short lifetimes (less than ~50 Myr, Meynet et al. 1994), they are non-negligible in terms of radiative, mechanical, and chemical feedback. Indeed, they are luminous ($10^4$-$10^6$ $L_\odot$), important sources of ionizing photons, inject momentum into their environment via winds and explosions, and are production sites of chemical elements (bulk of alpha-elements, some carbon and nitrogen, and some Fe-peak elements) and dust (via type II SN explosions and WC stars). Massive stars in binary systems can be strong sources of gravitational waves as demonstrated by the recent GW150914 detection (Abbot+2016).

**Why are they important?** There are three fundamental areas that would benefit from a deeper understanding of how massive stars work. 1) When did the first stars in the universe form and how did they influence their environments. 2) What are the cosmic origins of chemical elements, in particular those essential to life as we know it. 3) How exchanges of mass and momentum between massive stars and the environment shape the evolution of galaxies. In particular, understanding massive stars is crucial for developing the spectral population synthesis models that are used for interpreting observations of unresolved massive-star populations and star-forming galaxies across redshift.

**What is the problem?** The designs of past and current UV satellites are not optimized to observe i) large samples of massive stars or ii) stars beyond the nearest Local Group galaxies. **Large samples** are needed in order to separate the many competing factors that influence the observed properties of massive stars (mass, chemical composition, convection, mass-loss, rotation, close-binarity, magnetic fields, etc.), and which affect their nuclear reaction rates, the duration of their evolutionary stages, and their final stages. Massive-star evolution is particularly uncertain at initial stellar masses greater than 20 Msun, where evolutionary stages and supernova explosions are not well mapped. **Large samples** are also needed to obtain extragalactic UV extinction curves along many more sightlines. After 25 years of HST, the number of extragalactic UV extinction curves is extremely low (a few sightlines in LMC, SMC, M31). The mean UV extinction law for the SMC which is usually taken as template for low-z galaxies is based on a handfull of stars with diverse extinctions (with some showing the dust bump and some not; Maíz-Apellániz & Rubio 2012). Our scarce knowledge of UV extinction translates into large systematic uncertainties in stellar masses and star formation rates of galaxies (as large as one order of magnitude for the SFR, Bianchi+11). **Observations beyond the Local Group** are necessary in order to sample regimes of extreme low metallicity and star-formation, that build the bridge to higher redshift results. In particular, regions with metallicities well below 1/7 solar and approaching primeval conditions, and star clusters of initial mass greater than $\sim 10^5$ $M_\odot$, as during the cosmic peak of star formation. The dependency of stellar parameters with metallicity has hardly been addressed in part because we can't study individual stars in galaxies beyond 1 Mpc. The few studies in 1/7 solar metallicity dwarf irregulars, including IC1613, are controversial (Tramper+11, García+14; Bouret+15). The two nearest most metal-poor (1/30 solar) star-forming galaxies in the nearby universe are DDO 068 (7.9 Mpc) and I Zw 18 (18 Mpc), and a pair of galaxies like the Antennae (~20 Mpc) host young massive clusters of $\sim 10^5$ $M_\odot$ and greater (Whitmore+). Progress could be achieved by observing these targets.

**What is needed?** UV and optical spectroscopy are how main photospheric and wind parameters of massive stars are obtained. To characterize winds we need **UV spectroscopy, from 900-1800Å at R=10,000** (although 5,000) is probably enough. The UV resonance lines of OVI, SIV, PV, SiIV, CIV, NV and non-resonance CIII, OIV, OV, NIV constrain shocks in the wind, micro-clumping, macro-clumping and mass-loss, and provide in most cases the only diagnostic for terminal velocity. **Optical spectroscopy, containing Hα and HeII4686, at R=10,000.** If we want to explore Local Group galaxies (beyond the Magellanic Clouds) and redenned regions of the Milky Way, we would need to reach in the UV 1E-17 erg/cm$^2$/s/Å in 2h exposure time, with SNR=20 per resolution element at least.

To determine stellar properties (temperature and gravity), R=5,000 in the optical range (3500-7000Å) is enough. However, **R=20,000** would enable us to get: chemical abundances, projected rotational velocities, macroturbulent broadening. The latter three points are currently being investigated as indirect probes of the internal processes occurring during evolution. In particular, macroturbulence may be connected to stellar pulsations and/or turbulent pressure generated by convection zones. In this case we would need R=20,000 with SNR=100. If we want to comfortably explore Local Group galaxies, we need to reach V=20.5. With regards to extinction by dust, detailed maps in the Magellanic Clouds and beyond could be achieved with a multi-object FUV plus optical spectrograph with high sensitivity, via **panchromatic photometry** of resolved stars or **low resolution spectra**. We would certainly benefit from any multiplexing capability within **30"** (projected size of a Local Group OB association) to **10'** (rough diameter of a LG dwarf irregular galaxy), and **sub-arsecond spatial resolution**.

**What we would get?** Complete color-magnitude diagrams of massive-star populations down to 1 $M_{\odot}$ for testing stellar evolution tracks over a wide range of metallicities and star formation history scenarios. Simultaneous medium-resolution co-spatial UV+optical spectroscopy for ~100 massive stars in one exposure in Local Group galaxies for separating the astrophysics that determine the properties of massive stars and determining chemical abundances. Multiple exposures to improve S/N, address stellar variability, and monitor candidates for SN explosions and gamma-ray burst. Detailed dust extinction maps in the Magellanic Clouds and beyond. Large samples of chemically enriched wind bubbles around evolved massive stars and SNRs for constraining stellar nucleosynthesis and galaxy enrichment. Finally, emission and absorption spectra of the ISM in galaxies covering a wide range of conditions.

## 5.3 Nearby Galaxies Under Magnification: New Science with a Future UVOIR Telescope - Daniela Calzetti

Nearby galaxies provide unique test-beds for theories of star formation and galaxy evolution, because they can be dissected into elemental components: stars, star clusters, gas clouds, etc. which can be investigated and compared against models with exquisite detail.

The Hubble Space Telescope has produced a revolution in this field, by enabling the derivation of the histories and properties of both young and old stellar populations and the gas they ionize in galaxies up to a few Mpc. The combination of the James Webb Space Telescope and WFIRST will expand this legacy by pushing the investigation of the nearby Universe into new territory. For instance, we will be able to unravel the physical foundation of the star formation-gas relation (the Schmidt-Kennicutt Law) and the spatially-resolved star formation histories of galaxies over the 200 Myr-10 Gyr range out to about 6-8 Mpc. The low-end of the stellar Initial Mass Function and its variations will be investigated out to ~0.5 Mpc. We will be able to access locales where to study the physics of the dust processing and formation out to ~10 Mpc. Although the limits in accessible cosmic volume are somewhat fluid, as they depend on the diagnostic used and crowding of the systems under investigation, a distance limit of ~6-8 Mpc secures mostly dwarf galaxies, since only about a dozen giant spirals and at most one giant elliptical are present in this volume. Furthermore, the JWST and WFIRST are optimized for the near/mid-infrared regime, while young stellar populations in the age range ~2-100 Myr emit the bulk of their energy in the UV. Volume and wavelength constraints strongly limit the range of accessible environments by existing and planned space facilities. These are the environments where complex physical processes can be studied and models to describe them tested.

A UV-optical optimized telescope with an aperture ≥ 5 times that of the HST, will detect and resolve individual massive stars (≥10 $M_o$) and star clusters as low in mass as a few 1,000 $M_\odot$ and as old as ~100-200 Myr out to ~100 Mpc. Star formation histories as far back as ~10 Gyr will be obtained out to ~20 Mpc, i.e., out to and beyond the Virgo Cluster, with the same quality achievable today out to ~3-4 Mpc with the HST. The local ~100 Mpc volume contains several (Ultra) Luminous Infrared Galaxies (U/LIRGs). U/LIRGs are common denizens of the high-redshift Universe, becoming as common as normal star-forming galaxies, such as the Milky Way, above z~1 (Murphy et al. 2011).

At least three open problems will be addressed by such capability: (1) determine the upper end (slope, maximum stellar mass) of the stellar Initial Mass Function and establish any variation; (2) establish the modes of galaxy growth; and (3) reconcile the Local and Cosmic star formation histories. In order to answer all three questions, the UV/optical and large-aperture capabilities need to be coupled with large detector's Field-of-View (FoV, about 3'x3', in order to sample a sizeable portion of each galaxy) and, for the IMF, with multi-object (UV/optical) spectroscopy.

Any quantity related to star formation rates (SFRs), short-timescale (<300-500 Myr) star-formation histories, feedback, or chemical enrichment of galaxies, will involve the high-end of the IMF. This is because stars above a few solar masses emit, via stellar winds and supernovae explosions, virtually 100% of the ionizing photons, over 80% of the non-ionizing UV photons, and provide the vast majority of the energy, momentum, and chemical output from stars. Variations at the high end of the IMF can thus have a major influence on the interpretation of galaxy evolution. For instance, SFR measurements can be over- or under-estimated by as much as a factor ~2, depending on the strength of the IMF variation and the range of stellar masses considered. This impact can, additionally, be non-uniform across galaxy populations and across time.

Models can support both universal and variable IMFs, owing mainly to the complex physics and the variety of conditions under which stars form in gas clouds. Direct and indirect measurements of the shape of the IMF in a range of galactic environments are the only tool available to constrain potential variations. One example of indirect measurement of the high-end of the IMF involves probing the aggregate properties of the young star cluster populations in nearby galaxies (Calzetti et al. 2010; Andrews et al. 2013; Andrews et al. 2014). A stringent test for IMF variations requires covering a wide range of galaxy morphologies, masses, SFRs, specific SFRs, and external environments, including interacting and merging systems. This requirement is satisfied by the ~100 Mpc local cosmic volume probed by the large-aperture UV/optical telescope.

Models of galaxy growth predict that both minor mergers and inside-out star formation should contribute to the formation of the bodies of galaxies; however, the relative importance of each mechanism is still under debate (Brown et al. 2008, Wang et al. 2011). The outskirts of galaxies are dynamically 'quiet' environments, where the imprint of past events and structures persists for multiple Gyrs. These are perfect settings for testing models of galaxy growth, as minor mergers and inside-out star formation predict different population mixes. They are also excellent locales for investigating modes of star formation, owing to their extreme conditions of density, pressure, metal enrichment, dust content, and response to mechanical feedback. Given the characteristics of the environments, UV-optical capabilities need to be coupled with a large FoV (see above), and the sensitivity afforded by a large-aperture telescope, in order to acquire statistically significant numbers of outer disk stars and optimally complement the near-infrared surveys that will be performed with WFIRST.

The size of the Local Universe that is representative of the cosmic Volume, i.e. the local volume that yields the same average properties as the evolving Universe, is not a settled number. The Universe within the local 3-4 Mpc is over-dense relative to the cosmic average, and yields SFR densities that are about a factor ~2 higher than the cosmic value over the past 4-6 Gyr (i.e., since z~0.4-0.7, Drozdovsky et al. 2008; Williams et al. 2011). The number of Core Collapse Supernovae within the local 11 Mpc is higher by about a factor 1.7-2 than the number that would be predicted from the UV emission of all galaxies within the same volume, and it is even higher than the number predicted from the Hα emission (Botticella et al. 2012; Horiuchi et al. 2013). No such discrepancy is observed on cosmic scales for 0<z<1 (Dahlen et al. 2012). Solving these discrepancies and determining the Representative Local Volume will be the task of the next large-aperture UV/optical facility.

## 5.4 Some Science Cases for Space UV Polarimetry in the 21st Century - B-G Andersson

*"Polarization is the crucial third leg of the astronomer's light-analysis tripod" - D.P. Clemens*

Polarization provides a powerful, but all-too-often overlooked, tool of observational astronomy that allows us to probe magnetic fields in stars and the interstellar medium, reveals structures in unresolved, asymmetric objects, such as circumstellar disks, and enables the study of obscured or below the diffraction limit phenomena, including in the inner parts of Active Galactic Nuclei. Broad wavelength coverage, including the ultraviolet, provides critical constraints on many astrophysical processes. While many applications for the ultraviolet part of the spectrum can already be identified, UV polarimetry is poorly explored and provides significant discovery space. Because polarimetry is inherently photon-intensive, large apertures are needed for efficient observations. However, the Stokes-vector (Q,U,V) decomposition required for polarimetric analysis can be combined to yield the intensity (Stokes I) needed for photometry or spectroscopy, and hence no information is lost in polarimetry mode and the polarization optics transmission is typically very high. While some telescope designs can preclude the efficient acquisition of polarimetric data, careful considerations in the mission design can preserve the ability to include polarimetry in the science program for future (space-based) observatories. Below I outline a small number of concrete examples of science enabled by space-based UV/Optical polarimetry.

### 5.4.1 Interstellar Medium Studies

Dust-induced interstellar polarization allows studies of the interstellar magnetic field geometry and strength. A quantitative, empirically well-tested, theory of grain alignment is now available, allowing reliable quantification of the magnetic field, of the dust and other environmental parameters (Andersson, Lazarian & Vaillancourt, 2015). Radiative Alignment Torque (RAT) theory predicts that paramagnetic grains are aligned if exposed to a radiation field with wavelengths less than the grain diameter. This e.g. directly explains the shape of the UV-NIR polarization curve which requires that – for most diffuse ISM line of sight – grains smaller than ~0.045μm are unaligned. However, for some of the lines of sight probed in UV polarimetry with WUPPE or HST/FOS (e.g. Martin, Clayton & Wolff, 1999), significant UV polarization is seen, indicating alignment of grains down to ~0.01μm. Under RAT alignment this can be understood if some of the dust is exposed to radiation shortward of 912Å, providing a probe of the Galactic hard-UV radiation field. For two of the lines of sight with enhanced UV polarization, polarization in the 2175Å extinction "bump" is also seen. Further UV polarimetry can therefore address the nature of the carrier of this feature, allowing a better understanding of the extinction and mineralogy of the dust. An alternative interpretation of the enhanced UV polarization proposes that these lines of sight represent local enhancements in the strength of the ISM magnetic field. Under such conditions, very small grains can be efficiently aligned by paramagnetic relaxation alignment, and would allow a direct measurement of the magnetic field strength (Hoang, Lazarian & Martin, 2014). For atoms and ions with [hyper-] fine structure transitions, an effect similar to RAT alignment leads to line polarization, which can be used to probe the magnetic field geometry and strength. Many promising transitions occur at UV wavelengths (Yan & Lazarian, 2008)

### 5.4.2 Unresolved Asymmetric Objects

Polarization allows structures to be derived in unresolved asymmetric sources, by interpreting the scattered light or Zeeman effect lines. Because of the differing spectral dependence of electron (Thompson) scattering, and scattering by polarizable molecules or dust in the small particle limit (Rayleigh scattering), a significant amount of physical information can be gleaned from multi-wavelength polarimetry. For sources with rotational or orbital velocities, time-resolved polarimetry can be used to construct 3D models of structures and magnetic fields (e.g. Morin et al., 2008).

UV spectropolarimetry provides a probe of the physical state of circumstellar matter, unresolved by other techniques, for many stellar environments. Spectral-type Be stars are rapidly rotating massive stars that are laboratories for understanding the physics of hot circumstellar gas disks. The polarization from Be star disks can be used to derive disk temperature, density, and structure, using models of electron scattering modulated by line-blanketed depolarization (Bjorkman, Bjorkman & Wood, 2000).

Most massive stars occur in binary systems, influencing their evolution and possible end-states as supernovae or gamma-ray bursters. UV polarimetry of the wind collision regions of high-mass binaries allow detailed characterization of shock structures and provide constraints on the effects of binarity in massive stellar evolution (Lomax et al. 2015). Magnetized evolved binaries are likely precursors to SN Ia explosions. Polarimetry of Zeeman and Cyclotron lines from these objects provides unique information on the magnetic field characteristics in these systems, and its influence on their evolution. Stokes spectroscopy of the Cyclotron features allow full physical analysis, including temperature, density, accretion rate and white dwarf mass, as well as magnetic field strengths (Thomas et al., 2012).

While still somewhat controversial, the use of Rayleigh scattering from the extended atmospheres of exo-planets may provide a novel way to characterize their atmosphere and a unique way to establish the orientation of the orbital plane. Because of the $\lambda^{-4}$ dependence of Rayleigh scattering, sensitive, time resolved, UV polarimetry is needed to maximize the contrast at the low polarization levels expected and required to resolve the conflicting results to date (Berdyugina et al. 2011, Wiktorowicz, 2009)

### 5.4.3 Revealing Obscured Sources

As scattering redirects radiation, sources whose direct line of sight is obscured can be revealed in polarized light, differentiating between the scattered light and general diffuse emission. This can occur both in circumstellar environments and in active galactic nuclei (AGN). Polarized light from AGN originates as synchrotron emission from jets extending up to several kpc from the nucleus, and is due to scattering by dust or electrons in the vicinity of the AGN, obscured in direct light by the torus (Packham et al. 2011). The accretion disk emission in AGN peaks in the UV, and therefore polarimetry at these wavelengths is crucial to elucidate these sources. The scattering is by small particles yielding $p{\sim}\lambda^{-4}$. Therefore, the polarized component is more easily distinguished from other emission in the UV region. This enhances the contrast between AGN structures, such as scattering cones, and the unpolarized light from surrounding regions in imaging polarimetry (Kishimoto, 1999), affording key spatial information as to the true location of the AGN. The polarized spectrum from AGNs elucidates the nature and opacity of the scatterers and the dynamics of the circumnuclear disk. The large distances and photon needs of UV polarimetry make AGN observations stretch the capability even of HST. UV polarimetry with large UV space-based telescopes will allow studies of the 3D structure of the AGN central engines, accretion disks and surrounding dusty tori, and thereby the evolution of AGN.

### 5.4.4 Summary

A significant discovery space exists for UV polarimetry with large apertures, providing high total photon counts, particularly as part of a broad wavelength coverage. In addition to the examples given here, improvement in theory now guide observations and interpretation to expand the areas of research. Care must, however, be exercised in optical designs to not preclude the inclusion of polarimetry.

## 5.5 Opportunities for Star and Planet Formation Research with a Large-Aperture Optical/UV Telescope - P. Hartigan

Understanding how stars and planets form is a large and growing area of current astronomical research. For the latest HST proposal round (Cycle 22) if we count the panels of Extra-Solar Planets, Debris Disks, Resolved Star Formation, and half of Cool Stars and ISM/Circumstellar Matter as related to this topic, the subject area constitutes a bit more than 1/6 of all the submitted proposals, all in the optical/UV. Research related to extragalactic star formation commands a similarly high fraction. At mm-wavelengths the fraction of star-formation research is even higher, with the panels of ISM/Star Formation/Astrochemistry and Circumstellar Disks/Exoplanets/Solar System accounting for nearly half of the approved proposals for ALMA in Cycle 2. A space-based large-aperture optical/UV telescope would lead the research efforts in star formation in many ways. Figure 5.5.1 summarizes the principal physical processes associated with circumstellar disks and outflows. Understanding these phenomena constitutes much of the current research in this discipline. While we must always consider limitations imposed by dust extinction when assessing any specific star-formation proposal, this figure illustrates why the optical/UV spectral region is essential for investigating how stars and planets form.

Examples where optical/UV observations play key roles include:

- Physics of Accretion Columns: Stars accrete as much as 10% of their final masses from circumstellar disks, and this process affects the masses, coalescence, migration, and composition of protoplanets as they form. Through a process that is poorly-understood, material in the disk is lifted onto the stellar magnetic field and funneled to the stellar photosphere, where it encounters a strong accretion shock. Radiation from this accretion hot spot emits primarily in the UV, and high-spectral resolution of emission line profiles in the optical and UV show both inverse P-Cygni profiles indicative of infall, and normal P-Cygni profiles that result from outflow. A large UV/optical telescope would enable direct observations of this process for thousands of stars, and we may even be able to observe the effects of accretion onto protoplanets in the most nearby systems either with spatially-resolved imaging, or by observing periodic velocity signatures in the emission lines.
- Chromospheric Activity and Young Stellar Winds: Low-mass stars like the Sun are born with active photospheres and exhibit strong X-ray fluxes, chromospheric emission lines and starspots. Radiation from coronal and chromospheric activity and stellar winds strongly influence how atmospheres evolve around planets as they form. A UV/optical telescope that operates in the FUV between 10.2 eV and 13.6 eV is the only way to trace ionized species such as O IV, which bridge the gap between optical-emitting plasmas at $\sim 10^4$ K and X-ray emitting plasmas at $10^6$ K. Monitoring the UV emission lines over time will provide important constraints on flare activity and stellar cycles at the earliest ages.

- Jet Formation and Collimation: We know that accretion drives outflow, but the mechanism by which disks redirect infalling material from a disk into a collimated supersonic flow lies hidden within ~10 AU of the star, just below current angular resolution limits for the closest systems. The phenomena of accretion disks and outflows is universal throughout astrophysics, and is easiest to observe in young stars where the densities are high enough for the flows to radiate emission lines when they pass through shock fronts. In addition, jets from young stars are fast enough to facilitate observable Doppler motions with moderate spectral resolution and to exhibit detectable proper motions over a few years for the closest systems. Issues related to magnetic reconnection of tangled kink-unstable toroidal fields, detecting rotation in jets, learning what causes jets to precess, and understanding why jets eject pulsed flows will only be within reach using a facility like the proposed UV/optical telescope.
- Physical Conditions within Jet Shocks: We have learned a great deal about shock waves in jets in the last decade thanks to the stability and spatially-resolved spectra achievable with HST. The field of ISM plasma diagnostics is over a half-century old, and the same techniques are used to infer electron densities, temperatures, and ionization fractions within each pixel in young stellar jets. The physics of L-S (spin orbit or Russell-Saunders) coupling implies that most of the energy levels of common ions have energy levels of a few eV. For that reason, the best diagnostics of density and temperature are in the optical/UV. Measurements of these line ratios are the best way to compare with the output from modern 3D shock codes, and new theoretical shock models that optical emission line ratios can also be used to infer magnetic field strengths in flows as long as the observations have sufficient spatial resolution (0.2 arcsec for the closest systems) to separate collisionally-excited Balmer emission from the shock from forbidden-line radiation in the cooling zones. Emission line studies of this kind also determine abundance depletions, which clarify the initial conditions present when the star formed as well as constrain dust formation and destruction in disks. While HST has showed the way for this type of research, currently only a handful of jets are close enough and bright enough to resolve the cooling zones of jets and create detailed movies of the MHD evolution of shocked flows.
- Dynamics of Gas Disks: Thermal radiation from dust is the domain of IR and mm-wave observations. Scattering from dust in the optical/UV provides the only means to detect small grains present in the infalling envelope or that become entrained in the outflows. As heavier dust settles to the midplane of the disk, a dust 'wall' should form within a few tenths of an AU from the star where dust evaporates into gas. It is in the gas where most of the mass resides, and we must follow this component if we are to learn how gas-giant planets and their reservoirs evolve. The principal tool used to study $H_2$ in disks is with fluoresced UV emission and absorption lines. These mainly lie in the Lyman and Werner bands between 10 − 13 eV. When these lines are resolved spectrally, they determine whether or not a slow, possibly photoevaporative wind emanates from the disk (e.g. Figure 5.5.1). Such studies have only been done to date for the closest gas disks. A large-aperture UV/optical telescope should be able to resolve the region where photoevaporative flow is launched, and provide an actual image of the gas in another solar system on inner planet scales.
- Binaries: Roughly half of the stellar systems in our galaxy are binaries. Studying binary

fractions at close separations at the earliest ages is the only way to learn how binary separations, and thereby the angular momentum distribution of the systems, evolve with time. At large separations each stellar component can have its own disk, but these become truncated once the separations become on the order of the disk sizes. Numerical models have shown that eccentric systems can force pulses of accretion at periastron, and these close passages may also generate spiral waves and disk warps. The structure of binary jets is completely unknown on small scales. Research into these areas requires high-contrast imaging at the finest spatial scales possible, and is ideal for an optical/UV space-based telescope. There is even the possibility of using eclipse-mapping techniques once enough close binaries are found.

Figure 5.5.2 summarizes the impact that a 10-m class optical/UV telescope would have upon spatially-resolved studies of star and planet formation. The colors in Figure 5.5.2 correspond to the various scales depicted in Figure 5.5.1. For example, in the optical at the distance to Taurus/Auriga, HST currently resolves scales of ~5 AU, at the boundary between the envelope and disk scales depicted in Figure 5.5.1. A 10-m class telescope would push this boundary all the way out to Carina, where the number of objects increases from ~100 to over 30000, enabling statistically-significant studies of star formation phenomena. Pointing a 10-m optical/UV telescope at a disk in Taurus would resolve Earth-like orbital distances of ~1 AU. Note that the disk/jet emission of interest on these spatial scales will be located close to a very bright point source, a difficult task for any future ground-based AO system. The potential for new discoveries related to a particular object grows as the distance to the object decreases, both from the standpoint of improved spatial resolution and from being able to observe lower flux levels with better signal-to-noise ratios. Most stars form in regions that contain massive stars, so to understand star formation on a global scale we must extend our investigations to these types of regions, which, as shown in Figure 5.5.2, are more distant than their lower mass counterparts. Young stellar disks and planetary systems born in the vicinity of massive stars will evolve quite differently from those that develop in regions that are largely shielded from external radiation. Regions with massive stars emit prodigious amounts of UV radiation, create cavities and H II regions in their nascent molecular clouds, and drive powerful winds. All of these processes affect the amount of mass that is available to form disks and planets around the lower mass stars. The shape of the IMF implies that all high-mass star-forming regions produce large amounts of low-mass stars as well. Hence, statistical studies are best done in such regions. The takeaways here are (i) currently the best spatial resolution in the optical/UV stops ~5-10 AU for the nearest disks, just where planetary orbital phenomena become manifest: a large-aperture optical/UV telescope will bring the blurred, inaccessible inner planet regions of disks into focus for the first time; and (ii) new phenomena observed in the last decade at the limit of HST's capabilities have transformed how we draw the star-formation scenario in Figure 5.5.1: the proposed telescope will extend studies of these phenomena from a handful of systems to thousands, and enable studies of entire massive star formation regions that have to date only been possible for isolated, low-mass regions.

On the spectral frontier, higher spectral resolutions are required to probe rotational broadening at larger distances from the star (Figure 5.5.3). In this sense, spectroscopy and imaging are complimentary - it is more difficult to image close to the source, but easier to resolve orbital motions there. A spectrometer with a spectral resolution of ~$10^5$ will resolve thermal widths for all gas in the warm disk. Lower spectral-resolution suffices for gas in H II regions or behind strong shocks. Having a slit-spectrograph is crucial - most projects will want to observe how plasma processes vary across the disk and along flows.

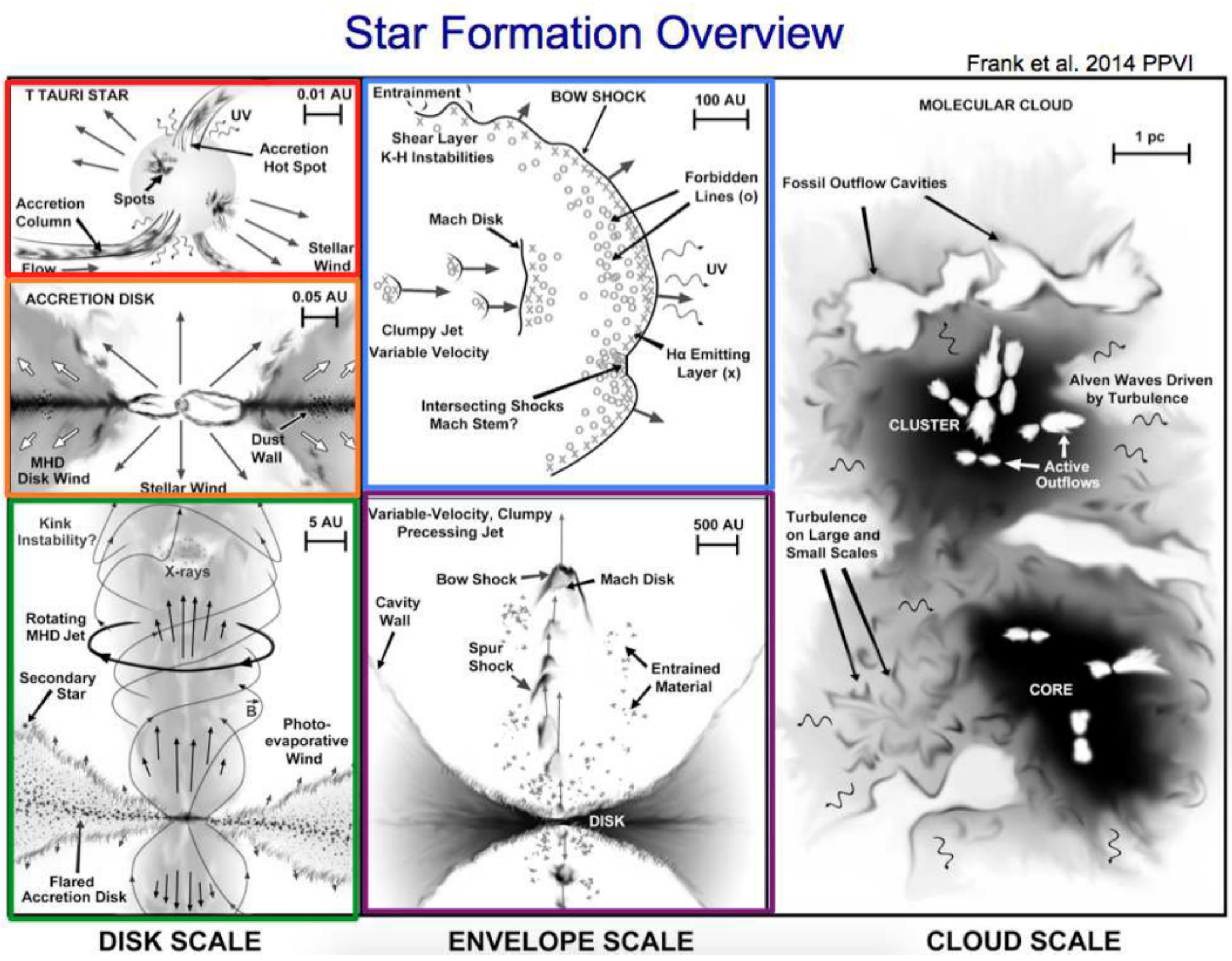

**Figure 5.5.1** — Star formation summary, adapted from Frank et al. (2014, Protostars & Planets VI). The drawing is to scale, and summarizes the main physical processes associated with accretion disks and collimated jets. The colors that outline each panel are also used in Figures 5.5.2 and 5.5.3.

## Spatial Scales

Increasing Mass/Size of Region →

Statistical Studies Possible →

| | Gas Disk | Cloud with Dusty Accretion Disks | Cluster with a few O stars | Cluster with >~100 of O stars | Starburst Cluster |
|---|---|---|---|---|---|
| Nearest Example: | TW Hya | Taurus/Auriga | Orion | Carina | 30 Doradus (LMC) |
| Distance: | 54 pc | 140 pc | 410 pc | 2.3 kpc | 49 kpc |
| # of Stars in Group: | Isolated | 50 | 1000 | 30000 | >30000 |
| 1.2 λ / D HST UV (2000 A): | **1.1** AU | **2.9** AU | **8.5** AU | **50** AU | 1000 AU |
| HST Optical (5000 A): | **2.8** AU | **7.2** AU | **21** AU | **120** AU | 2500 AU |
| 1.2 λ / D 12-m UV (2000 A): | **0.22** AU | **0.58** AU | **1.7** AU | **9.5** AU | **200** AU |
| 12-m Optical (5000 A): | **0.56** AU | **1.4** AU | **4.2** AU | **24** AU | **500** AU |

Considerations

- What emits/absorbs in UV/opt
- Extinction
- Spectral information can be more important than spatial resolution especially in red and orange zones

UV

Accretion Columns
Chromospheres
Gas Fluorescence/Absorption
Photoevaporation
Jet Bow Shocks
HII Region dust distribution

Optical

Accretion Columns
Inner Disk Wall
Jet Acceleration
Jet Collimation
Resolved Disk Imaging
Binary Disk Interactions
Photoevaporation
Jet Rotation
Cooling Distances in Jets
Jet Bow Shocks
Disk Imaging/Detection
HII region cavities

**Figure 5.5.2** — Spatial scales associated with star formation regions at different distances. Each example is the closest for its category. For example, the closest region with accretion disks is Taurus/Auriga. Regions to the right contain objects like those on the left. That is, Carina has several clusters of O-stars each of which resembles Orion, and Orion has many accretion disks like those in Taurus, and Taurus has gas disks like TW Hya. The colors refer to the depictions in Figure 5.5.1.

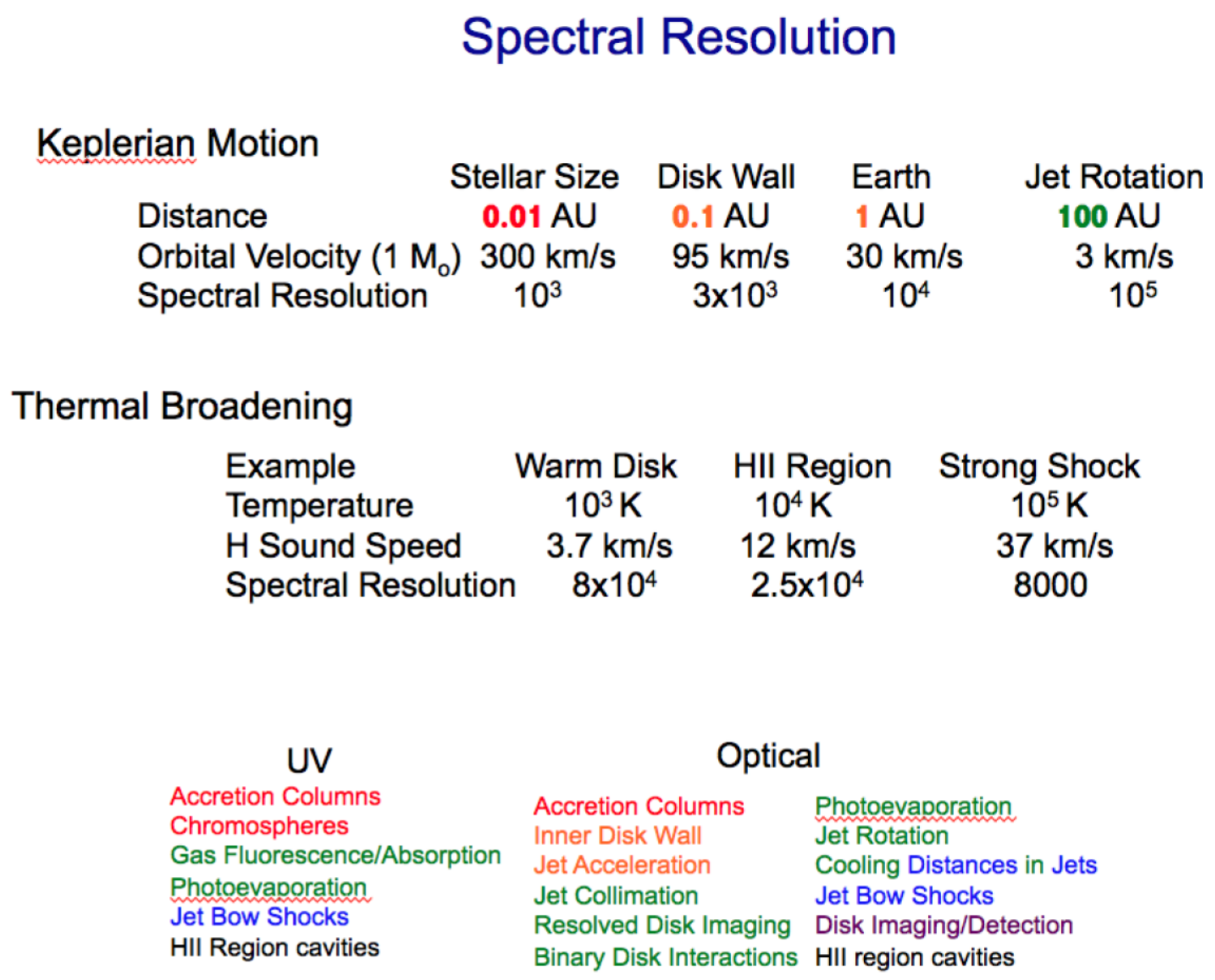

**Figure 5.5.3** — Similar to Figure 5.5.2, but for spectral resolution. The colors refer to size scales in Figure 5.5.1.

## 5.6 Solar System Astronomy with a Large Aperture Space Telescope – Walt Harris

Our solar system is the best-studied example for planet system evolution, starting from a protoplanetary disk to the end state planetesimal reservoirs, planetary types and location, and the evolution of the system via migration, tidal forces and solar forcing. Remote sensing has been enormously useful both for improved understanding of the individual objects at the system-level and for probing their origins and subsequent evolution. Study at the level of systems, including atmospheres, satellites, magnetospheres, and their mutual interactions with each other and the Sun, is key to narrowing the parameter space of planetary formation and how their modern states respond to their external environment. This can be summed up with a series of *grand questions*.

1) *How does the interaction with the Sun and Solar Wind affect the movement of mass, momentum, and energy across boundaries in a planetary system?*
2) *How do solar driven energetic processes affect atmospheric structure, composition, dynamics, and evolution?*
3) *How does the presence and/or release of internal energy affect surface process, atmospheric dynamics, and magnetospheric environment?*

4) *How does the solar driven Earth system compare with the other planets?*
5) *What was the initial volatile composition of planets, their satellites, and other small bodies and how have they changed with time?*

Space based remote sensing of the solar system is an important complement to robotic visitation by virtue of the scale of study it provides. The number and diversity of potential targets in the solar system means that there are far more objects of compelling interest than can be visited *in situ*. Thrust capacity restrictions limit the number of objects that can be visited quickly and/or co-orbitally, which means that most encounters must be either fly-by events (e.g. New Horizons) or require a long cruise phase (e.g. MESSENGER, Rosetta). Even for cases where spacecraft spend an extended period near an object, they are typically limited in the type and capabilities of the instruments provided, and often by their limited perspective relative to the entire system under study. Global monitoring of multiple bodies is enabled from remote sensing, with relative quality of the observations dependent on the aperture size of telescope used.

The ultraviolet to near-infrared spectral range is particularly valuable for remote sensing of solar system objects because it encompasses the spectral ranges dominated by the solar continuum (300-3000 nm) and emission line dominated energetic emissions from the Solar and planetary environments. Reflected visible continuum provides high signal to noise measurement of atmospheric dynamics, surface characteristics and composition, and near space volatiles in magnetospheres and comets. The continuum-free UV extends access to fainter emissions from atmospheric airglow, space weather, auroral processes, exospheric and cometary atomic species, magnetospheric plasmas, and upper atmospheric structure.

A diffraction limited 10+ meter class observatory would have 20x or greater sensitivity than the Hubble Telescope and would be capable of resolving features as small as 8 to 230 km over distances from 1 AU out to the orbit of Pluto. With this combination of features, such a telescope would be able to deliver image quality comparable to or better than that achieved with the Voyager spacecraft throughout the region of the solar system occupied by the major planets. The different regions of its spectral range would enable a variety of detailed studies that would exceed combination of temporal and spatial coverage that has been possible to date.

### 5.6.1 Plasma Processes:

Ultraviolet emissions from planetary aurorae, airglow, and magnetospheric plasmas has been a staple method for monitoring energetic processes in planetary atmospheres and their response to solar activity. The Space Telescope Imaging Spectrograph (STIS) is currently the workhorse of these studies, and will continue to provide supporting remote sensing data after existing missions (e.g. Cassini) are complete and as new missions (e.g. JUNO) begin. STIS has proven effective at monitoring emissions from Mars, Jupiter (Figure 1), and Saturn, but is insufficiently sensitive to emissions at Uranus, Neptune, and diffuse magnetospheric sources. Combined with improvements in detector efficiency and dynamic range, a large space telescope will increase sensitivity by 100 fold to enable rapid acquisition of auroral, airglow, and plasmasphere measurements throughout the outer solar system.

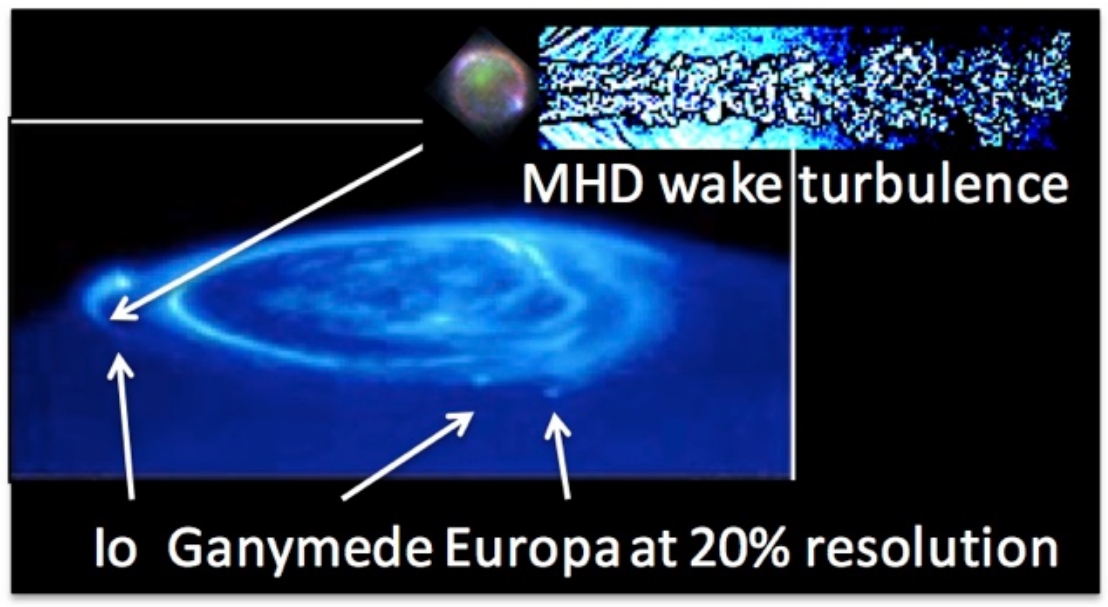

**Figure 5.6.1 -** STIS observations of the Jovian auroral at 20% of the predicted next generation resolution show dawn-dusk signatures of interaction with the solar wind.

In addition, field aligned current interactions with Io, Europa, and Ganymede are revealed from ‘footprints’ South of the main oval. Inset shows the auroral signature of Io’s interaction with Jupiter. Observable wake turbulence in the Io torus is simulated at right.

### 5.6.2 Surfaces and Activity:

Signatures of ices, minerals, and activity on planet, satellite, and small bodies can be observed spectroscopically across the full spectral range of a large aperture telescope. Current detection limits are limited to small bodies in the inner solar system and unresolved structures on larger satellites of the giant planets. The volatile surface composition of small bodies has been only sparsely examined with current technology.

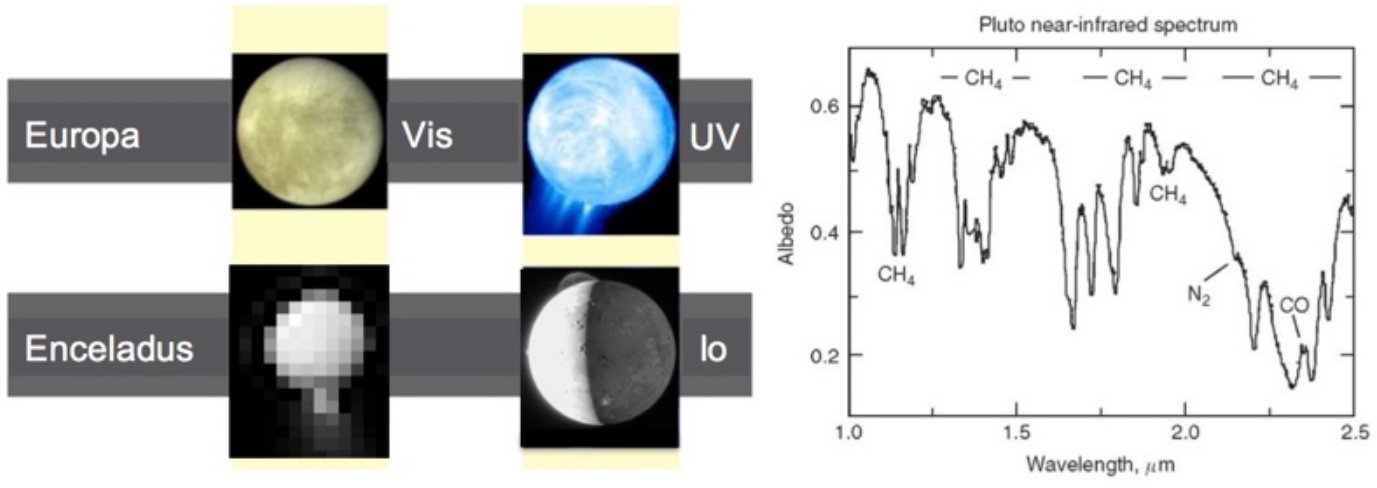

**Figure 5.6.2 -** Examples are shown of the surface detail detectable on several planet satellites, along with geyser/volcanic activity. NIR observations can obtain surface ices on objects as small as 10 km in radius (Grundy et al. 2013).

### 5.6.3 The Outer Solar System Inventory:

The Hubble Telescope is currently the most sensitive instrument for small body detection in and beyond the Kuiper Belt. A future 10 m class observatory should retain that advantage, even over 30 m ground based facilities because of the reduced background. Such an observatory would enable detection of ~ 1 km sized objects throughout the Classical Kuiper Belt and spectroscopic characterization of objects up to 10 km. The range over which Dwarf planets (R ~ 500 km) would be detectable would extend to the perihelion of the Sedna-class objects. Possible larger planets ranging in size from Super-Earths to Neptune-sized could be detected out to distances extending into the inner regions of the Oort Cloud (5000-15000 AU).

## 5.7 The Galactic Baryon Cycle: CGM Drivers for a Large UV-Optical-IR Space Telescope – Chris Howk

The flows of matter between galaxies and their surroundings are thought to be critical to the overall evolution of a galaxy. A galaxy’s star formation and even AGN activity is coupled to the rates and manners in which gas flows away from and into the system. This includes exchanges of matter within the disk-halo interface of a galaxy, between the central regions of a galaxy and its extended gaseous corona, between the galaxy’s halo and its satellites or group, and between a halo and the filamentary IGM beyond.

Understanding gas flows through the circumgalactic medium (CGM) is thus a crucial piece of the whole picture of galaxy evolution. We have seen over the last few years the huge impact new capabilities for studying the CGM can have, as the unprecedented UV sensitivity and good resolution of the Cosmic Origins Spectrograph (COS) have led to some extraordinary new insights into (and unanswered questions about) the CGM about $z \lesssim 0.5$ galaxies and its role in their evolution:

1. The CGM about z $\lesssim$ 0.2 hosts a large share of galactic baryons, with $M_{CGM} \sim M_\star$ for ~L* galaxies (Stocke et al. 2013; Werk et al. 2014). When were those baryons first put in place?
2. The CGM is a massive reservoir of galactic metals, with galaxies having ejected at least as much metal mass as they have retained (Tumlinson et al. 2011; Peeples et al. 2014; Bordoloi et al. 2014). What mechanisms drive these metals outward? are they destined to be recycled back to the ISM (Keeney et al. 2013; Ford et al. 2014)?
3. Dense, ionized streams of gas in the low-redshift CGM of galaxies have a bimodal metallicity distribution, with roughly half having metallicity ~20-100% solar and half exhibiting extremely low metallicities of <5% solar (Lehner et al. 2013, Wotta et al. 2016). The low-metallicity gas resembles the long-sought "cold accretion" entering galaxies from the IGM (e.g., Kereš et al. 2005, Nelson et al. 2015). However, does the presence of metal-poor streams affect the star-forming properties of galaxies?
4. Galaxies can drive complex multi-phase superwinds containing as much mass as the cold ISM with $T\sim10^4$–$10^6$ K that flow deep into their halos (Tripp et al. 2011). How common are such superwinds? do they require recent AGN activity? How are these related to the more quiescent winds seen in z ~ 0.5 – 1 and higher galaxies (e.g., Du et al. 2016, Rubin et al. 2014).
5. There is significant cold, bound gas surrounding passive galaxies at z~0.2, perhaps as much as seen in their star-forming counterparts (Thom et al. 2012). What keeps this gas from forming stars? how does the CGM affect (or is affected by) the quenching of star formation? Why are the covering factors of metal lines so much lower (Tumlinson et al. 2011, Huang et al. 2016)?

A future large telescope with FUV capabilities provides a path to answering many of these questions. With such a facility, we would:

**Map the ionization state and metallicity of the CGM for various types of galaxies**: Our current "maps" of the CGM are largely statistical (e.g., Tumlinson et al. 2011, Borthakur et al. 2016); very few individual galaxies having multiple probes of individual galaxies (Keeney et al. 2013, Chen et al. 2014, Lehner et al. 2015). At the same time, our ability to study even statistically the CGM of rare, but critical, types of galaxies such as extremely massive systems, post-starburst galaxies, or AGN host galaxies is limited by the small surface densities of QSOs on the sky that are bright enough for UV absorption line spectroscopy with COS. Figure 6.1 demonstrates that non-linear gains can be made in the number of QSO-galaxy pairs able to be targeted just by moving to larger apertures and/or enhanced system throughput. A future LUVOIR would facilitate "mapping" the CGM of individual galaxies by using absorption line spectroscopy toward multiple QSO or even background galaxy sight lines; it would push our ability to study the CGM into the epoch of galaxy transformation, testing the putative connection between the CGM and galaxy quenching; it would hopefully allow emission line imaging or multi-object spectroscopy to elucidate the morphology of the flows of energy and matter through galaxies' CGM (see §5.10). As shown in Figure 6.2, the majority of ionization, metallicity, and mass diagnostics are found in the FUV or EUV. Capturing the CGM diagnostics over a full range of temperatures/ionization states for a broad redshift range requires good UV sensitivity over the range 1000 Å $\leq \lambda \leq$ 3000 Å, if not to the Lyman limit. Point source spectroscopy should reach $R \geq 20{,}000$; a capability for R ≥ 75,000 – 100,000 for local sources (Local Group) is preferred.

**Connect the CGM to IGM filaments and the general environment of galaxies:** What is sometimes left unsaid is the assumption that flows in and out of galaxy halos imply the transport of matter across the virial radius. A future LUVOIR facility with a moderate resolution (*R* ~ 5,000 to 10,000) survey mode, preferably coupled to a multi-object capability, could provide the path to connecting the CGM properties of galaxies directly to the cosmic web through absorption line spectroscopy, as well as crudely map the CGM itself. One goal of such work would be to study the interaction at the CGM/IGM interface, which may reside near the virial radius, depending on galaxy mass and coronal structure (Nelson et al. 2015).

**Probe the structure, metal content, and energetics of galactic winds from their origins in the disk-halo interface of galaxies to the radius at which they stall (if they do):** Understanding the driving of flows through the disk-halo interface (the interstellar thick disk), to the galactic corona/CGM, and potentially further will take a combination of multi-object absorption line spectroscopy of the potential launch points of disk-halo flows in nearby galaxies (OB stars/associations) and absorption+emission spectroscopy of wind diagnostics in cosmologically-distant galaxies (both “down-the-barrel” and toward background sources). Understanding the dynamics of such outflows, especially through the CGM, likely requires that we both probe the driving fluids (seen in the very high ions – e.g., Ne VIII and Mg X – and through X-rays with *Athena*) and trace the energetics (both the thermal balance through cooling as well as the kinematics).

Ultimately the groundbreaking observations required to make better sense of the role of the CGM in galaxy evolution involve studying specific stages of galaxy evolution with a full range of ions, which drives the need to do absorption line spectroscopy in the FUV, down to at least 1000 Å, and perhaps to the Lyman limit. Falling short of this would make it impossible to study the highest-ionization (in some cases hottest) CGM gas and the probes of energy balance in the CGM. Perhaps the most impressive leaps, however, will come from an imaging or spectral slicing capability that allows us to understand the morphologies of streams about galaxies, as discussed in Section 5.10.

## 5.8 Motivations for a future UV-visible telescope: High-redshift Galaxy and Deep-field Studies - Marc Rafelski

When considering high redshift galaxy surveys in the context of a future UV-optical space telescope, it is important to consider what other dedicated missions will already have accomplished. The two largest such missions include the James Webb Space Telescope (JWST) and the Wide-Field Infrared Survey Telescope (WFIRST). The JWST will survey small areas of the sky with a large collecting area at high resolution in the near-infrared through mid-infrared, while the WFIRST will likely survey a large area of the sky at low resolution in the near infrared. Therefore, future high redshift galaxy survey missions must consider this area of parameter space as already addressed, and consider questions that cannot be accomplished in the infrared or at the resolution of these telescopes.

A large optical-ultraviolet (UV) space telescope will have 5 times the field of view (FOV), 5 times the resolution, 25 times the sensitivity of the Hubble Space Telescope (HST), and cover the UV through optical wavelengths not accessible with other telescopes. These properties will enable such a telescope to answer questions not possible with JWST or WFIRST, as well as new questions that will be asked with their future scientific results. JWST and WFIRST will excel in studying galaxies at the earliest times approaching reionization, and the older stellar populations during the peak epoch of star formation by measuring the rest-frame optical light of galaxies. However, these telescopes are unable to measure the rest-frame UV light from galaxies in the last 10 billion years of the universe.

The rest-frame UV light measures a different stellar population – the young massive stars from the star-forming regions of galaxies at the peak epoch of star formation and thereafter. Measurements of this light would enable us to address many important questions, including four addressed here: 1) the role of sub-galactic clumps in the build-up of galaxies 2) the properties of low-mass galaxies as a function of environment 3) the Lyman continuum escape fraction and 4) the role of neutral HI gas via Damped Lyman-alpha systems (DLAs) in the formation and evolution of galaxies.

The resolution of the proposed telescope enables the measurement of sub-galactic clumps at scales of ~100 pc to learn about their role in the assembly of galaxies. We wish to measure the lifetime of these clumps, whether they form in the centers or edges of galaxies, if they migrate over time, and if they are connected to the accretion from the cosmic web. These clumps can only be studied in the rest-frame UV, as they are young star forming regions. In addition, the rest-frame UV light is necessary to break degeneracies in physical parameters obtained from stellar population modeling, such as mass, age, dust extinction, and star formation rates (SFRs). While JWST will be able to measure them at the earliest times in the universe, those galaxies are not yet sufficiently developed, and due to the cosmological scale factor and a smaller mirror, their effective resolution is significantly worse. Moreover, studying the galaxies at lower redshifts increases the sample time frame for measuring the evolution of the clumps over time. Measurements of the clumps later in time will in turn enable interpretations of galaxies measured by JWST earlier in time.

The large FOV and large collecting area enables the proposed telescope to survey a large area of the sky down to exquisite depths, which will for the first time allow studies of lower mass galaxies at the peak epoch of star formation. The diverse star formation histories (SFHs) of dwarf galaxies suggest that environment plays a key role in the development of low mass dwarfs, and therefore we need large samples of these very faint galaxies in different areas on the sky. While we will obtain redshifts for massive dwarf galaxies with JWST and WFIRST, large samples of low-mass dwarf galaxies will still rely on photometric redshifts, for which rest-frame UV light is essential to remove redshift degeneracies. In addition, simulations predict that low-mass dwarf galaxies have very bursty star formation, which would result in a large scatter in the rest-frame UV to H-$\alpha$ ratio. Therefore the SFRs of such galaxies measured purely in H-$\alpha$ with JWST and WFIRST will not reveal the SFHs of these galaxies, and the physical parameters from stellar population modeling that depend on the SFHs. On the other hand, measurements of the rest-frame UV to H-$\alpha$ ratio and the dispersion in the rest-frame UV will inform us of the SFH of the galaxies, and enable the measurement of individual star-forming regions within the dwarf galaxies.

JWST will do an exquisite job of studying galaxies just after reionization, yet it will likely not be able to determine what actually reionized the universe. The light that reionized the universe is at wavelengths bluewards of the Lyman-limit, and therefore in the rest-frame UV. While JWST measures this light from galaxies right after reionization, the opacity of the intergalactic median (IGM) absorbs the light before it reaches us. Therefore, the Lyman continuum escape fraction necessary to understand reionization is best measured at later times, which requires an UV to optical telescope. Currently, HST has measured only limits on the escape fraction, suggesting that it is small, and therefore requires deeper imaging. In addition to a detection of the escape fraction, we wish to characterize the physical parameters of galaxies as a function of their escape fraction. For instance, low-mass dwarfs may have a much higher escape fraction than their more massive cousins. This therefore requires very deep imaging over a large FOV.

The role of neutral atomic hydrogen gas in the formation and evolution of galaxies is currently poorly constrained, yet at early times in the universe it provides most of the fuel for star formation. At this time, most of this gas resides in DLAs, yet we know relatively little about their sizes, morphology, and SFRs since they are only observed in absorption to background quasars rather than in emission. These quasars then prevent direct imaging of DLAs by providing a very bright object right at the location one would wish to observe the faint DLAs. One method to overcome this challenge is to select systems with two DLAs, where the first one blocks all light from the quasar blueward of its Lyman limit, enabling direct measurements of the second DLA before its Lyman limit. The emission from these systems is extremely faint and has not been detected with HST, and the wavelength window enabled by this physical setup shrinks earlier in time. Therefore, a UV-optical telescope capable of measuring very low levels of emission is needed. Such measurements would for the first time inform us of the sizes, morphology, and SFRs of the DLA host galaxies, which would tell us about the cold gas reservoirs fueling star formation.

These four scientific questions require a telescope sampling UV through optical light at high resolution with a large FOV. These requirements are met with a large (~10m) UV-optical space telescope, which would enable large imaging surveys providing new insights into the formation and evolution of galaxies at the peak epoch of star formation.

## 5.9 The First Stars and Heavy Metal Production - Ian U. Roederer

Understanding the origin of the elements is one of the major challenges of modern astrophysics. This is expressed in several of the Cosmic Origins science goals. The approach advocated here is based on obtaining high S/N and high spectral resolution data of late-type (F-G-K) stars to infer their detailed abundance patterns. This chemical information can then be related back to the physics of element production in early supernovae or other stellar sites. This approach was not really covered by the AURA report.

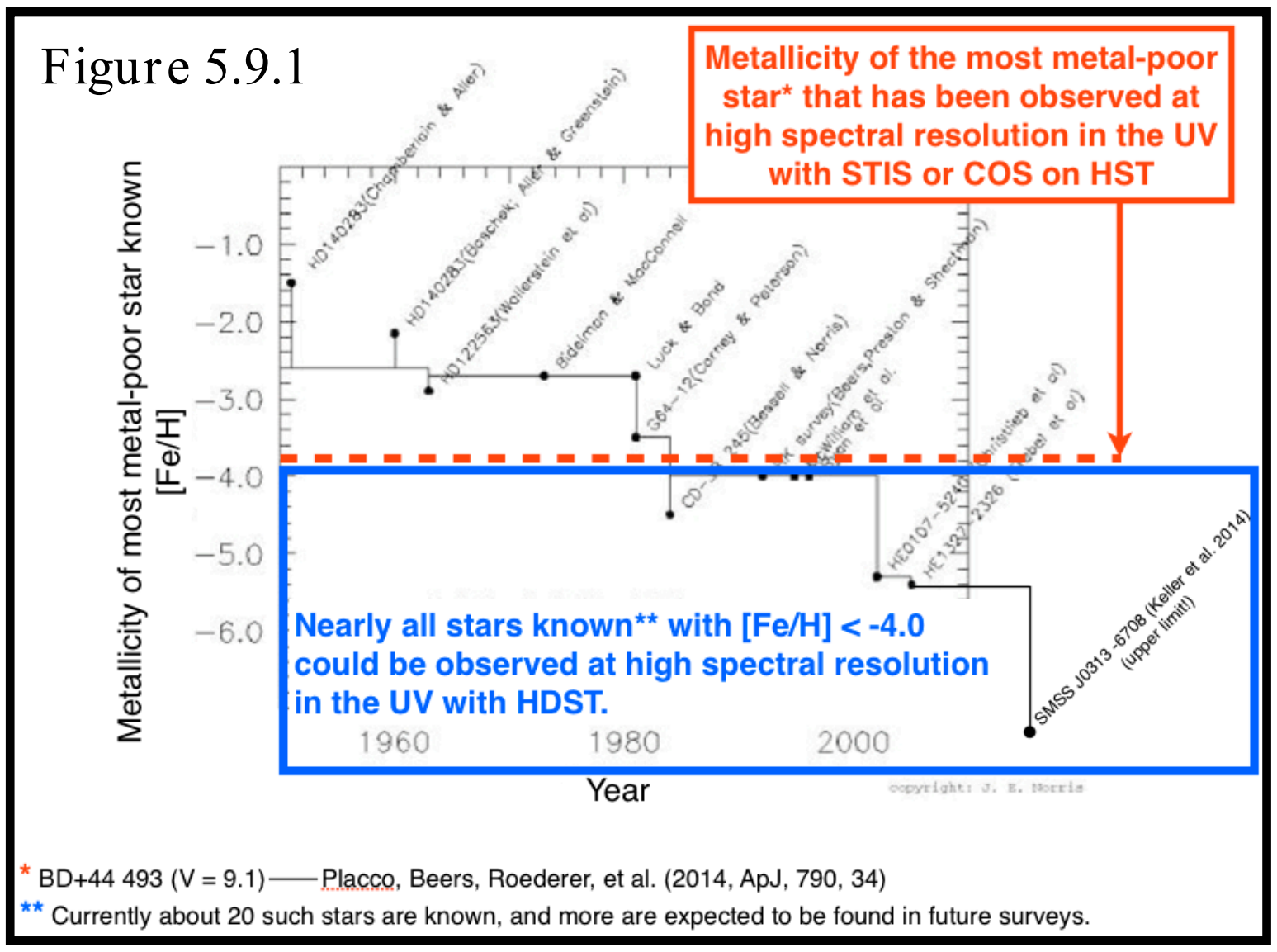

Our group and others have used HST to demonstrate that UV data can be collected and that they are useful for understanding stellar nucleosynthesis and Galactic chemical evolution. The challenge, of course, is that with HST we are limited to the very brightest stars. Figure 5.9.1 illustrates the situation. The lowest-metallicity star that has been observed at high spectral resolution with HST in the UV sits at the red dashed line. All of the metallicity regime below that will likely remain completely unexplored during the era of HST. But, with something like the High Definition Space Telescope (HDST), you could observe nearly all of the lowest-metallicity stars known. There are about 20 such stars known at present, and we expect that more will be found in future surveys.

Why is the UV important? The physics of stellar atmospheres dictates that most of the absorption lines are found in the UV, short of the atmospheric cutoff. Figure 5.9.2 illustrates this point for just two narrow spectral ranges, one in the optical on the left, and one in the UV on the right. The top panels show the spectrum of a solar-metallicity G-dwarf; the middle panels show the spectrum of a normal low-metallicity subgiant, and the bottom panels show the spectrum of one of the second-generation stars. The bottom right panel is a simulation. Notice how rich the UV spectrum is. In the star shown at the bottom here, a total of 18 metal lines can be detected in the optical; our estimates indicate that approximately 300 metal lines could be detected in the near-UV spectral range. This order-of-magnitude increase more than doubles the number of elements that can be detected. This is how you make serious progress in understanding the end states of Pop III stars and the distribution of the first metals in the proto-Milky Way.

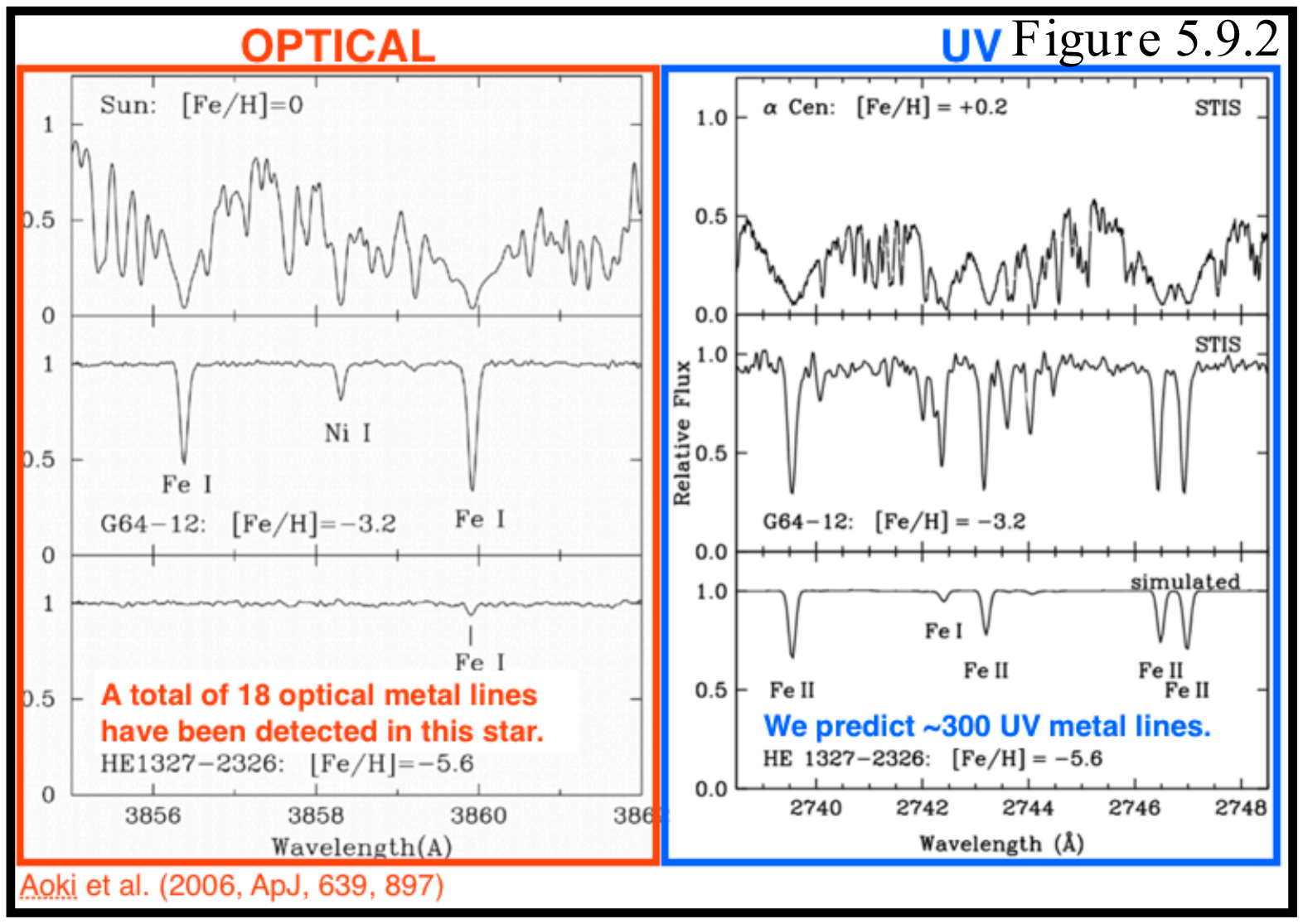

Figure 5.9.2

The second science case is more focused on understanding the origins of the elements heavier than iron, those formed by neutron-capture reactions like the r-process and s-process. One way to characterize the physical conditions at the nucleosynthesis sites is to study the chemical products. Access to the UV spectrum enables a 40% increase in the number of heavy elements that can be detected (from ~25 to ~40 elements). This includes some of the elements that provide the most sensitive constraints on the nucleosynthesis models. Examples may be found in Roederer & Lawler 2012.

Currently, this science is limited to stars with V < 10 or so, and most viable candidates have been exhausted. In an ideal world, we could observe the "gold standard" stars, the ones that have been extensively studied from the ground and have the most potential to expand our understanding of the neutron-capture physics. Figure 5.9.3 illustrates the situation. With HST, we are limited to observing solar-type dwarfs within about 100 pc and giants to a little beyond 1 kpc. With HDST, you can observe dwarfs to 2 kpc and giants out to nearly 20 kpc in areas of low extinction. That allows you to observe most of the stellar halo, numerous stellar streams including parts of Sagittarius, and dozens of globular clusters. In other words, HDST would give you access to the UV spectrum for just about any star you could observe in the optical today from the ground. That finally puts us in a regime where we get to choose the targets based on astrophysical significance, as opposed to observing whatever happens to be available in the solar neighborhood. With HDST, for the first time we would allow nature to dictate the targets, rather than the limits of our technology.

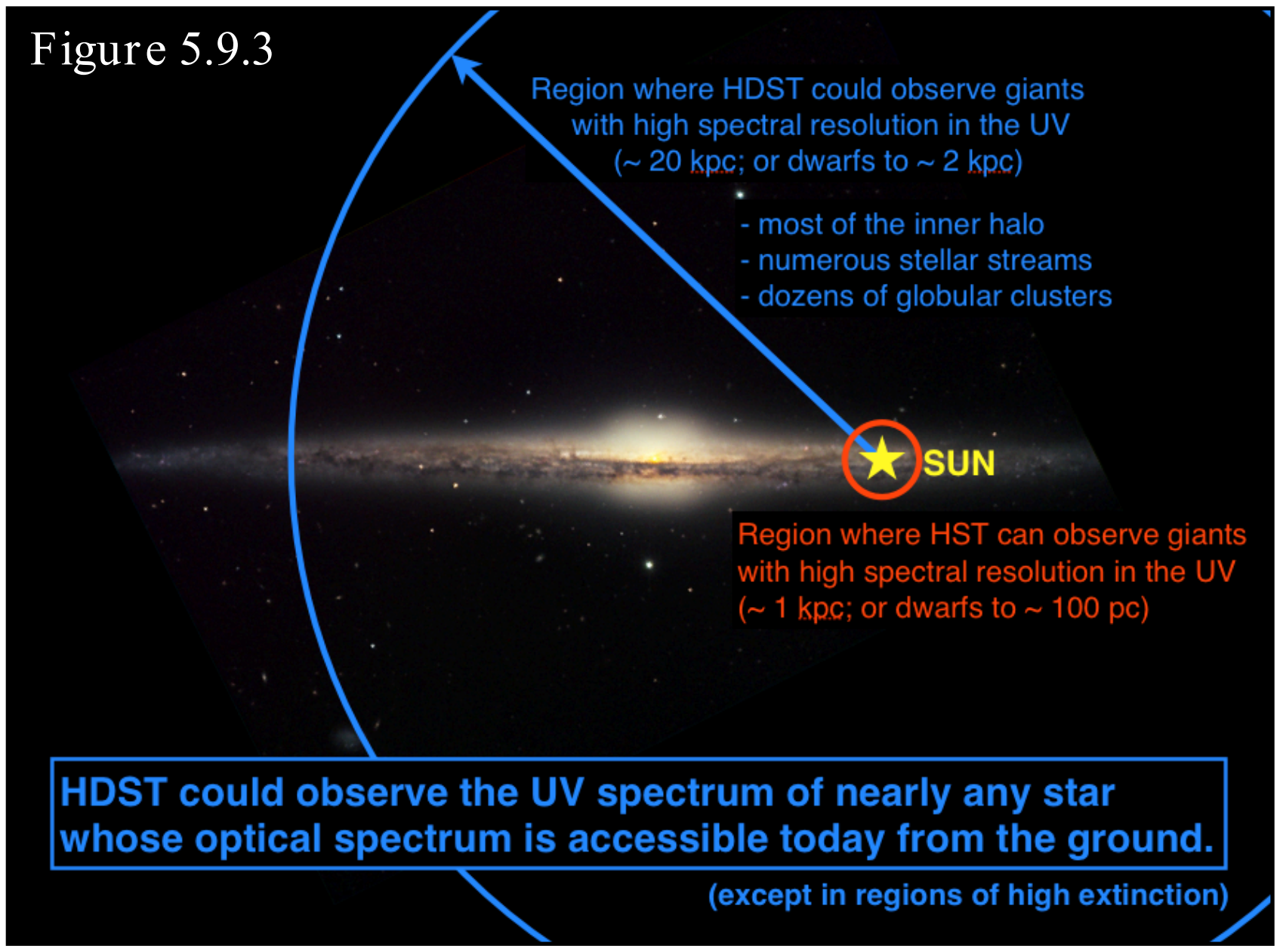

Figure 5.9.3

### 5.10 Galaxy Fueling and Quenching: A Science Case for Future UV MOS Capability - Jason Tumlinson and David Schiminovich

The gas flows that drive galaxy accretion and feedback are critical, but still poorly understood, processes in their formation and evolution. One of Hubble's successes has been in characterizing the circumgalactic medium (CGM) that spans 30 times the radius and 10000 times the volume of the visible stellar disk. Thanks to Hubble and its ground-based optical collaborators, we know roughly how much matter the CGM contains, but the extremely low densities make it difficult to ascertain its exact role in galaxy evolution. The critical question is how this gas enters and leaves the galaxies: galactic star formation rates are limited by the rate at which they can acquire gas from their surroundings, and the rate at which they accumulate heavy elements is limited by how much they eject in outflows. Much of the still-unknown story of how galaxies formed comes down to how they acquire, process, and recycle their gas.

Hubble can make only crude statistical maps by sampling many halos with one absorption-line path through each. The future challenge is to "take a picture" of the CGM using wide-field UV spectroscopy that is > 50x more sensitive than Hubble. A moderate resolution (R~5000), wide-field (3-5 arcmin) multi-object UV spectrograph that can detect this CGM gas directly would revolutionize this subject. This is intrinsically a UV problem since most of the energy transferred by diffuse gas on its way in or out of galaxies is emitted or absorbed in rest-frame UV lines of H, C, O, Ne, and other metals, including rest-frame extreme-UV lines that redshift into the 1000-2000 Å band at $z$ > 0.5 (Bertone et al. 2013; Figure 5.10.1).

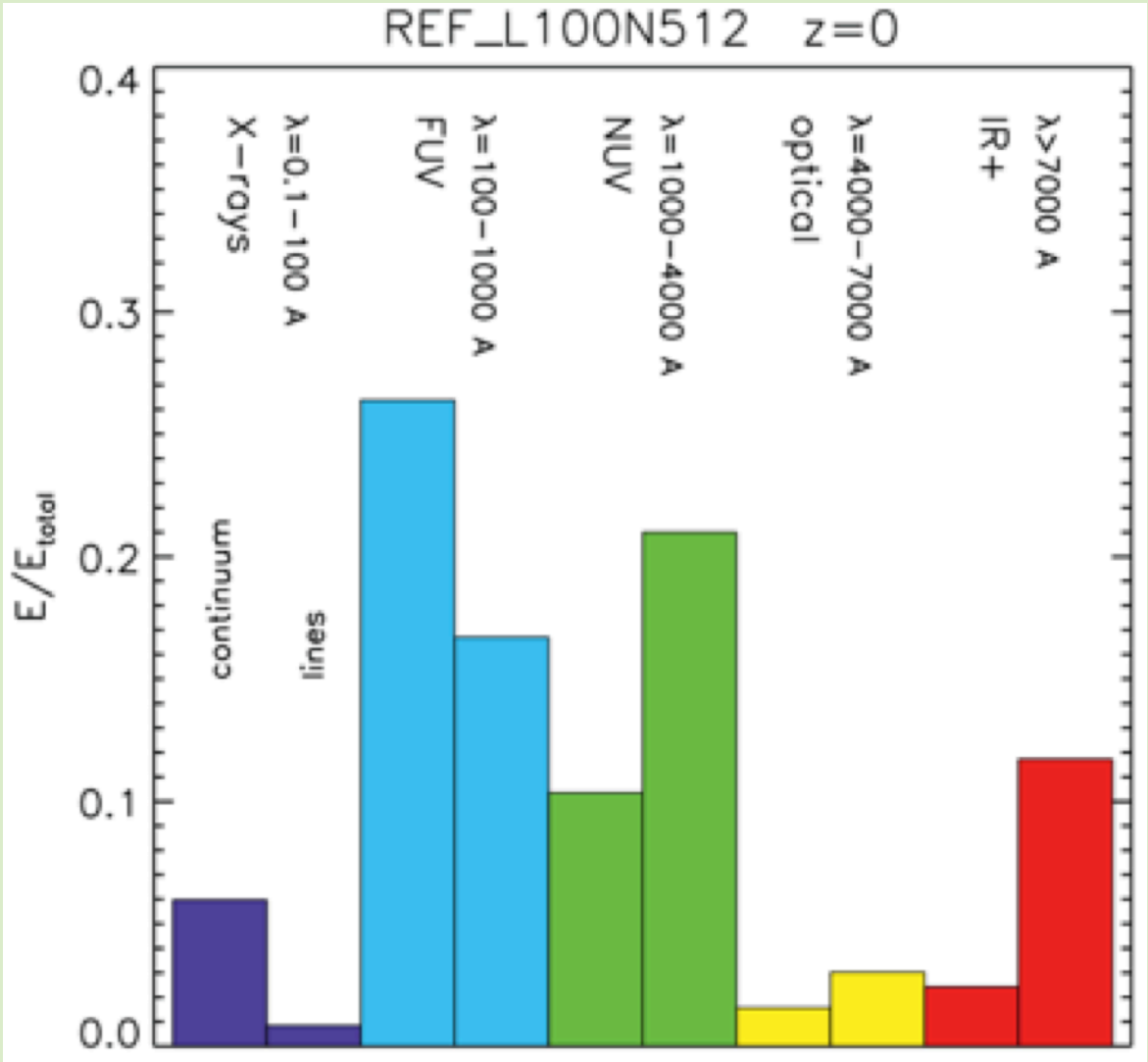

**Figure 5.10.1:** From "How the Diffuse Universe Cools", by Bertone, Aguirre, & Schaye (2013). The bars show the fraction of energy emitted in various rest-frame bands by gas cooling radiation in a hydrodynamical galaxy formation simulation. More than half of the radiation is emitted in the rest-frame FUV (100-1000 Å) and most of the remainder is in the NUV (1000-4000 Å).

If sufficiently large (10-12 meters) and properly equipped, NASA's LUVOIR Surveyor will be able to map the density, temperature, and mass flow rates of the CGM, directly, using the UV radiation emitted by CGM gas as it cycles in and out of galaxies. Observing up to 50-100 sources at a time, LUVOIR could map the faint light ($S_B \sim 100$ photons cm$^{-2}$ s$^{-1}$ sr$^{-1}$) emitted by gas entering and leaving galaxies, to count up the heavy element content of this gas, to watch the flows as they are ejected and recycled, and to witness their fate when galaxies quench their star formation, all as a function of galaxy type and evolutionary state. LUVOIR could map hundreds of galaxies in fields where its deep imaging identifies filaments in the large-scale structure, and where ground-based ELTs have made deep redshift surveys to pinpoint the galactic structures and sources of metals to be seen in the CGM.

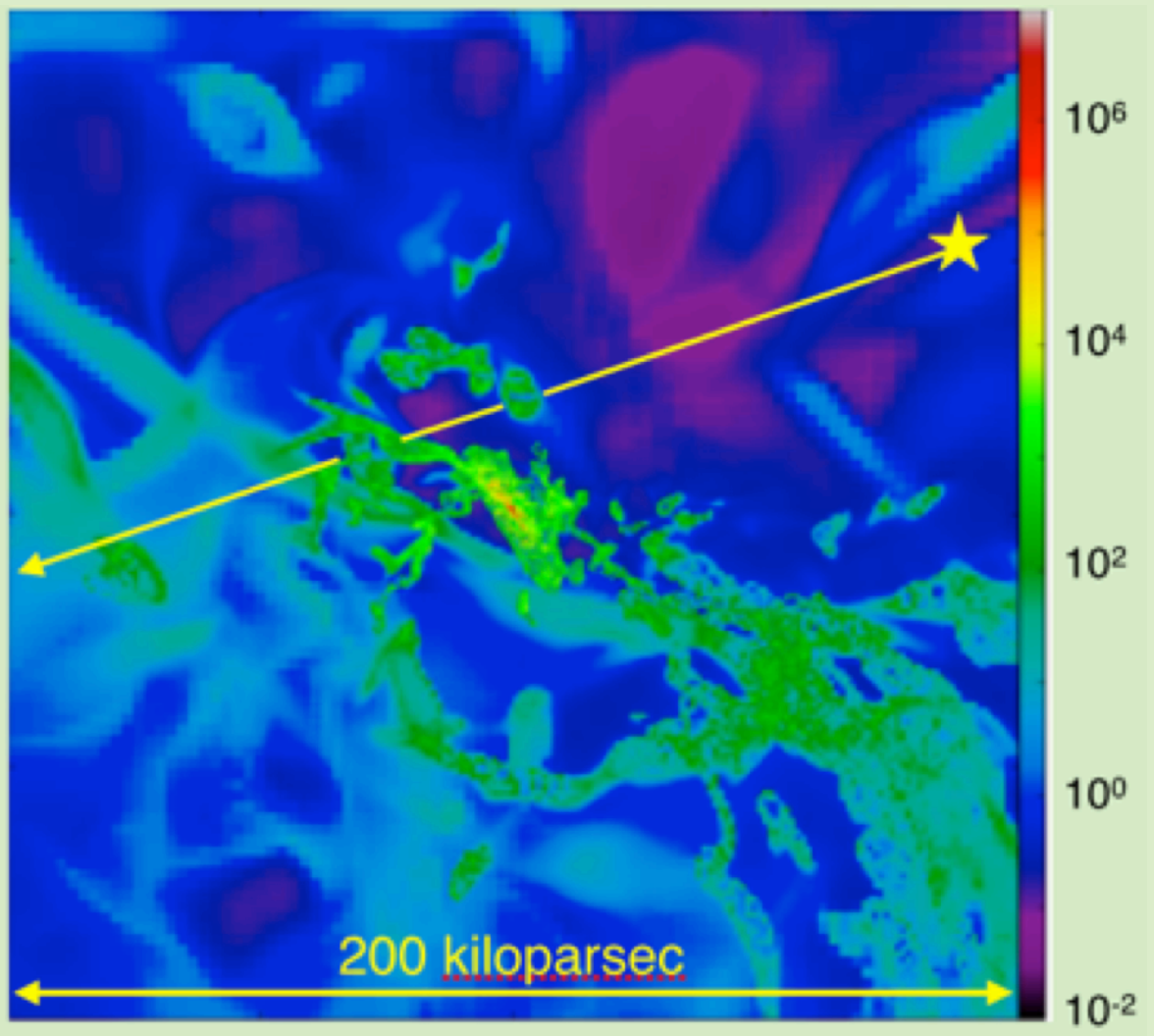

**Figure 5.10.2:** A visualization of the Circumgalactic Medium (CGM) gas fueling a Milky-Way like galaxy at z ~ 0.25. The color renders the strength of UV emission by C IV in the CGM gas in terms of photons $cm^{-2}$ $s^{-1}$ $sr^{-1}$. For real 2D characterization of the CGM, the LUVOIR Surveyor must reach CGM gas with $S_B \gtrsim 100$ units. The contrast between the 2D image of the CGM and the single hypothetical QSO sightline demonstrates the power of imaging the CGM over continuing to develop statistical maps from single sightlines. Image from Joung, Fernandez, Bryan, Putman and Corlies (2012, 2015).

Because this radiation is far weaker than local foreground radiation, ground-based telescopes seeking it at redshifts where it appears in the visible ($z > 2$) must perform exquisite sky foreground subtraction to reveal the faint underlying signal. These foregrounds will be considerably lower from space (by 10-100x), shortening required exposure times by an equivalent factor. Even apart from lower sky backgrounds, accessing the UV provides far better access to the relevant line diagnostics over most of cosmic time: the key lines HI Lyα, CIV 1550, and O VI 1032 are inaccessible from the ground over the last 10 billion years of cosmic time. This includes all cosmic star formation since its z ~ 2 peak, opening for view the complete co-evolution of galaxies and their gas supplies over the period when ~80% of the cosmic stellar mass density was formed (Madau & Dickinson 2014).

This unique UV capability will also address the mystery of how galaxies quench and remain so. The number density of passive galaxies has increased 10-fold over the 10 Gyr interval since $z \sim 2$ (Brammer et al. 2011). Galaxies undergoing quenching are the ideal laboratories to study the feedback that all galaxies experience: the galactic superwinds driven by supernovae and stellar radiation, the hot plasma ejected by black holes lurking in galactic centers, and the violent mergers that transform galaxy shapes while triggering the consumption or ejection of pre-existing gas. Only a 10-12 meter LUVOIR Surveyor would have the collecting area to support deep, wide-field UV MOS searches for CGM gas at the line emission fluxes that are expected, *and* the spatial resolution to observe the transformation of star forming disks to passive spheroids at 50-100 pc spatial resolution and closely examine the influence of AGN on this process. For galaxies identified as quenching, emission maps of the surrounding CGM will determine the fate of the gas that galaxies must consume or eject and powerfully elucidate the physical mechanisms that trigger and then maintain quenching. Only a 10-12 meter space telescope can achieve such spatial resolution in the optical and observe the rest-frame UV light necessary to witness the co-evolution of stars and gas in galaxies undergoing this transition. As most of the development of the present-day red sequence occurred since $z \sim 2$, and the key diagnostics are rest-frame UV lines, this critical problem is a unique and compelling driver for a 10-12 meter LUVOIR Surveyor mission in future decades.

## 5.11 Lyman Continuum Luminosity Function Evolution of Star-Forming Galaxies – Stephan McCandliss

**Science Motivation** – The timing and duration of the Epoch of Reionization is crucial to the subsequent emergence and evolution of structure in the universe (c.f. Madau et al. 1999, Ricotti et al. 2002, Robertson et al. 2015). The relative role played by star-forming galaxies, active galactic nuclei and quasars in contributing to the Metagalactic Ionizing Background (MIB) across cosmic time remains uncertain. Deep quasar counts provide some certainty to their role, but the potentially crucial contribution from star-formation is highly uncertain due to our poor understanding of the processes that allow ionizing radiation to escape into the intergalactic medium (IGM). Moreover, the fraction of ionizing photons ($f^e_{LyC}$) that escape from star-forming galaxies is a fundamental free parameter used in models to "fine-tune" the timing and duration of the reionization epoch that occurred somewhere between 13.4 and 12.7 Gyrs ago at redshifts between $12 > z > 6$.

Galaxy luminosity functions at high redshift (Bouwens et al. 2006; Labbe et al. 2010; Gonzalez et al. 2010; Finkelstein et al. 2014) along with a host of assumptions for the clumping factor, the ionizing output and the initial mass function of the first stellar assemblages, have been used to constrain $f^e_{LyC}$ to 0.2, fulfilling the requirement to power the EoR – provided contributions to the LyC from the unobserved population of galaxies at the faint end of the luminosity function are included. Of all these assumptions, the uncertainty in $f^e_{LyC}$ is universally acknowledged as the least understood parameter (Ellis 2014), requiring observation for quantification.

Ionizing radiation escape is a mysterious process. Heckman et al. (2001) have pointed out that mean galactic column densities for H I range from $10^{21}$ cm$^{-2}$ for normal galaxies to $10^{24}$ cm$^{-2}$ for nuclear starbursts. Yet it only takes a H I column density of $1.6 \times 10^{17}$ cm$^{-2}$ to produce an optical depth $\tau = 1$ at the Lyman edge. Escape from such large mean optical depths requires that the galaxy interstellar medium (ISM) to be highly inhomogeneous, peppered with low neutral density, high ionization voids and chimneys created by supernovae or the integrated winds from stellar clusters. This implies that escape will be extremely dependent on local geometry, requiring resolutions of star cluster sized structures with typical diameters ~30 to 100 pc. Such objects subtend angles of 0.003 to 0.010 arcseconds at redshifts of 2 or 3.

Direct observation of Lyman continuum (LyC) photons emitted below the rest frame H I ionization edge at 911.7 Å becomes increasingly improbable at redshifts $z > 3$, due to the steady increase of intervening Lyman limit systems towards high z (Inoue & Iwata 2008). A key project of the James Webb Space Telescope (JWST) is to search for those sources responsible for reionizing the universe. However, neither JWST nor the Wide-Field InfraRed Space Telescope (WFIRST), can address the key question of, "How Does Ionizing Radiation Escape from Star-Forming Galaxies?"

The far-UV and near-UV bandpasses provide the only hope for direct, up close and in depth, detection and characterization of those environments that favor LyC escape. By quantifying the evolution over the past 11 billion years ($z < 3$) of the relationships between LyC escape and local and global parameters such as: metallicity, gas fraction, dust content, star formation history, mass, luminosity, redshift, over-density and quasar proximity, we can provide definitive information on the LyC escape fraction that is so crucial to answering our key question. Our goal is a definitive determination of $L_{900}$ galaxy luminosity functions over a redshift range from 0 to 3 and will allow us to test whether the escape fraction of low luminosity galaxies is luminosity dependent.

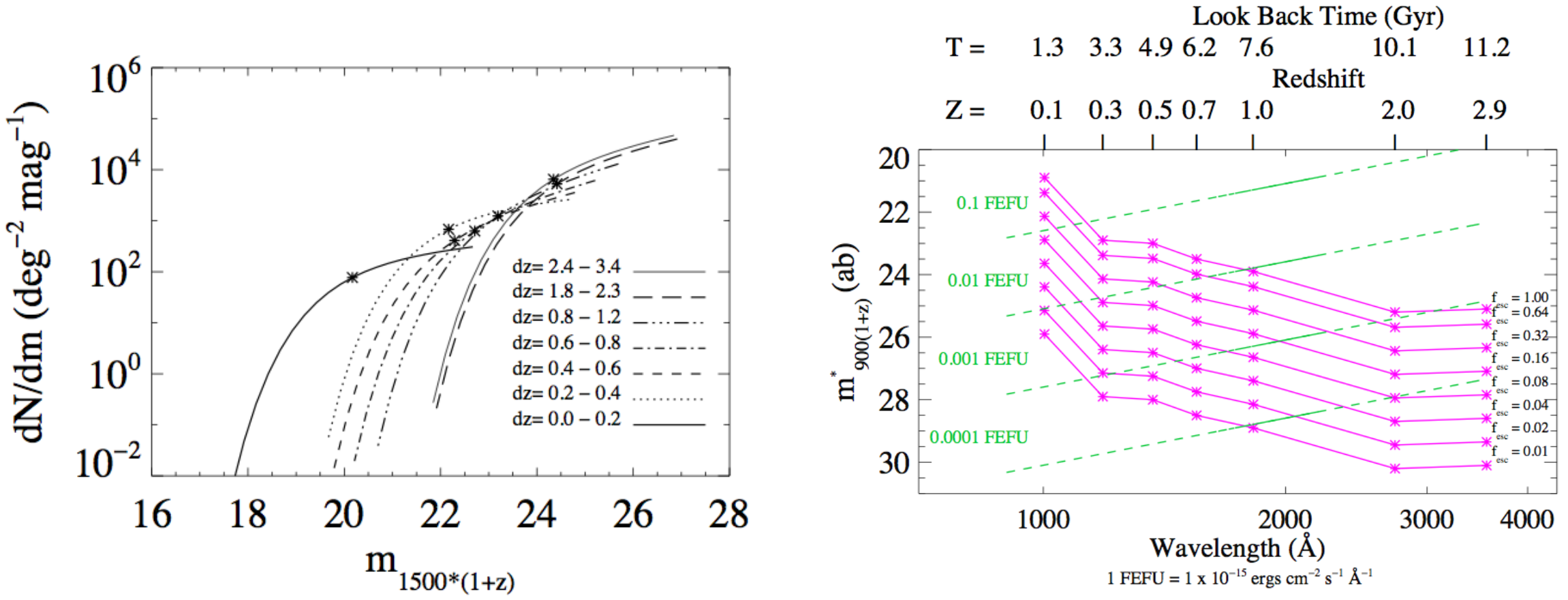

**Figure 5.11.1** - Left panel – Surface densities as a function of observer's frame apparent magnitude for galaxy populations with redshifts between 0 – 0.2, 0.2 – 0.4, 0.4 – 0.6, 0.6 – 0.8, 0.8 – 1.2, 1.8 – 2.3, 2.4 – 3.4, estimated following Arnouts et al (2005). There are 100s – 10,000's of galaxies per square degree. Right panel- the purple asterisks show the characteristic apparent LyC magnitudes (ab) m∗900(1+z) as a function of look back time, and in redshift and wavelength space, for different escape fractions. Contours of constant flux units are overplotted as green dashes marked in FEFU fractions defined in the abscissa subtitle.

**Observational Requirements** – We have undertaken a study (McCandliss et al. 2008; McCandliss 2012) to estimate the flux detection requirements for escaping Lyman continuum photons from star-forming galaxies, as a function of redshift, guided by the galaxy luminosity functions of Arnouts et al. (2005), shown in the left panel of Figure 5.11.1. In the right panel of Figure 5.11.1 we provide ionizing continuum flux estimates for "characteristic" ($L^*_{UV}$) star-forming galaxies as a function of look back time and escape fraction. We find ab-magnitudes for $L^*_{UV}$ galaxies of 30 having escape fractions of 1% between redshift of 2 to 3. We note that the faint end of the higher redshift luminosity functions are 10x fainter than $L^*_{UV}$, so the detection requirements for the faintest galaxies with similar escape fractions will be 10x lower (although some theorist argue that small galaxies should have high escape fractions).

We will take $f_{900} = 10^{-20}$ erg cm$^{-2}$ s$^{-1}$ Å$^{-1}$ as a representative flux. This is a challenging flux level to reach, requiring a product of effective area, time and bandpass ~2.5 x 10$^9$ cm$^2$ s Å at 2000 Å to reach a S/N of 5. A telescope with an effective area of ~15,000 cm$^2$, observing for 5 hours with either a filter or spectrograph bandwidth of 10 Å can satisfy this requirement, assuming no significant background. Additional requirements include a sample size exceeding 25 objects per luminosity bin per redshift interval to yield an rms deviation of < 20% for each point. The total angular area of the sample should exceed > 1 degree (the characteristic angular scale for BAOs), by a fair margin to beat down cosmic variance. Redshifts are required for each object.

**Instrumental Requirements** – These observational requirements can be met with a 10 – 12 m class UV telescope with multi-object spectroscopic capability with a spectral resolution of 200 – 1000. The diffraction limit for a 12 m telescope at 2000 Å is 0.003 arcseconds satisfies the spatial sampling requirement. A 2 arcminute wide focal plane at f/24 and 12 m requires a 170 mm detector FOV. The TRL for such UV detectors and multi-object spectrographs is TRL 5; for 12 m space qualified mirrors diffraction limited at 2000 Å is likely TRL 1. Such a telescope could be compatible with a Habitable-Exoplanet Imaging mission.

## 5.12 Characterizing the Habitable Zones of Exoplanetary Systems with a Large Ultraviolet/Visible/Near-IR Space Observatory – Kevin France

Understanding the surface and atmospheric conditions of Earth-size ($R_P \approx 1 - 1.5\ R_\oplus$), rocky planets in the habitable zones (HZs) of low-mass stars is currently one of the greatest astronomical endeavors. The Astro2010 Decadal Survey ranks the **"Identification and characterization of nearby habitable exoplanets"** as the premier Discovery Goal for astrophysics at present, with **"can we identify the telltale signs of life on an exoplanet?"** as a specific focus question. The nearest known Super-Earth mass planets ($2 - 10\ M_\oplus$) in the HZ orbit late-type stars (M and K dwarfs). In addition, all of the HZ planets found by *TESS* will be around M dwarfs, making these systems prime targets for spectroscopic biomarker searches with *JWST* and future direct spectral imaging missions (Deming et al. 2009; Belu et al. 2011).

The planetary effective surface temperature alone is insufficient to accurately interpret biosignature gases when they are observed in the coming decades. The UV stellar spectrum drives and regulates the upper atmospheric heating and chemistry on Earth-like planets, is critical to the definition and interpretation of biosignature gases (e.g., Seager et al. 2013), and may even produce false-positives in our search for biologic activity (Hu et al. 2012; Tian et al. 2014; Domagal-Goldman et al. 2014). Therefore, our quest to observe and characterize biological signatures on rocky planets *must* consider the star-planet system as a whole, including the interaction between the stellar irradiance and the exoplanetary atmosphere.

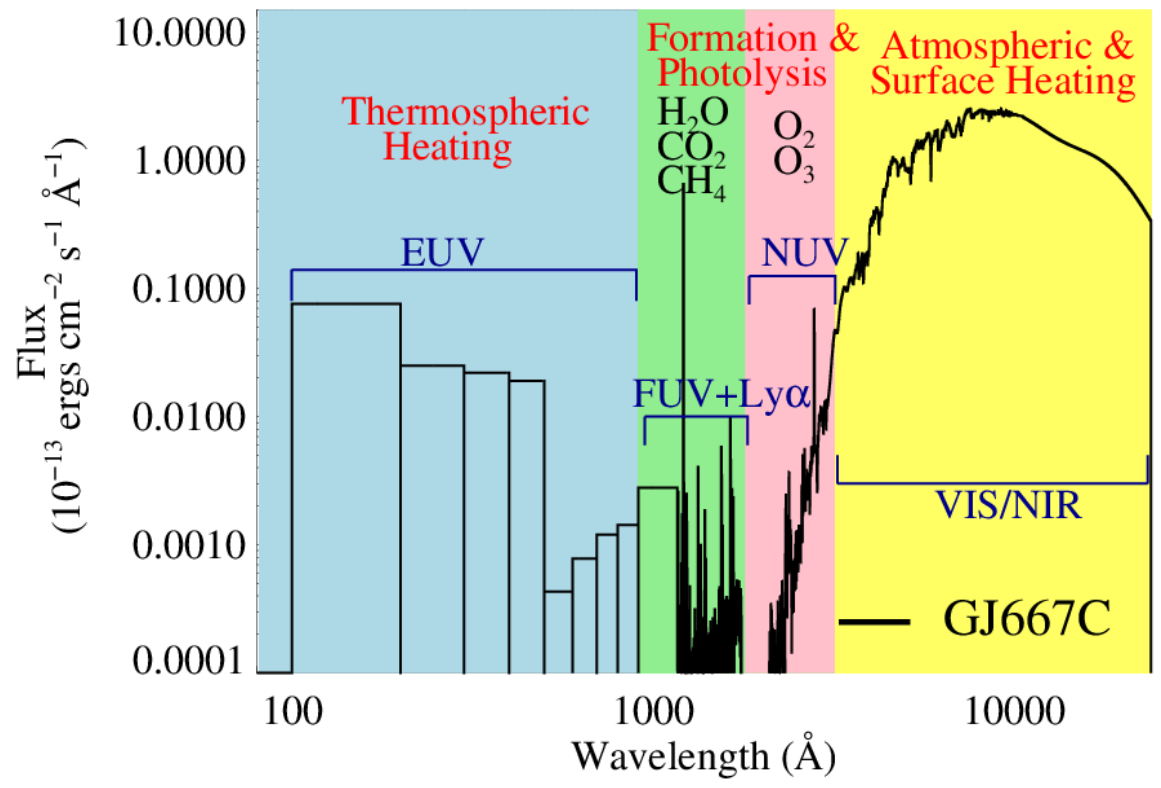

**Figure 5.12.1**: HST spectrum of GJ 667C, illustrating the influence of each spectral bandpass on the atmosphere of an Earth-like planet orbiting this star (France et al. 2013, Linsky et al. 2014). GJ 667C hosts a super-Earth mass planet located in the HZ.

Spectral observations of $O_2$, $O_3$, $CH_4$, and $CO_2$, are expected to be important signatures of biological activity on planets with Earth-like atmospheres (Des Marais et al. 2002; Kaltenegger et al. 2007; Seager et al. 2009). The chemistry of these molecules in the atmosphere of an Earth-like planet depends sensitively on the strength and shape of the host star's UV spectrum. $H_2O$, $CH_4$ , and $CO_2$ are sensitive to far-UV radiation (FUV; 100 – 175 nm), in particular the bright HI Lyα line, while the atmospheric oxygen chemistry is driven by a combination of FUV and near-UV (NUV; 175 – 320 nm) radiation (Figure 5.12.1). Furthermore, the FUV and NUV bandpasses may be promising spectral regimes for biomarker detection (Bétrémieux & Kaltenegger 2013). High levels of extreme-UV (EUV; 10 – 91 nm) irradiation can drive hydrodynamic mass loss on all types of exoplanets (e.g., Koskinen et al. 2013; Lammer et al. 2014). Ionization by EUV photons and the subsequent loss of atmospheric ions to stellar wind pick-up can also drive wide-scale atmospheric mass-loss on geologic time scales. As the opacity of the ISM precludes direct observations of the EUV irradiance, FUV observations provide the best constraints on the EUV emission from cool stars (Linsky et al., 2014; Shkolnik & Barman 2014).

The temporal variability of these sources must also be considered as even weakly-active, "old" (several Gyr) M dwarfs show extreme flare events in their UV light curves (factors of greater than 10 flux increases on time scales of minutes, see Figure 5.12.2; France et al. 2013; Loyd & France 2014). Impulsive UV events are also signposts for energetic flares, those that are associated with large ejections of charged particles. Energetic particle deposition into the atmosphere of an Earth-like planet during a large M dwarf flare can lead to significant atmospheric $O_3$ depletions (> 90% for large flares; Segura et al. 2010). This alters the atmospheric chemistry and increases the penetration depth of UV photons that could potentially sterilize (or catalyze) surface life. Given that particle fluxes cannot typically be directly measured for stars other than the Sun, UV observations offer the best estimates of these important particle environments.

NASA's next large ultraviolet/optical/infrared flagship mission, a LUVOIR Surveyor, could carry out both the direct detection of atmospheric tracers on these worlds and the essential characterization of the star-planet system. The transformative science enabled by a LUVOIR Surveyor to the field of astronomy as a whole, including exoplanet (ExoPAG) and cosmic origins (COPAG) themes, is made clear in NASA's 2013 Astrophysics Roadmap (*Enduring Quests, Daring Visions: NASA Astrophysics in the Next Three Decades*): a LUVOIR Surveyor (or ATLAST/HDST-like mission; Tumlinson et al. 2015) directly addresses the largest number of core NASA science priorities in a single mission in the next 30 years, working toward six "primary goals", ***twice as many core science goals as any other intermediate-term mission concept*** (Section 6.4 in 2013 Roadmap).

There are a set of baseline requirements for the characterization of the energetic radiation environments around low-mass exoplanet host stars. Spectral coverage to wavelengths as short as 102 nm is crucial to characterizing the star in important activity tracers (e.g., O VI and Lyα), as well as peak absorption cross-sections for relevant molecular species in planetary atmospheres (e.g., $O_2$ and $H_2$). We would like to characterize typical mid-M dwarf exoplanet host stars out to 30 pc with sufficient signal-to-noise to carry out time-resolved, $R \approx 3000$ spectroscopy of UV emission lines (e.g., N V, Si IV, C IV; $F_\lambda \sim 1 \times 10^{-16}$ erg cm$^{-2}$ s$^{-1}$) on the scale of minutes (S/N ~ 10 per 60 s). These sensitivities can be achieved with a 10-meter class mirror, optical coatings with broadband UV reflectivity ≥ 90%, and photon-counting detectors with UV DQE ≥ 75%. These technology goals can be achieved with a reasonable investment in laboratory development efforts and hardware demonstration flights (via suborbital and small satellite missions; e.g., France et al. 2012) in the next 5 – 10 years, enabling the start of a LUVOIR Surveyor mission in the 2020s.

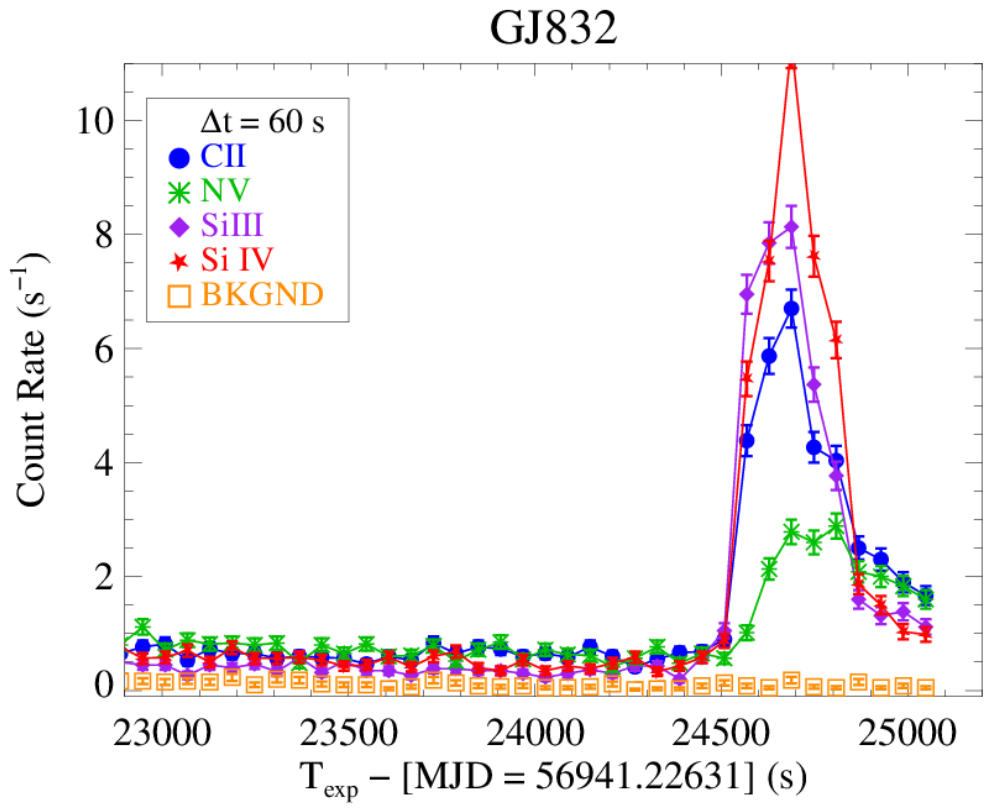

**Figure 5.12.2**: FUV flare on the "optically inactive" M dwarf GJ 832 (d = 4.9 pc), illustrating a flare event in FUV emission lines on sub-minute timescales (HST-COS data, binned to 60 s intervals).

# 6 Technology

## 6.1 Overview – Matt Beasley

1. Development of improved coatings for space telescopes allows increased capabilities in wavelength coverage, improved wavefront control, and overall efficiency. New materials with improved control of the deposition process are critical improvements that enable more science by adding targets and creating more capable instruments.

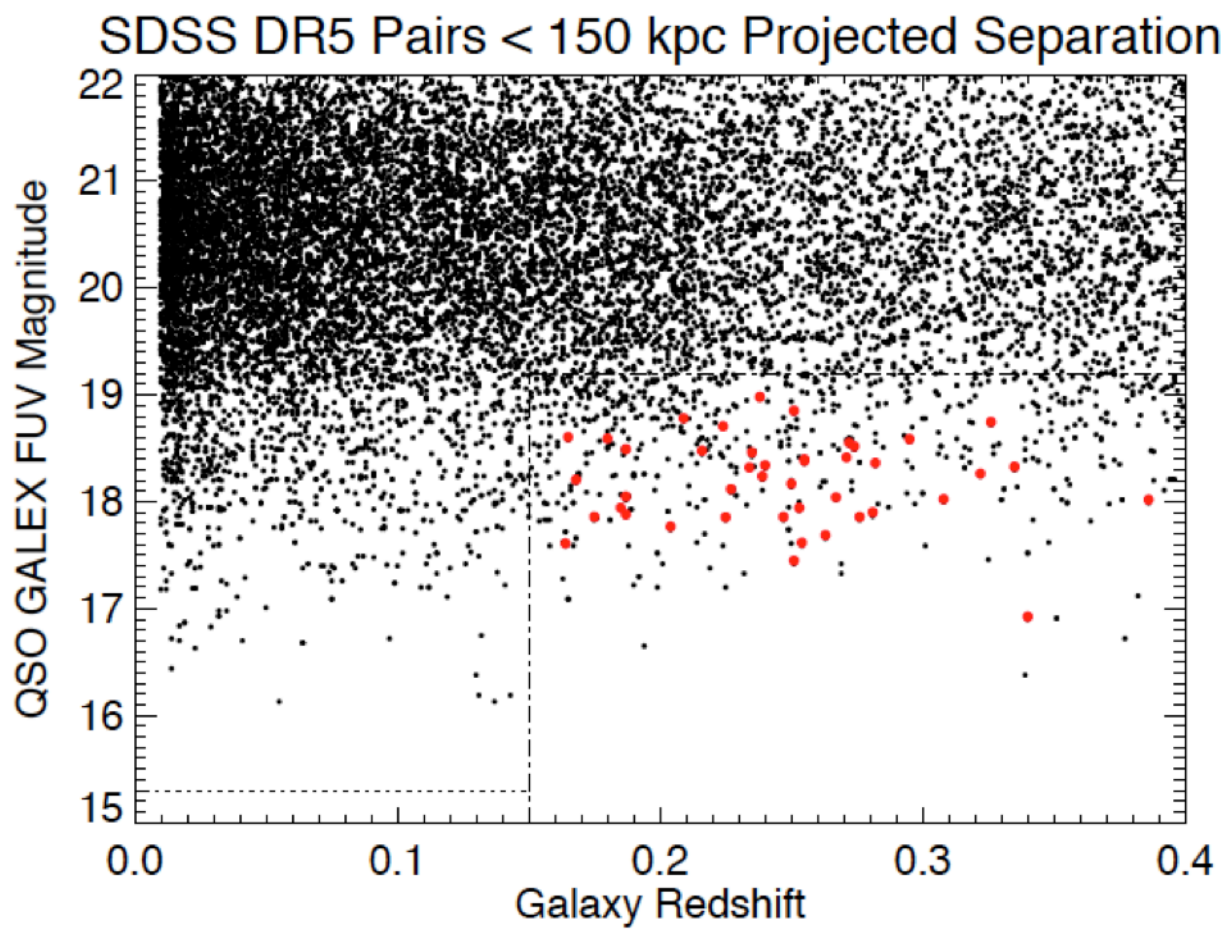

**Figure 6.1**. From Jason Tumlinson. The dotted line describes the available targets with Hubble-COS and the FUSE spacecraft. The red dots are the targets that have been observed. Clearly, extending Hubble-COS sensitivity toward the blue allows vastly more targets.

Improvements to mirror coatings suitable for large mirrors would contribute to the largest possible gains in scientific return from a Cosmic Origins Mission. First, if ultraviolet reflectivity can be improved compared to existing missions, it will add to the number of scientifically viable targets. In general, the number of targets increases dramatically as a function of minimum detectable flux. (See Figure 6.1) Second, improved coatings improve wavefront error (WFE) budgets. With an optical coating requiring less of the overall WFE, more error can be allocated to fabrication and alignment, which reduces risk and therefore cost of a mission. Third, opening the ultraviolet further toward the blue, changes and improves the fundamental science that can be accomplished. Access to wavelengths currently unavailable (such as 90 – 110 nm) provide data that simply cannot be accessed without the increased bandpass. (Figure 6.2) Finally, and it cannot be stated strongly enough, improved reflectivity creates more capable instruments beyond a simple increase in effective area. For example, far-ultraviolet spectrographs can be cross-dispersed without a significant loss of effective area, allowing high spectral resolution and broad bandpass simultaneously. That combination, compared to existing systems, reduces the observing time necessary on a target by the multiplex advantage of an echelle – well beyond the simple improved effective area throughput. Moving to multi-object systems could increase the advantage even further.

Beyond improvements to ultraviolet capability, improved coatings would be the path forward to developing systems that would produce acceptable wavefront errors for exoplanet observations. Specifically, dielectric protective coatings may have too much polarization for some current coronagraph designs. Bare metal coatings will have a smaller overall polarization effect, but there are only a few metal coatings suitable for coating a large mirror with associated handling practicalities.

2. Optical / NIR detector development: Optical detectors are nearly perfect for general astrophysical detections. High time resolution is critical for a small number of science cases. Improved detector radiation tolerance will extend the usable lifetime of missions, particularly for missions not in Low Earth Orbit. Leveraging the commercial developments in detectors will add capability to future missions beyond what the limited budget for technology investment can provide. Projects that improve the TRL of new detector technology can supply the large focal planes that future origin missions will need to maximize their science return.

Making sure NASA leverages commercial detectors, which spends vastly more on R&D than NASA does, will allow the development of large, high resolution, high dynamic range detector systems. Alternate read schemes combined with the low noise of available sCMOS (scientific CMOS) detectors could allow cosmic ray tolerant imaging, improving the data yield, particularly for missions beyond Earth's magnetosphere.

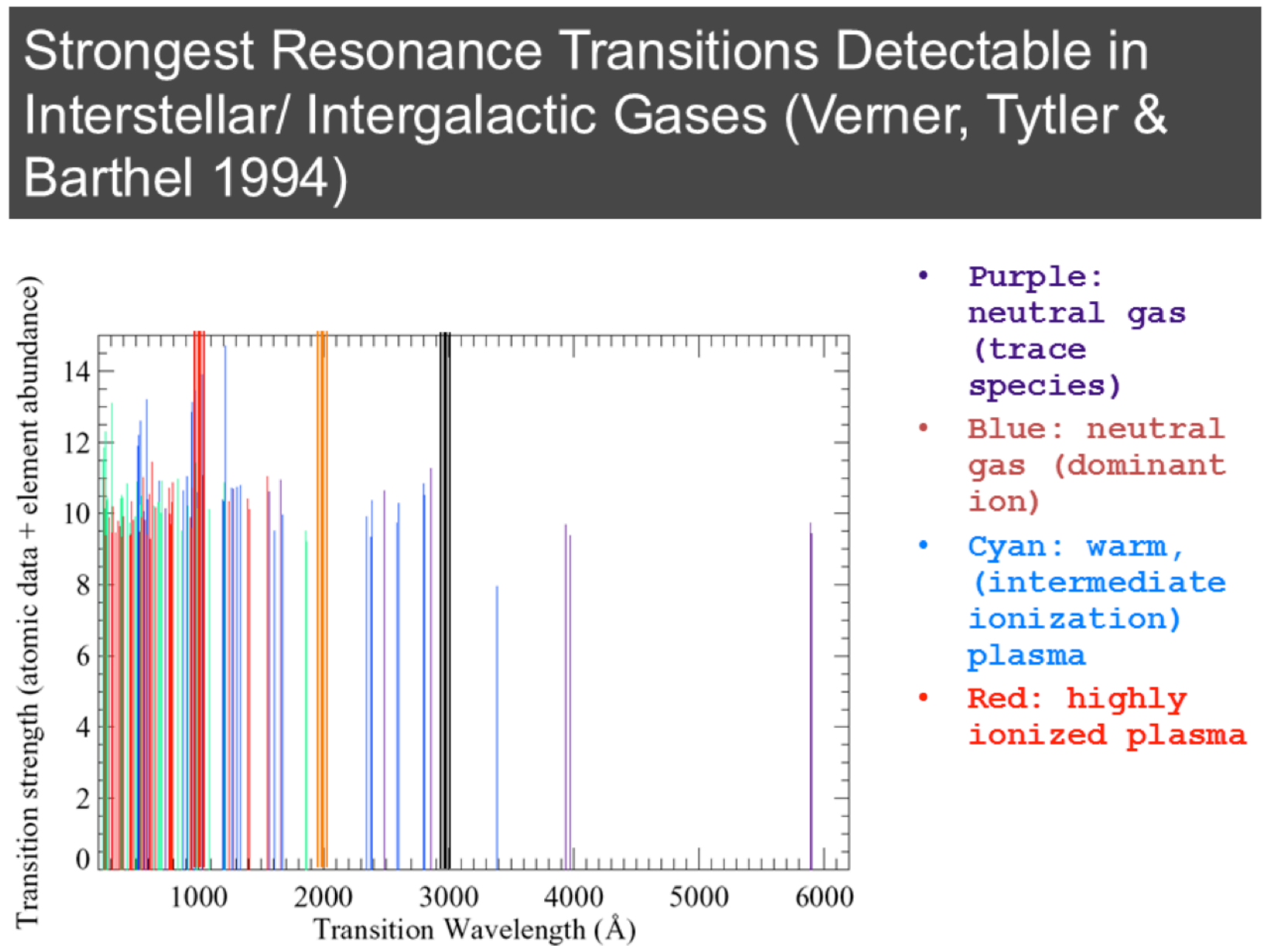

**Figure 6.2**. Astrophysically relevant lines: there are more below 1000 angstroms than above it (Source: Todd Tripp)

3. Ultraviolet detectors can be improved for quantum efficiency, resolution, and dynamic range. Extending these capabilities over large area detectors increases the science return for any ultraviolet mission.

More capable UV detectors will leverage the capability of more sophisticated instrumentation afforded by improved ultraviolet reflectivity. Moreover, even without improved reflectivity, moderate to large gains in detector efficiency are still possible for ultraviolet detectors.

4. Mirror technology improvements decrease the cost of missions and improve system capability. Development work that demonstrates technologies for fabrication and integration of mirrors will reduce cost and risk for future missions.
5. NASA should seek to improve opportunities for instrument builders – Nancy Roman fellowships, other design opportunities. While not a direct technology development, increasing the number of practitioners and more novel designs will enable future missions.

## 6.2 Detectors for UV-Visible Flagship Missions - John Vallerga

Choosing the best detector technology for the next UV-Vis. Flagship mission is not obvious, given the recent advances in their performance and technology readiness. However, I will try to summarize the performance of existing technology and mention candidates that have exciting new capabilities if they can establish the resilience of a flight qualified detector.

First some criteria that most proposed detectors must meet. Larger telescopes with large fields of view require larger detectors (assuming a fixed f-number). For example, an 8 meter f/3 telescope with a 30 arc minute field of view requires a 20 cm detector. High quantum efficiency is always desired as its impact on the science scales directly as the telescope area does. The UV sky is dark, so the signal to noise ratio of many science programs is limited by the detector background and/or readout noise. Imaging detectors that have the ability to "photon count", which means detecting the arrival position of individual photons, effectively eliminate readout noise. A detector with high dynamic range is needed to broaden the choice of science targets and overlap previous mission capabilities.

Of course, there are other desirable attributes not directly related to the science performance. The detector should have a long lifetime with respect to the mission duration. It and its support electronics should be radiation hard and preferably have a low operational complexity (e.g. high voltage or cryogenic requirements, vacuum pumps etc.). Response should be uniform and stable for low signal to noise observations. Also, low mass and low power requirements are always desired.

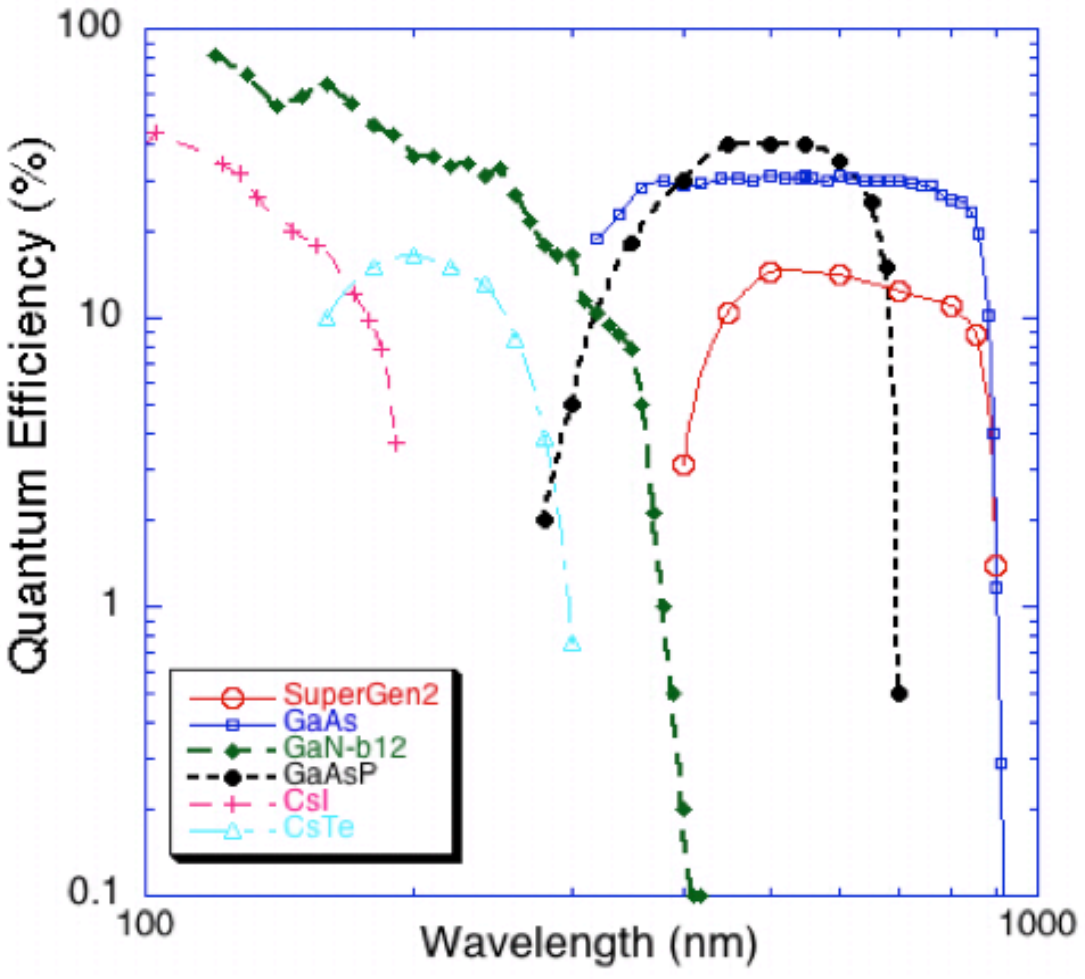

**Figure 6.2.1** - Sample of photocathode efficiencies

### 6.2.1 The Candidates

*Microchannel Plate (MCP) Detectors*

MCP detectors have been the detector of choice for many recent UV instruments (FUSE, COS and STIS on Hubble). MCP detectors count individual photons by amplifying the initial photoelectron by many orders of magnitude and detecting the location of this charge on a patterned anode. The initial photoelectron was released by the incoming photon from a photocathode, either deposited on the MCP itself or on a window directly above the input surface. Hence the QE of such detectors is limited by choice and quality of existing photocathodes (Figure 6.2.1) and tends to be smaller in the optical/IR.

MCP detectors are photon counting, can be made in areas up to 20 x 20cm (Figure 6.2.2), and are inherently rad hard (made of glass). The also can be curved to match curved focal planes, have excellent spatial (<20 micron FWHM) and temporal resolution (sub-nanosecond), low background rates < 0.02 ct $cm^{-2}$ $s^{-1}$. Most UV photocathodes are "solar blind", i.e. have a poor red response.

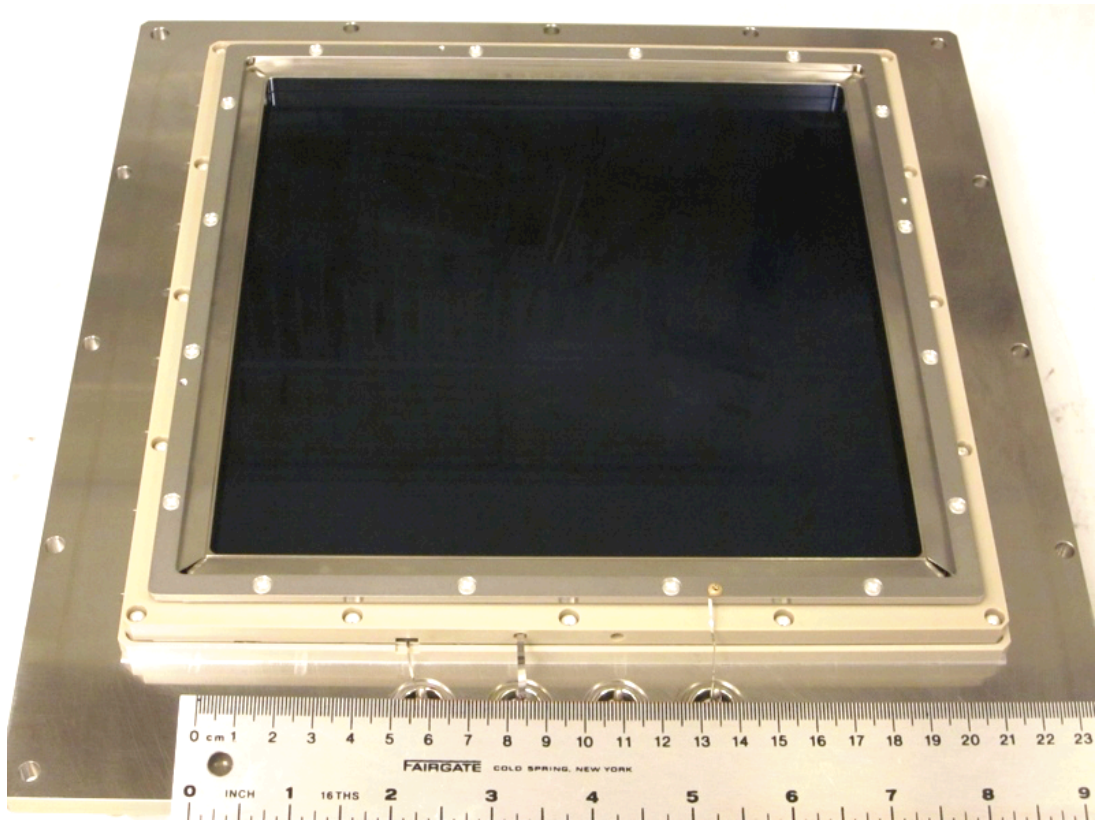

**Figure 6.2.2** - A 20 x 20 cm photon counting detector using atomic layer deposited MCPs

Some noted weaknesses of this detector technology have been the low NUV/optical QE, limited lifetime (local gain loss) to extracted charge, limited count rate of the electronics, non-uniform response, spatial distortions, and high voltage requirements. Yet recent NASA investments in this technology through the ROSES APRA and SAT programs have achieved (Siegmund et al 2014; Tremsin et al 2013; Vallerga et al 2014) 30% QE at 200nm, lifetimes of 7 Coulombs $cm^{-2}$ (order of magnitude higher than predicted mission lifetimes), uniformity better than 3%, count rates > 5MHz global and 40kHz local, and spatial distortions < 6 microns. Given this progress, and its heritage, this is the detector most likely to be qualified for the upcoming UV portion of a UV-Vis flagship mission.

*Silicon Detectors*

Detectors based on the absorption of input photons into the silicon bulk where an electron-hole pairs are generated are ideally suited for the optical to near-IR because of their extremely high QE (Figure 6.2.3). Various integrating detectors such as CCDs and CMOS arrays are mature technologies with significant flight heritage (WFC3 on Hubble) and can be tiled into large arrays as well. Until recently, however, they have not been used in the deep UV because of the inability of UV photons to penetrate through the surface layers into the bulk silicon. They must operate at cryogenic temperatures to minimize dark current and must be read out often to minimize cosmic ray hits, thereby taking a readout noise penalty.

Surface treatments, such as delta doping and atomic layer deposition of anti-reflection coatings can overcome the lack of UV sensitivity in specific bandpasses (Hennessey et al 2015; Hamden et al 2014; Hamden et al 2012). Not only will this work on charge integrating CCDs and CMOS devices, but also other silicon technologies that offer potential photon-counting capabilities such as avalanche photodiode (APD) arrays and electron multiplying CCDs (EMCCD). Surface treated EMCCDs will soon be launching on the sub-orbital rocket mission "CHESS" (Hoadley et al 2014) and the FIREBALL-2 balloon payload (Hamden et al 2014).

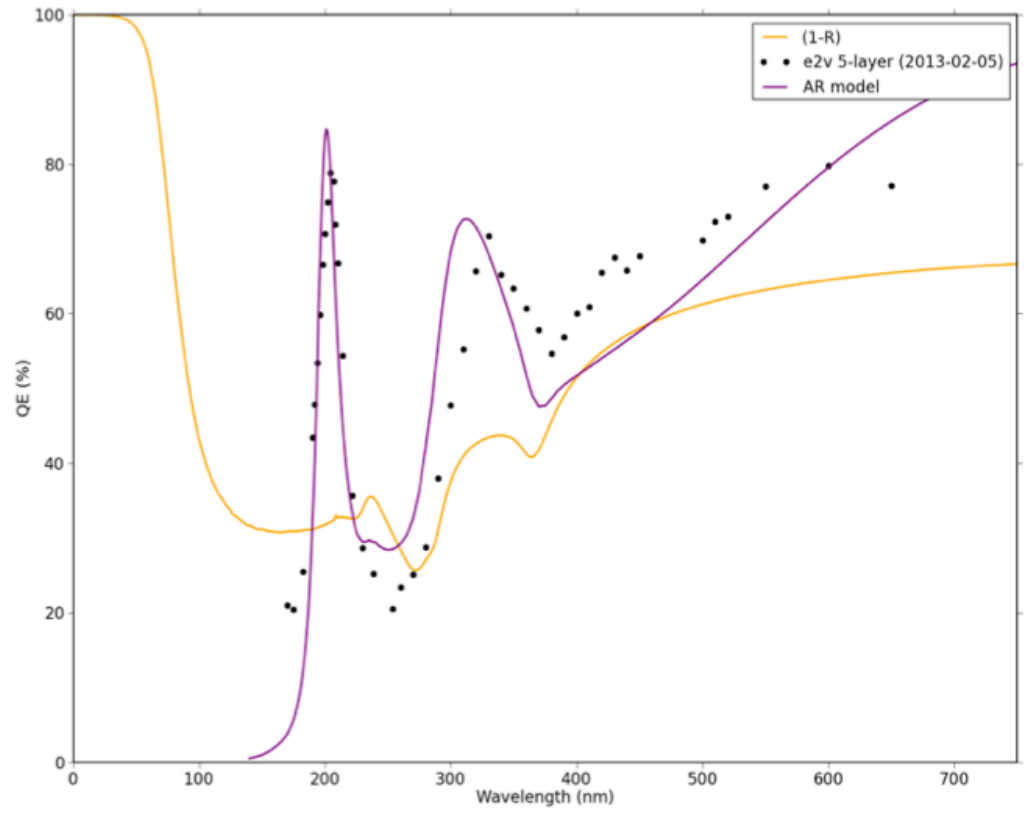

**Figure 6.2.3** - QE of photon counting silicon EMCCD optimized for balloon UV window. (Hamden et al 2014)

*Microwave kinetic inductance detectors (MKIDs)*

MKID detectors (Mazin et al 2012) can be thought of as an array of superconducting resonant microwave receivers kept at 100mK operating temperature. When a photon is absorbed in the pixel, quasi-particles are generated and the inductance change moves the receiver off-resonance. The amplitude and phase of this effect is proportional to the energy absorbed, so the wavelength of the individual photon can be determined to a spectral resolution of ~20. This brings the advantage of non-dispersive spectroscopy to the optical and NIR where 20 colors can be determined in a single pixel with its low background. The MKID detector is photon-counting, has no read noise or dark current and has good QE. Unfortunately, array sizes of 10-20 kpixel are just beginning to arrive at ground based telescopes, though formats of megapixel range are possible. Flight qualified 20cm arrays will probably not be ready for the next Flagship mission, in any bandpass.

### 6.2.2 Summary

Quantum efficiency trumps most performance parameters in detector technology. In the far UV (< 150nm), photocathode efficiencies are large enough that MCP detectors with their improved performance in photon counting, large format, resolution, background, lifetime and radiation hardness make them the detector of choice. In the visible/NIR, a silicon CCD array will be chosen, as usual, with its exquisite QE, uniformity and stability. In the near UV, the crossover wavelength from silicon to MCP will depend on the latest QE performance of GaN photocathodes vs. surface treated and coated silicon EMCCDs operated in photon counting mode, assuming both technologies demonstrate their flight worthiness.

## 6.3 Primary Mirror Parameters and Features – Gary Matthews

The primary mirror for a large UV system such as LUVOIR, ATLAST or HDST will most certainly have to be segmented and one major trade will be the segment size and configuration. For JWST, it turned out that a 1.3m flat-flat hexagonal segment was the right answer. Shortly after the award to Northrop Grumman, there were 36 segments of less than 1m in size. As the design and analysis process progressed, it was determined that the size was just too small and drove a lot of actuator mass and complexity due to having 7 actuators per segment. By going from 36 segments to 18 segments, the number of actuators was cut in half.

For future UV/Visible systems with a coronagraph in a 12m class size, a new trade is required. From a manufacturing perspective, something like the 1.4m class segment is advantageous since the segment is easy to handle yet large enough that the edges effects are kept to a minimum.

It is important to first evaluate the science needs with respect to segmentation:

- How do segment edges impact the PSF of the image?
- Does the system need a round outer diameter (Figure 6.3.1) or is a JWST-like (Figure 6.3.2) outside edge acceptable?
- Are there any primary mirror obstructions that are more troublesome than others?
- Does segment size offer significant science return advantages? For example, what if the segments were 5m each?

Though there are certainly other factors that need to be considered those listed above should be given initial consideration during the system architecture conceptualization since they may drive the technology development required for the mission. Figures 6.3.1 and 6.3.2 below provide a range of possible solutions.

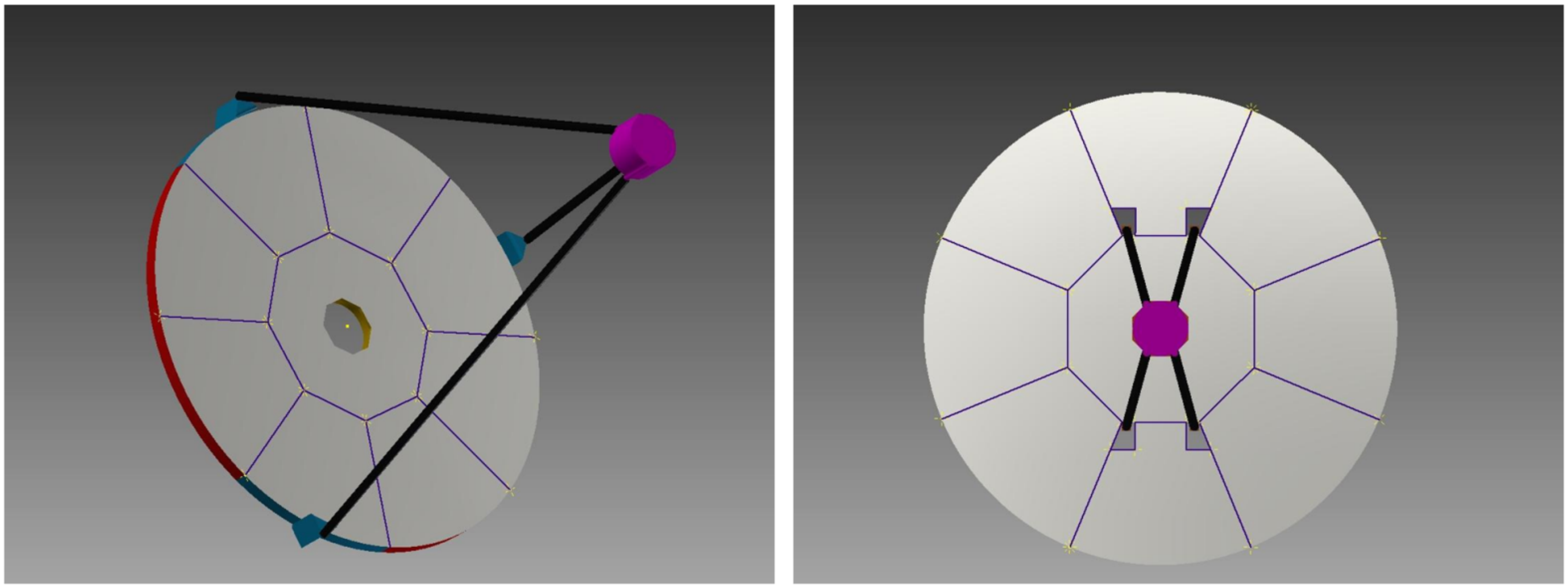

**Figure 6.3.1** – These are two examples of 5m segments in a 12.7m telescope. On configuration has the secondary struts on the outer diameter and the other configuration shows the secondary with in-board struts. There are advantages and disadvantages of each configuration.

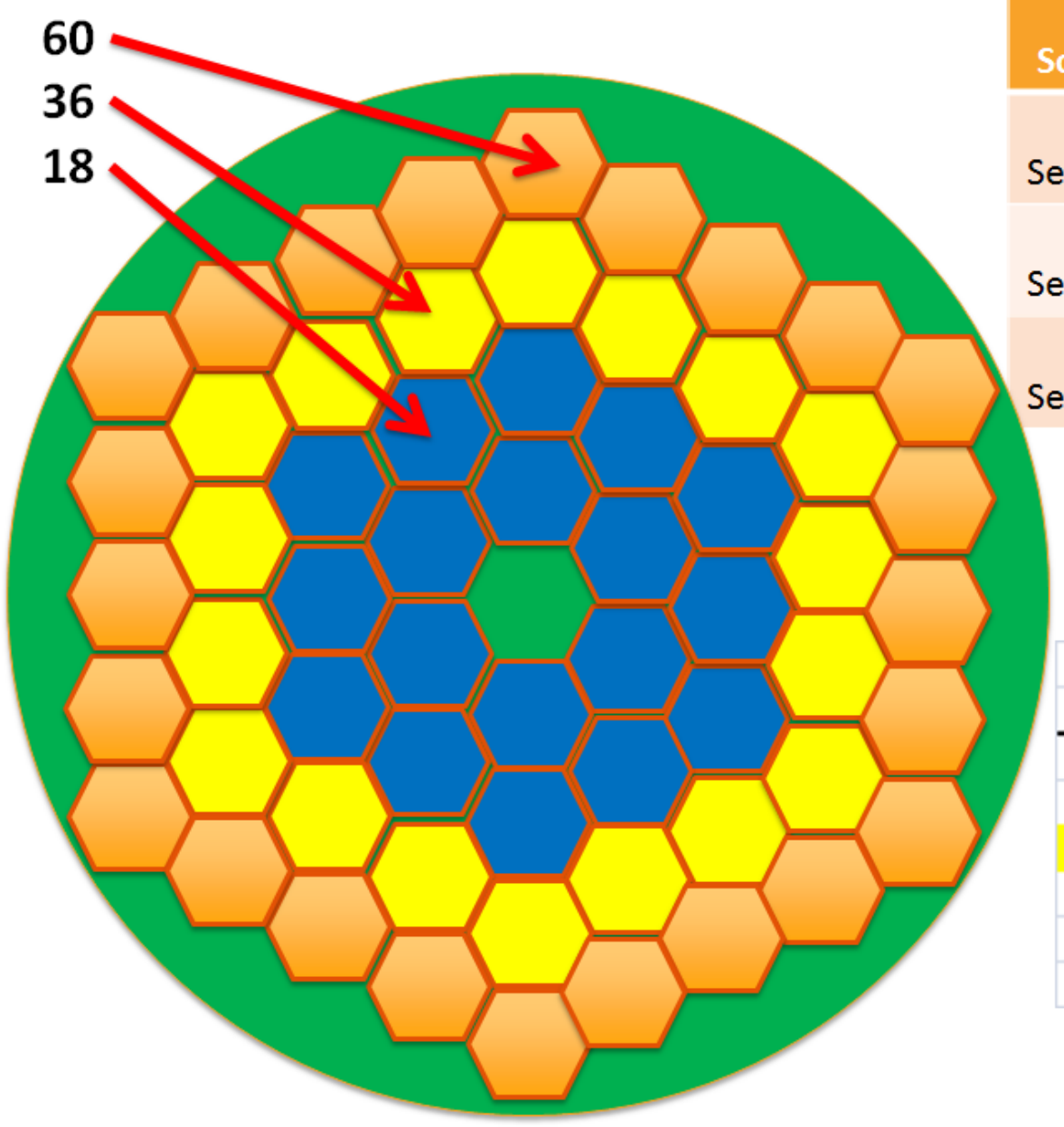

| 12.7m P-P Solution | Segment Size (F-F) | Segment Size (P-P) |
|---|---|---|
| 60 Segments | 1.41m | 1.63m |
| 36 Segments | 1.81m | 2.10m |
| 18 Segments | 2.54m | 2.93m |

Note: JWST Segments – 1.31m F-F

| PM Diameter (m) | Area (m^2) | Hubble Scale Factor |
|---|---|---|
| 2.4 | 4.5 | - |
| 9.2 | 66.5 | 14.7 |
| 12.0 | 113 | 25.0 |
| 12.7 | 127 | 28.0 |
| 13.0 | 133 | 29.3 |
| 14.0 | 154 | 34.0 |

Note: Does not include obstructions or center hole

11

**Figure 6.3.2** – This is a JWST-like segmented design. The size of the segments for various segment numbers are shown as compared against JWST. The overall primary mirror collecting area is also shown against HST. As can be seen, a 12m system has 25X the collecting area. Note that the secondary strut obstruction is ignored for this first order analysis.

For a segmented system, some level of mirror actuation will be required for segment phasing (rigid body) and possibly figure control. There are several issues with processing a set of matching segments that will be phased and used as a large primary mirror:

- Radius of Curvature Matching - matching the radius of Curvature (RoC) in gravity to one part in several thousand poses a metrology problem that is likely past the current state-of-the-art for mirror testing. If it can be measured, it can be ion figured into the mirror, but the metrology error is very difficult. On a 10m system, the RoC of the primary mirror is likely to be close to 20m. Matching that RoC to say 100 microns is an error of 5 parts per million.

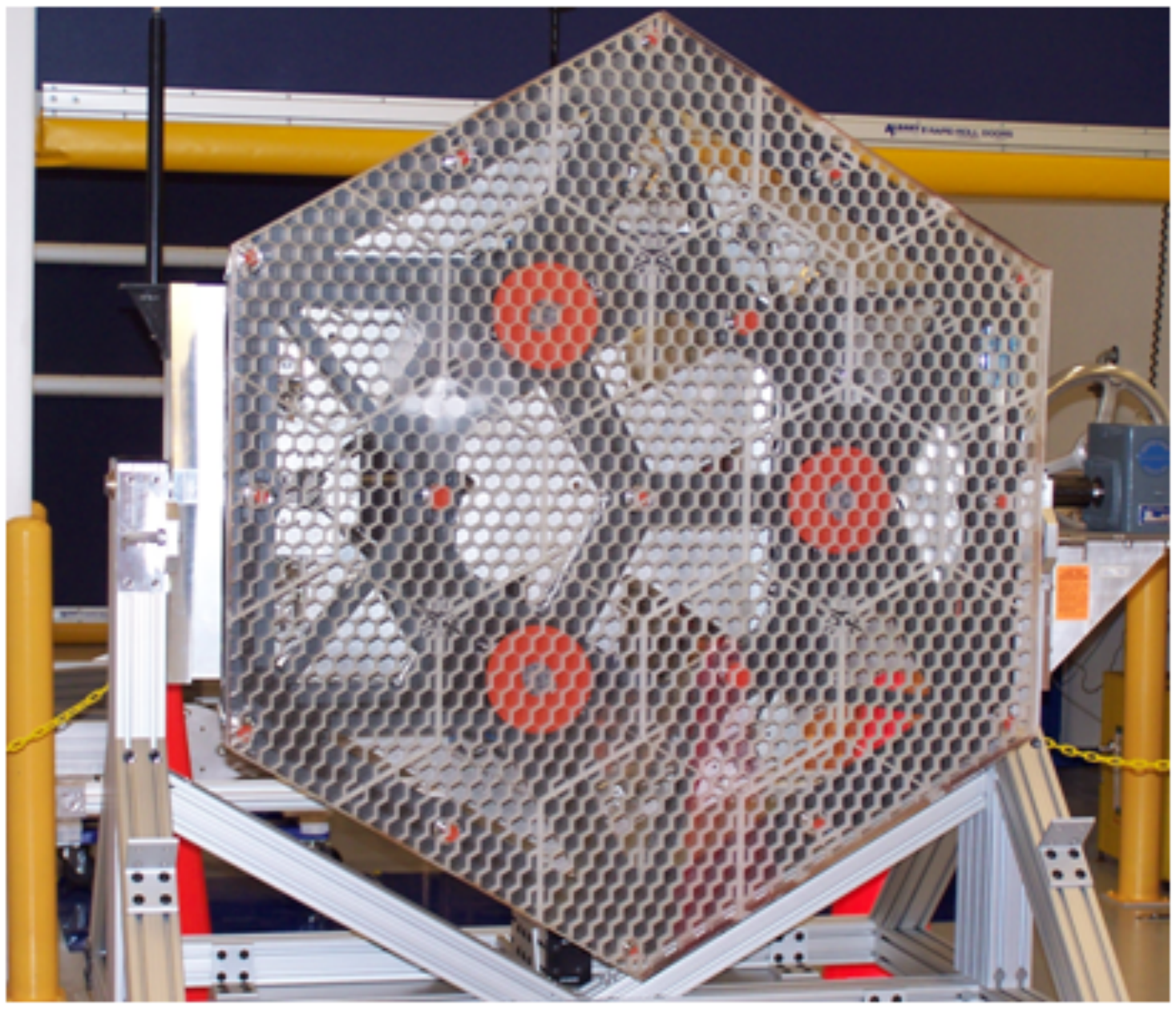

**Figure 6.3.3** - A 10 kg/m$^2$ mirror with a front and back face sheet separated by a lightweight core.

- Metrology Errors and Mounting Errors - these errors will likely be driven by astigmatism. And given that this is the first mode of a plate, small errors in mounting strains will manifest themselves as mainly astigmatism. Small errors in placement of the segment will also drive segment level astigmatism and coma.

- On-orbit Thermal Changes - thermal changes tend to be low order. A pure front to back gradient will cause a RoC change. Other thermal gradients will likely impose trefoil (three points) or astigmatism.

All of these will degrade the performance of the optical system. But it also points to the fact that a large number of actuators are not required for a stiff mirror with a front and back faceplate (Figure 6.3.3). Unlike a dense bed of actuators on a thin face sheet, these stiffer mirrors resist the mid-spatial frequency errors that occur with thin face sheet actuation. As can be seen in Figure 6.3.4, a small number of force actuators, potentially 10 or so, are more than enough to drive in global, low spatial frequency error corrections.

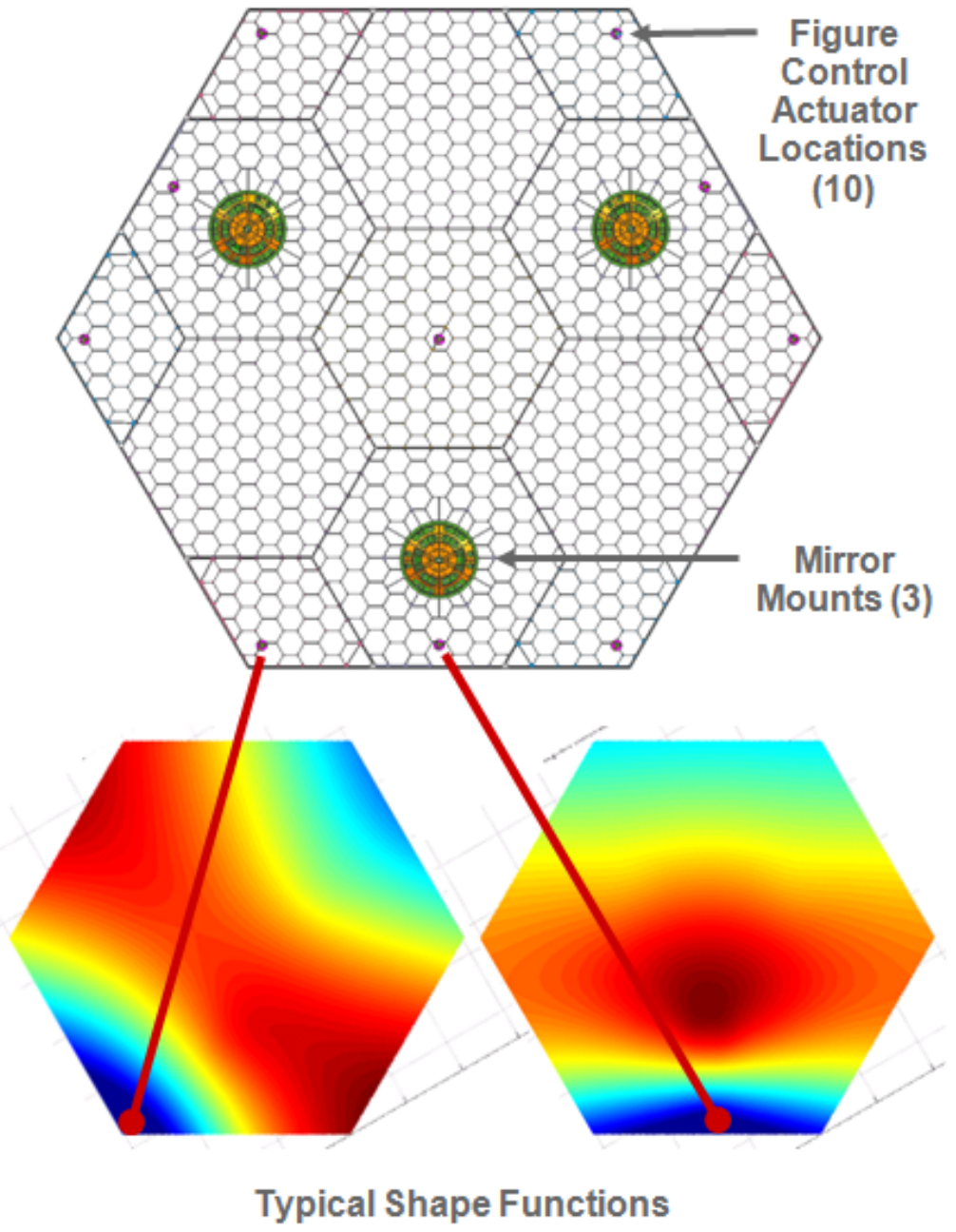

**Figure 6.3.4** - Force actuator influence functions are shown for two of the 10 actuators.

As a demonstration, Figure 6.3.4 shows the influence function for two actuators on the mirror shown in Figure 6.3.2. These influence functions can then be analytically combined to produce many low order error corrections with limited mid-spatial frequency error creation.

The other advantage of the force based system with sparse actuators is that the system has graceful degradation if there is an actuator failure. Some loss in correctability is lost at the 20% level or so.

Figure 6.3.5 shows how the lightweight test mirror can be corrected from 15nm RMS wavefront to around 8nm RMS wavefront.

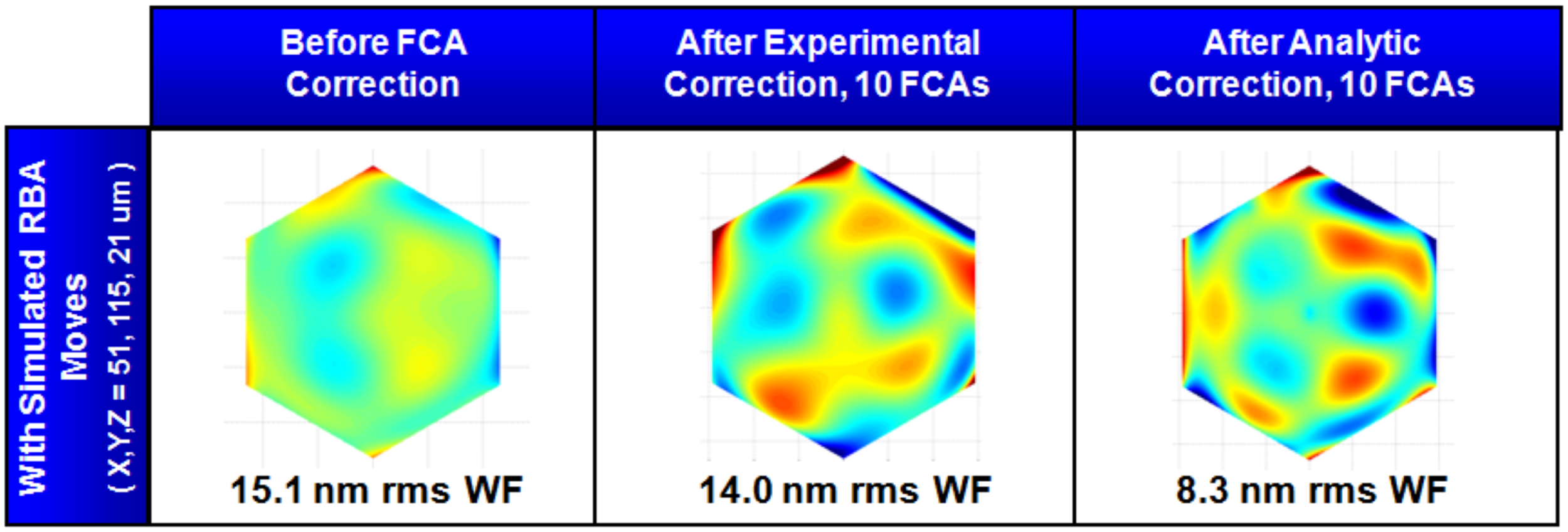

**Figure 6.3.5** - Correction of a lightweight mirror for 8nm RMS. Note that the color scales are different in each picture.

Finally, it should be kept in mind that any mirror actuation should be kept to a minimum. The idea of making a low quality mirror and then actuating it to UV/Visible quality will likely result in the creation of an unacceptable amount of mid and high spatial frequency errors. Correction values must be kept to a minimum so that only small on-orbit errors need to be corrected. In fact, potentially only rigid body correction should be considered to enhance the overall stability of the primary mirror. This would require very stiff segments with the addition of segment level precision thermal control to create a very stable primary mirror system.

As we begin to plan for a very large UV/Visible telescope that will most likely contain a coronagraph as a primary scientific instrument, mirror quality and stability will be crucial performance aspects of the telescope. Even beyond the 10 picometer stability required by the coronagraph, the overall primary mirror quality will still need to be extremely good in order to provide the scientific return required. On a segment basis, it is reasonable to assume that the mirror will need to be finished to ~5nm rms surface quality. If you break this down into various errors by spatial frequency, it might look like the budget below:

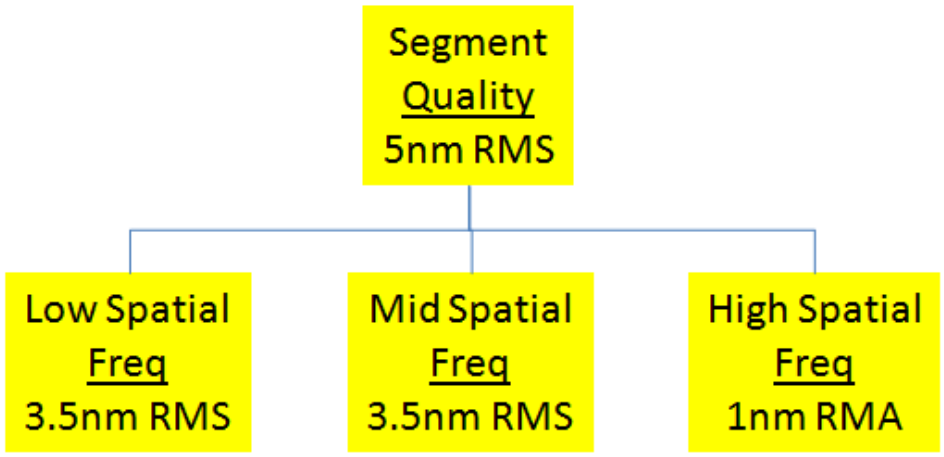

Once the primary mirror is assembled, additional error sources would increase the total error of the system. These are typically low spatial frequency errors caused by telescope structural changes and operational thermal environments which result in primary wavefront errors.

The purpose of this white paper is not to provide an error budget for the primary mirror, but more to provide an assessment of where the state-of-the-art is for producing lightweight segments that would be applicable to a segmented UV/Visible optical system.

There are several choices for mirror materials that have a high TRL status. There are metals or metallic like materials such as Beryllium and SiC of various configurations. In the glass and glass ceramics, there is Zerodur, ClearCeram®-Z, ULE®, and Fused Silica.

One clear advantage of the glass and glass ceramics is the ability to ion beam figure (IBF) to create the final high quality surface figure required for this mission. With a PVD (Physical Vapor Deposition) or CVD (Chemical Vapor Deposition) overcoat, the SiC mirrors can also use the IBF process, but this is an extra process which adds to the complexity of the mirror.

The downside of the metallic materials such as SiC is that they have a CTE that is 100 times that of the glass/glass ceramics which would drive mirror figure error. The argument is that since SiC has 100 times the conductivity the mirror will always have a uniform temperature and minimal front to back thermal gradients which drive mirror figure changes. The issue is that transients will still cause a much larger wavefront error even with the higher conductivity than the glass materials. This will drive thermal control parameters to be very difficult. In essence, a near zero CTE makes sense for a UV/Visible optical system. A rigorous trade assessment regarding material selection needs to be conducted to insure that all aspects of material selection (thermal, dynamic, and operational performance and constraints) are considered.

In 2010, Harris completed a program called the Multi Mirror System Demonstrator (MMSD) shown in Figure 6.3.3. In this program, 1.4m hexagonal mirror blanks were manufactured in a production environment. Harris demonstrated that $10 kg/m^2$ mirror blanks could be fabricated on 3 week centers using existing capability.

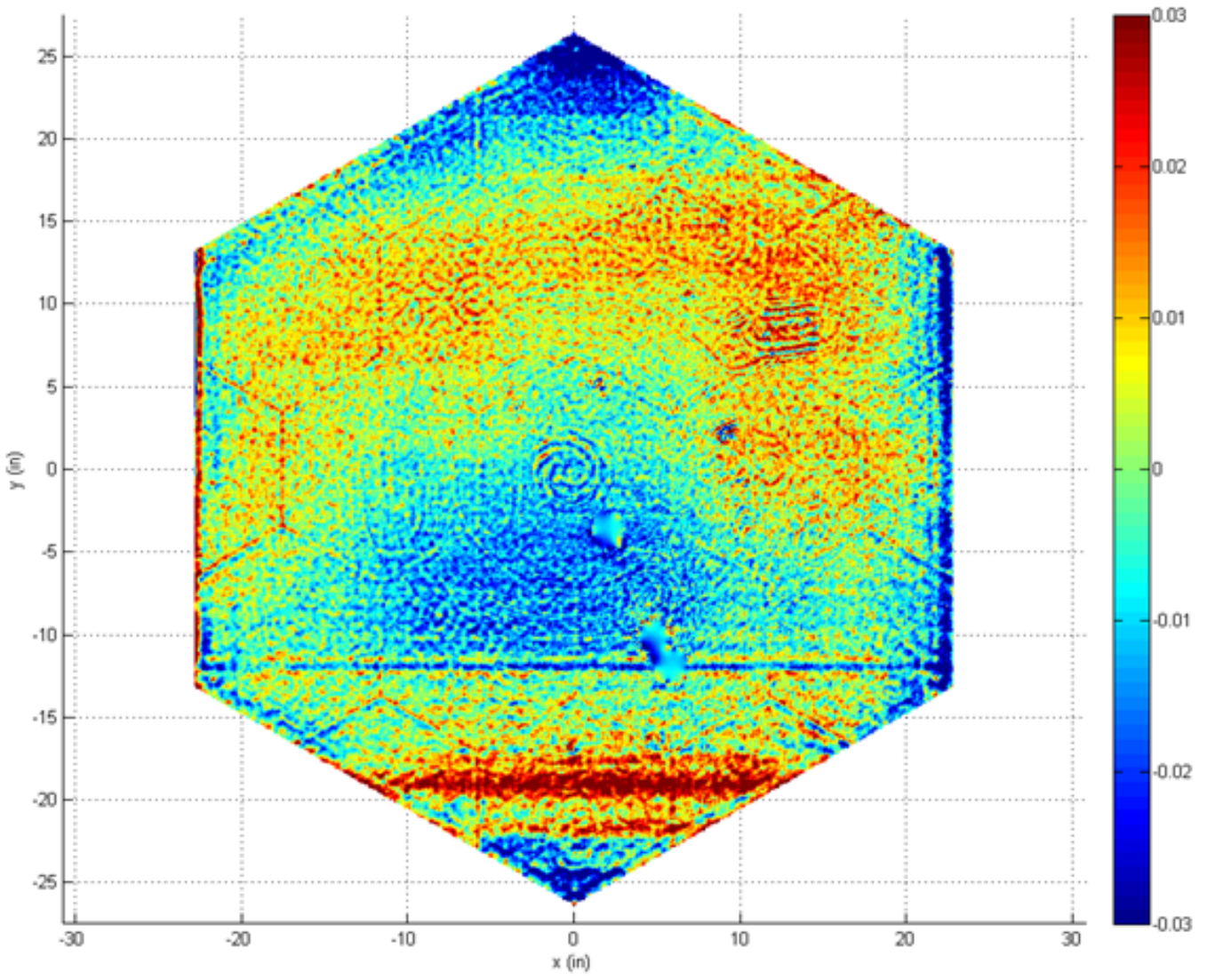

**Figure 6.3.6** - MMSD Final wavefront error of 16.3nm RMS.

The mirrors were then processed and used for qualification testing (random vibration and shock). One of the mirrors was finished to the specified final wave front error quality of 19nm RMS wave front (9.5nm RMS surface) specification. The MMSD mirror was finished to a 15.1nm RMS WFE (7.5nm RMS surface) quality (Figure 6.3.6). It should be noted that there was no PSD type specification levied on the program so the mid spatial content is likely higher than would be acceptable for a dedicated UV/Visible mission. However, IBF could easily be used to address mid spatial frequency errors for mirrors of this type of construction.

Even though the mirror figure quality is approaching the 5nm rms notional requirement discussed earlier, there is additional development work before this type of mirror would be applicable to a true UV mission. There is still some core print through which needs to be evaluated against the requirements for this new mission. The other issue will be edges. As can be seen in Figure 6.3.6, the edges need to be treated such that there will be minimum edge relief or potentially no edge relief. These are areas for further development work. Finally, there will need to be metrology measurement development to create a repeatable test to enable very high quality segments to be reliably manufactured.

On the infrastructure side, TMT will provide an excellent precursor for metrology of off-axis segments. This same type of technology may be able to be used on this strategic mission.

In summary, the ability to process a very lightweight segment is close to the high level requirements potentially required. Some level of technology development will fully demonstrate production capability to the LUVOIR specification.

## 6.4 Alternate Architectures for Future Astronomy Missions - Ronald S Polidan and the Evolvable Space Telescope and Rotating Synthetic Aperture Teams

In our presentation to the COPAG we discussed the analyses we have conducted of two alternative architectures for future large astronomical observatories: An Evolvable Space Telescope (EST) concept that starts with a modest size (4 x 12 meter, off axis equivalent to a ~7-meter telescope collecting area) and evolves to a 16 to 20 meter aperture observatory through assembly and servicing in space and a Rotating Synthetic Aperture (RSA) concept that utilizes a rotating large rectangular aperture as its primary mirror.

A primary goal of this effort is to demonstrate to NASA and the science community the *absolute need* to explore alternate architectures to mitigate costs, adapt to new technologies, and better integrate into future in-space assembly and servicing. The space observatory environment has changed substantially over the past decades: both positively, such as advances in mirror and wavefront control technologies (e.g. Stahl et al 2014) and a growing in-space operations capability; and negatively, as budgets have been greatly reduced. The big question is how we can use the significant investments in technology over the past 10 to 15 years, to create a larger aperture, more productive space telescope at less cost. The alternate architecture approaches discussed in this paper are very viable and apply to all of the future science goals. A culture change will be required to implement one of these alternate architectures, but we believe we should not constrain ourselves to a future of rebuilding our grandfather's space telescope.

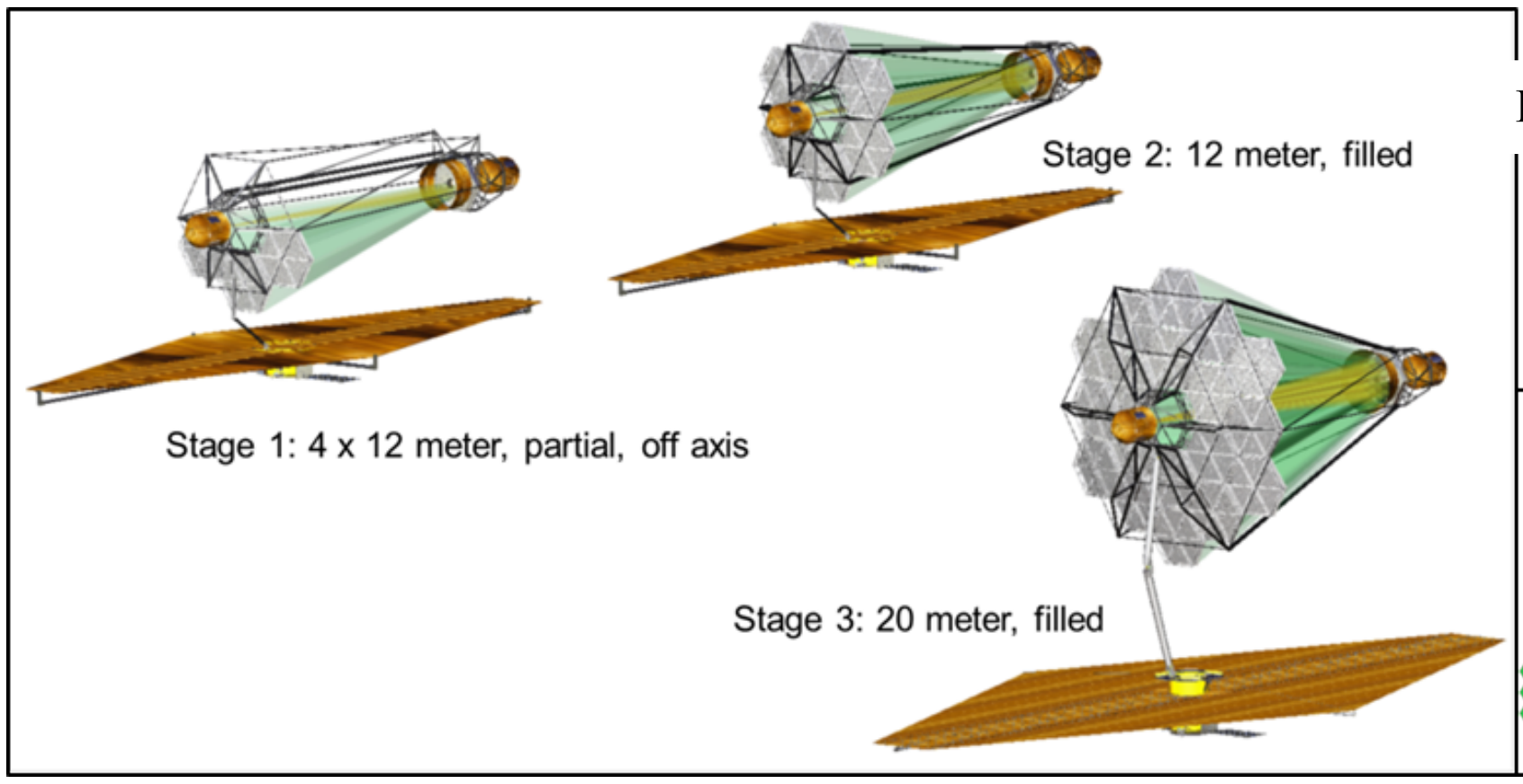

**Figure 6.4.1** EST stages: Primary mirror (below) and full observatory configuration (left)

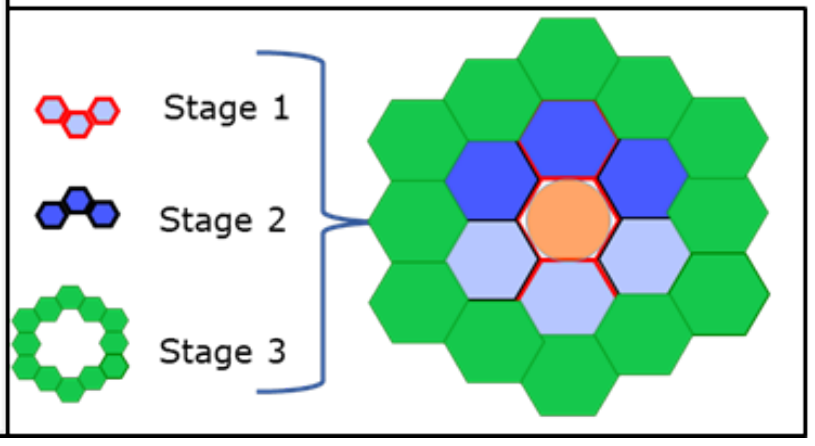

### 6.4.1 Evolvable Space Telescope (EST)

The EST concept focuses upon how to build a large space telescope in a flat science budget era by mitigating big cost peaks, and by encouraging and embracing the development of in-space assembly and servicing infrastructure. EST starts with a modest size off-axis telescope (equivalent to a ~7-meter filled aperture telescope) that is launched as a fully functional telescope with instruments (EST Stage 1) performing first rank science. After the passage of time (~5 years) an augmentation mission sends additional mirror segments, instruments, and any other hardware to the observatory to grow it, in space, into a ~12-meter filled aperture observatory (EST Stage 2). Future augmentation missions would again increase its size and add new instruments and the support hardware to create a ~20-meter filled aperture observatory (EST Stage 3). In short, EST is building upon the Hubble Space Telescope (HST) heritage of service and upgrading to support a space telescope for a long lifetime while increasing its capabilities in accordance with new science issues and advancing technologies. After EST Stage 3 additional augmentations are also possible either to maintain or upgrade the 20-meter telescope for decades or to grow it to even larger sizes with added mirror elements. Other upgrades, such as cryogenic systems to enable coverage of the longer wavelengths available to the James Webb Space Telescope (JWST), may be added to these later stages if the technology advances sufficiently.

### 6.4.2 Rotating Synthetic Aperture (RSA)

The RSA concept offers a very different alternate architecture. This concept was developed for non-astronomical applications and has had a substantial amount of technical investment. Since the technology is relatively mature, our focus has been on analyzing its use for astrophysics, rather than developing a technical solution. RSA is a rectangular aperture telescope that rotates to fill the UV-plane. It offers astrophysics a lower cost, greater resolution, better stability, approach that is also able to meet the starlight suppression needs for exoplanet astrophysics. An assessment of the image quality expected from RSA is in progress.

A full description of both EST and RSA along with additional information on the needed in space infrastructure and a concept for precursor testbed (MoDEST: Modular Demonstration of an Evolvable Space Telescope) on the International Space Station can be found in Polidan et al 2015.

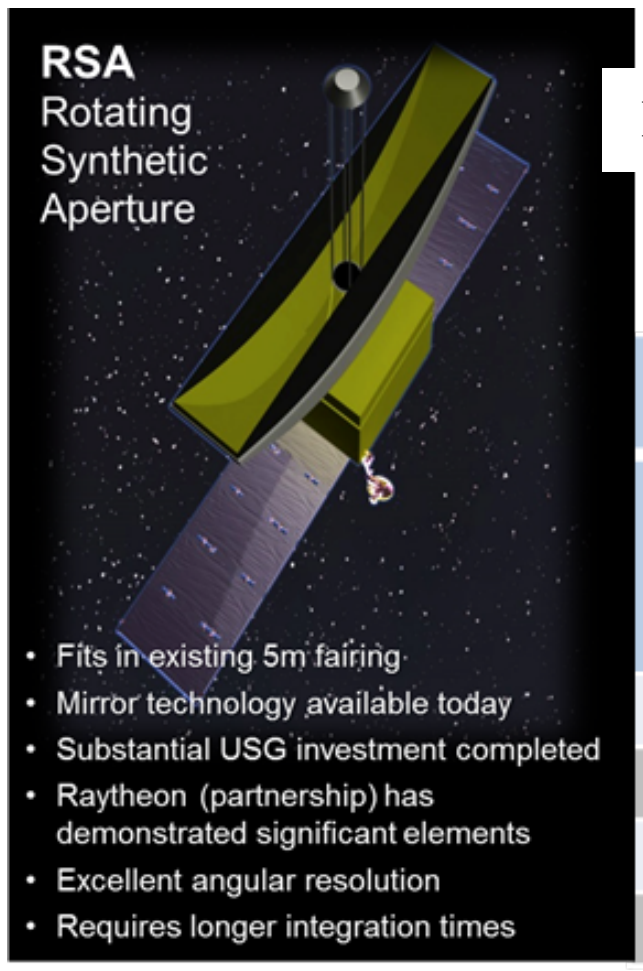

**Figure 6.4.2** RSA Observatory configuration (left) and representative physical parameters and performance comparison to 12 meter circular aperture (below)

| Rotating Synthetic Aperture Parameters | | | | 12-meter Aperture Comparison | | Diameter of a filled aperture with equivalent light gathering capability (m) |
|---|---|---|---|---|---|---|
| Mirror Aspect Ratio | Mirror Length (m) | Mirror Width (m) | Mirror area ($m^2$) | RSA Fill Factor | RSA Integration Time factor | |
| **4:1** | 16 | 4 | 64 | 56.6% | 3.12 | 9.03 |
| **6:1** | 18 | 3 | 54 | 47.7% | 4.39 | 8.29 |
| **8:1** | 20 | 2.5 | 50 | 44.2% | 5.12 | 7.98 |
| **9:1** | 32 | 3.5 | 113 | 100% | 1 | 12 |

### 6.4.3 Summary

We encourage NASA and the astrophysics community not to be constrained to only traditional space telescope approaches, but rather to fully explore and evaluate alternate architectures for these space telescopes. The technologies that enable these different approaches are either mature or maturing rapidly, so the risk levels are very manageable. The development of in-space infrastructure for spacecraft assembly and servicing is also being actively pursued by many government and industrial organizations, which bodes well for the availability of the needed capabilities in the mid-2030's timeframe. These alternate approaches can offer cost savings and performance enhancements over traditional methods and can enable a more capable astrophysical observatory earlier in time than the traditional approach. They do, however, require a culture change.

## 6.5 Far UV Mirror Coating Technology Advances - Juan Larruquert

### 6.5.1 Broadband mirrors

By far the most efficient broadband mirror coating extending to the FUV is a layer of Al protected with $MgF_2$. It has a reflectance of ~80% or more all the way from 120 nm to the infrared and more. It was developed about 50 years ago and it has been basically unsurpassed, except for small reflectance enhancement due to cleaner vacuum deposition systems. It has been largely used in space instrumentation. When high reflectance is required down to ~100 nm, another classical coating is available: Al protected with LiF. The latter material being hygroscopic results in that this coating is only used when there is a real need to reach wavelengths below 120 nm.

A significant reflectance increase both for Al/$MgF_2$ and for Al/LiF in the short FUV has been recently reported (Quijada et al. 2014). It consists in depositing Al normally and initially protect it with a thin film of $MgF_2$ or LiF, both Al and fluoride deposited at room temperature (as for standard Al/$MgF_2$ coatings); then the mirror is heated to 220-250 °C to complete the $MgF_2$ or LiF film thickness. This results in a reflectance above 90% at 120-130 nm. This deposition procedure might not be easily applicable for instance on a large telescope mirror for the difficulty to uniformly heat a large-size substrate. Long-term ageing data, and a better knowledge of the coating structure may still be required for space applications. A possible replacement for $MgF_2$ may be $AlF_3$, since both have a similar transparency cutoff wavelength; even though no large reflectance differences can be predicted over the two materials, other properties like stability, porosity or roughness might vary over them.

The aforementioned fluorides are deposited by evaporation techniques for highest transparency; this results in films with large porosity, where water and contaminant molecules are embedded in the film. A prospective technique to deposit compact layers of fluorides is Atomic Layer Deposition (ALD). ALD's property of growing a film monolayer by monolayer has suggested that an Al film could be just protected with an ultrathin $MgF_2$ film of few monolayers. This way, not only coating reflectance could increase to almost bare Al reflectance, but its operation range could be extended down to ~83 nm. As another prospective technique, a patent was filed on a procedure to transform the native aluminium oxide grown on an Al film with a conversion of aluminum oxide into aluminum fluoride by proper flow of fluorine or a fluorine-containing gas in the chamber. These methods to protect Al with an ultrathin fluoride film are under investigation; if they are demonstrated to be operative, they would revolutionize FUV coatings. We need to be cautious on the expectations for these techniques because of the high reactivity of Al in presence of oxidizing agents, like water vapour; any oxide on Al would strongly degrade Al FUV reflectance.

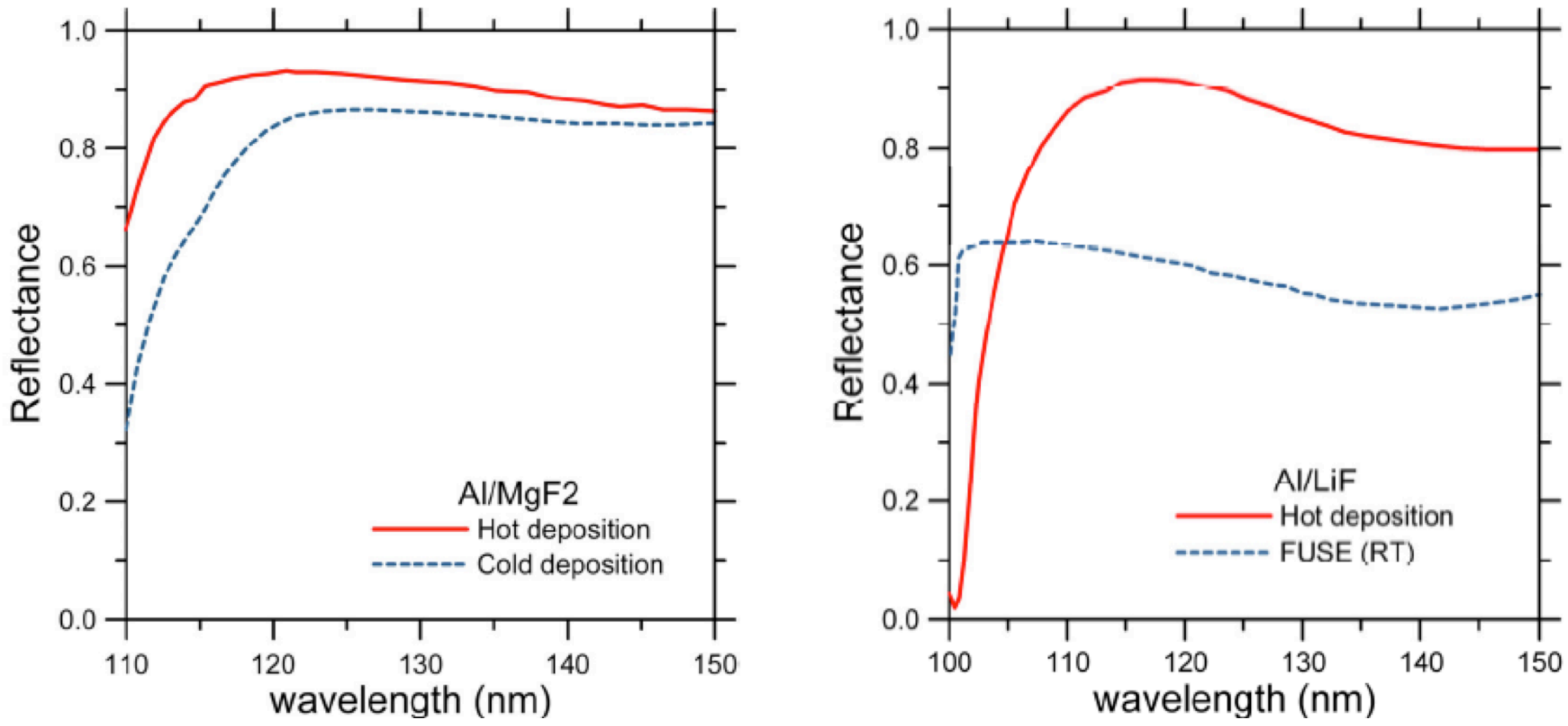

**Figure 6.5.1** - FUV reflectance of Al+$MgF_2$ and Al+LiF mirror coatings. Al and an overcoat of 4-5 (8) nm layer of $MgF_2$ (LiF) deposited at room temperature, later heated to 220ºC (250ºC) to complete the suitable $MgF_2$ (LiF) film thickness (Quijada 2014). For comparison, measurements on coatings at room temperature are also displayed.

Below ~100 nm, reflectance is much smaller than above, since there is no transparent material such as to protect an Al film. CVD SiC (Choyke et al. 1977) provides a reflectance of ~40-50% down to 80 nm; CVD deposition is a high temperature process that requires careful polishing. To deposit at room temperature we can use, single layers of SiC (Keski-Kuha et al. 1988) or B4C (Blumenstock and Keski-Kuha 1994), with a reflectance of ~30% in the spectral range down to ~70-80 nm. A reflectance increase can be obtained with an Al/$MgF_2$/SiC or Al/$MgF_2$/$B_4C$ coating (Larruquert and Keski-Kuha 1999).

### 6.5.2 Narrowband Coatings

Narrowband transmittance filters based on Al/$MgF_2$ multilayers are another classical coating, which have been used in space instrumentation. They can be designed to peak at any wavelength down to 120 nm. Their main drawback is that peak transmittance is relatively low.

Reflective multilayer coatings based on two dielectrics have higher peak performance but lower out-of-band rejection than Al/$MgF_2$ transmittance filters. The main material combinations are ($MgF_2$ or $AlF_3$)/($LaF_3$ or $GdF_3$). Narrowband mirrors based on these materials have been obtained for a wavelength as short as 121 nm (Rodríguez de Marcos et al. 2015) and ~135-140 nm (Zukic and Torr 1992; Gatto et al. 2002).

For still shorter wavelengths, a narrowband reflective coating peaked at ~100 nm based on an Al-LiF-SiC-LiF multilayer has been developed (Rodríguez-de Marcos 2013). Its band is suitable to image H Lyman β (102.6 nm) and O VI lines (103.2-103.8 nm) and to reject longer wavelengths like H Lyman α (121.6 nm).

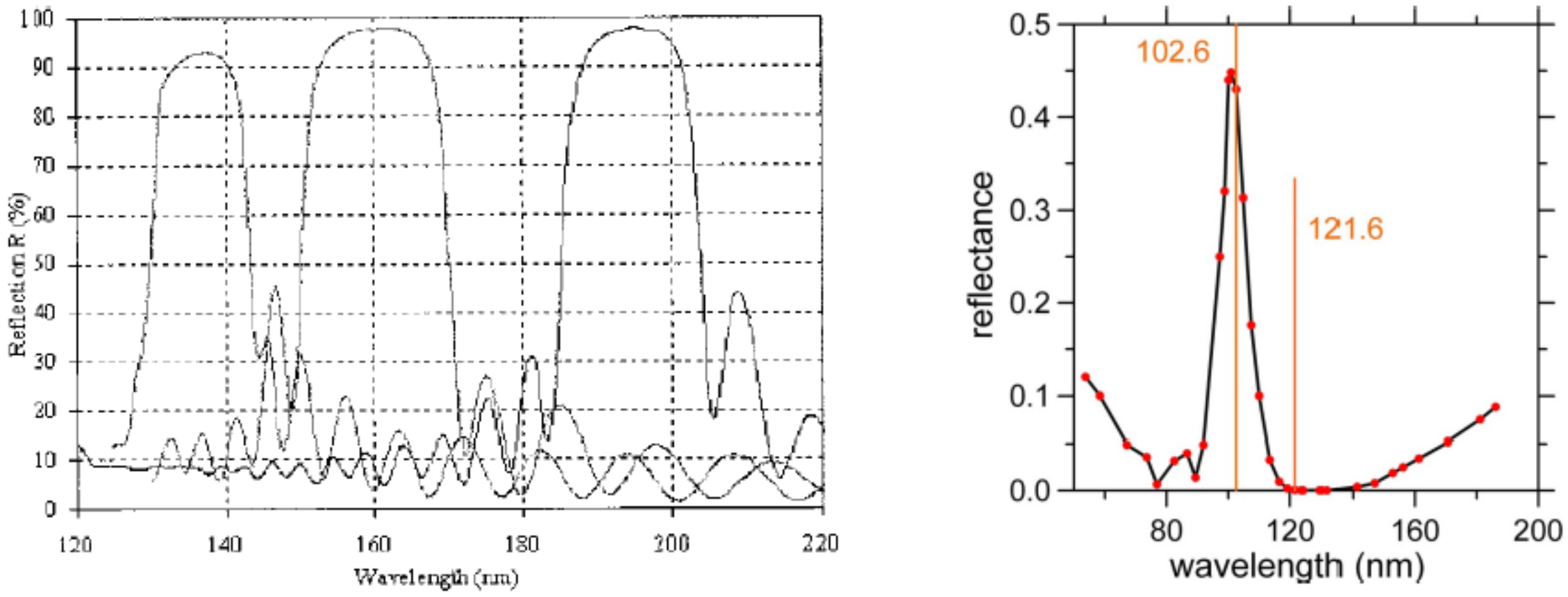

Left: FUV narrowband multilayers (Gatto, Appl. Opt. **41**, 3236, 2002). Right: Al-LiF-SiC-LiF narrowband coating peaked at ~100 nm (Rodríguez-de Marcos, Proc. SPIE **8777**, 87771E, 2013)

# 7 Missions

## 7.1 Summary of Presentation and Discussion: Probes and non-Flagships - David R. Ardila

As part of the work for the SIG #2, the Subcommittee on Mission Diversity discussed mission platforms other than Flagships. The goals were to understand the institutional landscape, and the role of non-Flagships in the context of Flagship plans. Non-flagship missions include Suborbitals (Rockets and Balloons), Cubesats, Missions of Opportunity, Explorers (Small and Medium), and Probes.

We had contributions from Martin Barstow and Ana Gomez de Castro, on the status of European UV efforts, John Hutchings, on Canadian efforts, Mario Perez, on the NASA perspective, and Charles Lillie, on the suborbital experience. Those contributions are summarized elsewhere. The meeting also included wide-ranging discussions on funding and mission cadence.

It was agreed that the augmentation of the Explorer program as recommended by the Decadal Committee has had a strong positive impact on the health of the field. The stable cadence of mission opportunities, not only at the Explorer level, but also at smaller price caps, has contributed to maintain a diverse mission portfolio. Most non-Flagship missions are developed and launch relatively quickly, ensuring knowledge preservation and updated technology.

The smallest missions contribute crucially to the advancement of NASA's goals and are fundamental to train the next generation of space scientists. Sounding rockets, balloons, and now cubesats represent the foundation in which other space missions are built. In particular, cubesats remain under-exploited from the point of view of astrophysics, and can be used for time-domain observations in the UV.

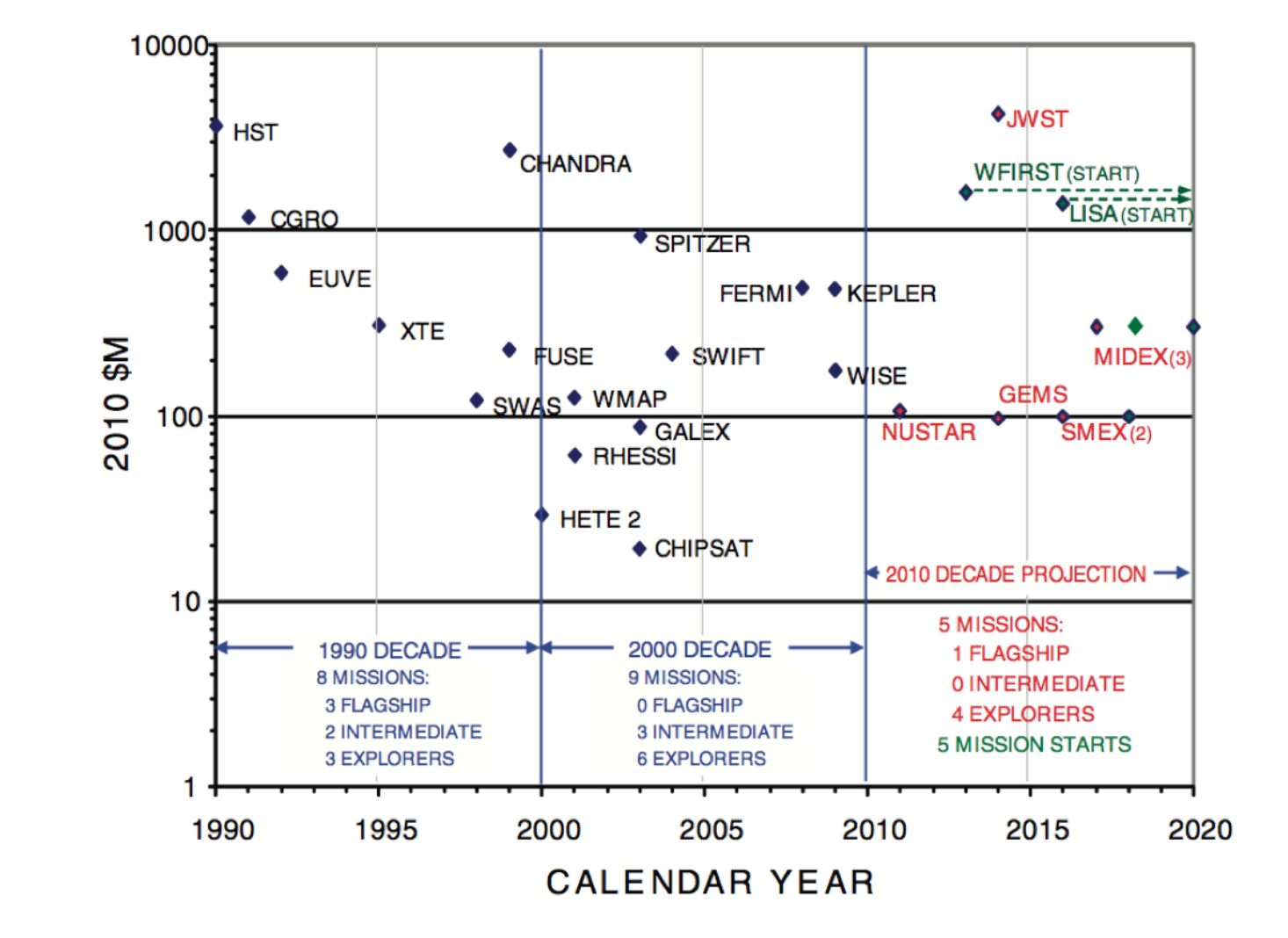

**Figure 7.1.1** - The cadence of Flagships and Explorers, from the perspective of the Astro2010 Decadal Survey (pg. 168). GEMS and LISA were cancelled. JWST's launch is 2018. TESS was selected as one of the MidEx missions.

International co llaboration remains very important for the health of the smaller missions, although it is unclear whether or not it results in cost reductions to any of the partners. Europe and Canada are in a situation analogous to the US regarding the UV: no immediate mission prospects, but considerable interest in new missions. Arago, a UV-Vis mission to perform spectropolarimetry, was submitted to ESA in response to the M5 call. M-class missions are €500 M in cost. If approved, it will launch in the 2030's.

The Subcommittee enthusiastically supported the need to foster public-private partnerships between NASA and private foundations or Universities, in order to reduce costs for the Agency. These partnerships are common for ground-based telescopes but have yet to be implemented in space missions.

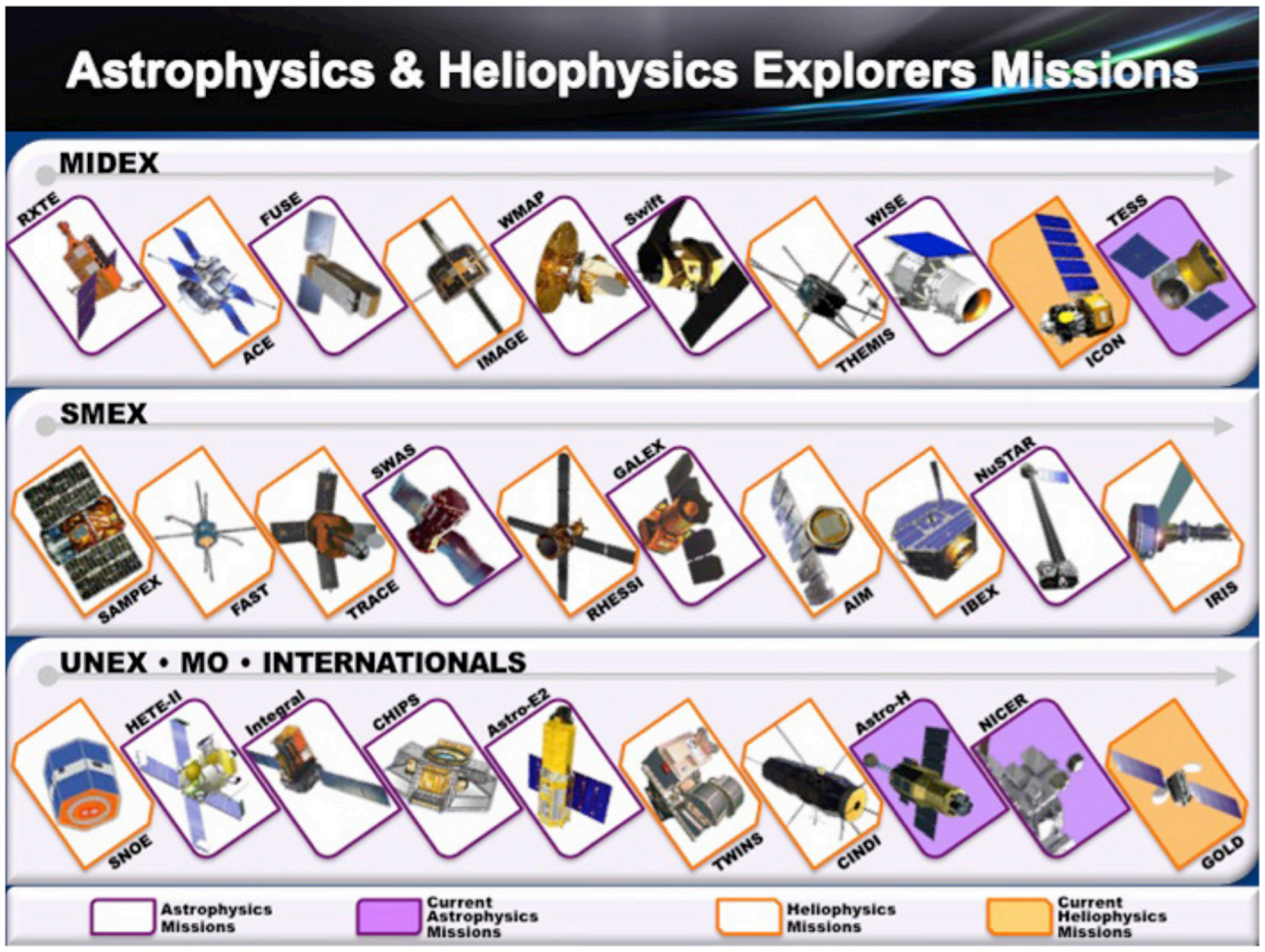

**Figure 7.1.2** - Astrophysics and Heliophysics Explorers. From http://explorers.gsfc.nasa.gov/missions.html

Both at the subcommittee and at the larger SIG level there was strong support for the creation of a Probe line, in the $1B range. However, there was significant concern as to the consequences of a Probe-class line in the cadence of Explorer and smaller missions. A Probe would cost the same as ~3 MidEx-es. MidEx AOs currently have a 4 year cadence. Furthermore, the Subcommittee agreed that if Probes are going to become part of NASA's portfolio,

they need to be developed and launched quickly, within the span of ~5 years.

## 7.2 UV Astronomy in Europe - Ana I. Gómez de Castro and Martin A. Barstow

### 7.2.1 The Network for Ultraviolet Astronomy

There is a broad interest in Europe both in UV science and instrumentation. The community is coordinated through the Network for Ultraviolet Astronomy (NUVA, *www.nuva.eu*) which created a road map for European UV astronomy/technology in 2008 and continues to coordinate community activity (for instance through the organization of the NUVA workshops every three years). A UV astronomy series has been produced from the workshops and are usually published as special volumes of the Journal: Astrophysics and Space Science (the Springer Verlag professional journal in the field). The publications have been widely distributed and include contributions not only from European but also world-wide activities. Figure 7.2.1 shows the active European members of the NUVA by country.

The community scientific interests cover all possible areas of astrophysics: planetary atmospheres and aurorae, minor bodies in the solar system, exoplanets, star formation and young planetary disks, stellar activity, stellar winds, stellar atmospheres, white dwarfs research, massive stars, stellar populations, interstellar medium (from turbulence or the Local Bubble physics to planetary nebulae and supernovae remnants), interacting binaries and cataclysmic variables, Novae and SNe, star formation at galactic scales and the history of star formation in the Universe at $z<2$, chemical evolution, intergalactic medium, gravitational lensing and astrochemistry. However, the stellar astrophysics (from young stars to cataclysmic variables) community dominates NUVA. Main contributors to all these fields can be found in the NUVA publication series. In 2008, the NUVA submitted its road map for UV astronomy to ASTRONET[30]: exploit HST, collaborate in the WSO-UV Russian-led mission and prepare for a NASA led 8-10m class UV mission.

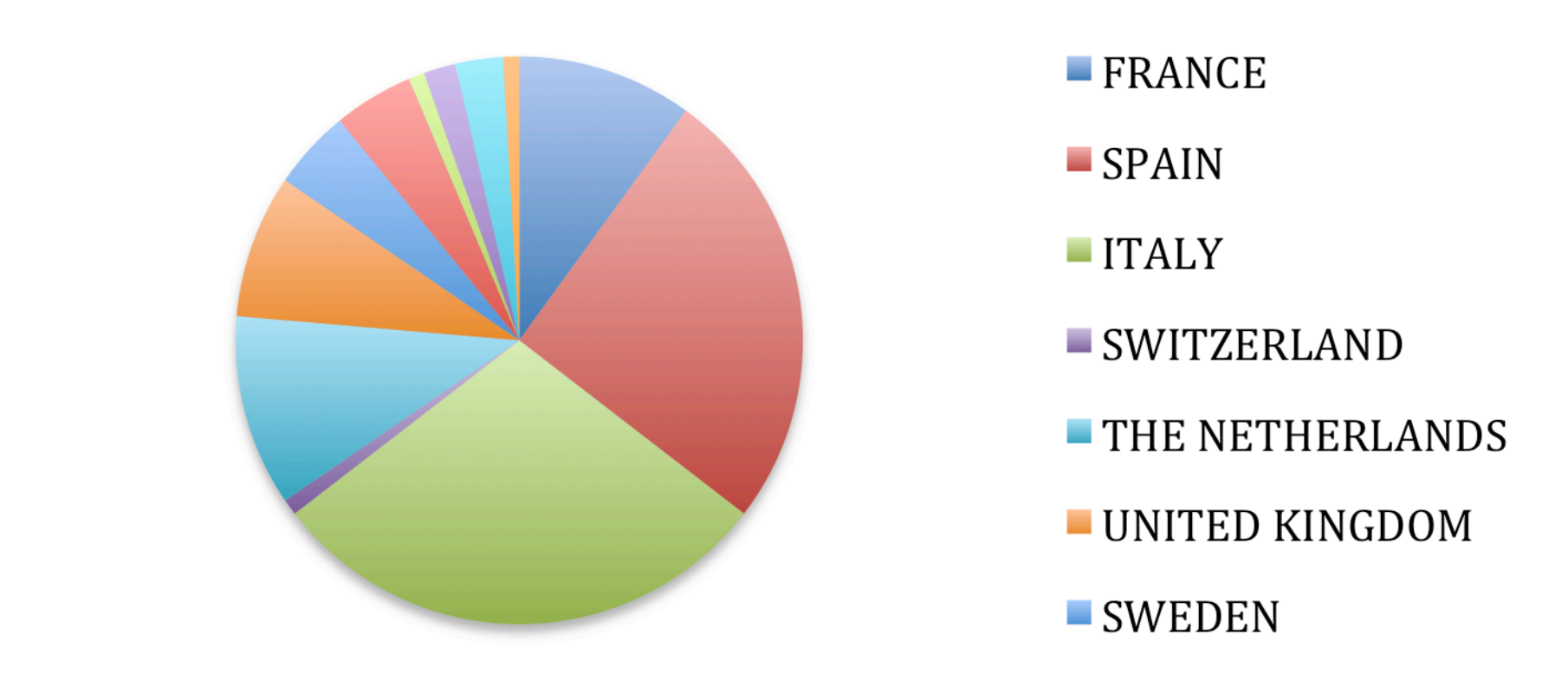

**Figure 7.2.1** - Distribution of active NUVA members per European country.

Recently, NUVA activity has been moved into the new IAU working group of UV astronomy within Division B (Facilities, Technologies and Data Science) to develop a road map for UV astronomy on a global scale that can exploit the synergies between all interested agencies and academia to coordinate the development of UV mission at all funding levels (from mini, cubesats, to Moon based telescopes or large coordinated missions).

[30] ASTRONET is the panEuropean Network elaborating the European Astronomy Infrastructures Road Map through the coordination of the National bodies.

### 7.2.2 The European Space Agency

The European Space Agency (ESA), currently has 20 member states (AT, BE, CH, CZ, DE, DK, ES, FI, FR, IT, GR, IE, LU, NL, NO, PT, PL, RO, SE, UK) and is likely to continue to grow. There are international cooperation agreements with Estonia, Slovenia, Hungary, Cyprus, Latvia, Lithuania, Malta and the Slovak Republic, while negotiations are underway with Bulgaria and Croatia. Canada also takes part in some ESA programmes under a long-standing arrangement.

The ESA main science programme is a mandatory one, in that all countries contribute through their annual subscription to the agency. The value of the contributions is calculated as a proportion of the national GDP of each partner. The main objectives of the programme are to:

- Provide best space tools possible for scientific community to achieve & sustain excellence, leading the world with discoveries and innovation.
    - Choice of projects by scientific excellence
    - Selection by competition in a bottom-up process (peer review)
    - Stability for scientific research teams
    - Reference science framework for the community, national agencies and international partners
- Contribute to sustainability of space capabilities & infrastructures in Europe
    - Provide continuity to industry
    - Foster technological innovation
    - Attract bright minds to space activities
    - Provide perspective to launch services and operations

The programme is driven by scientific excellence through a bottom up process where the scientific community provides broad input through open calls for proposals and competitive peer review. For each mission selected, ESA funds the spacecraft, launch, spacecraft operation and part of the science operations. Member states fund the payloads and the remaining science operations. Therefore, the missions are always partnerships between ESA, a range of member states and institutions within them.

While the community proposes science ideas, ESA provides a strategic framework within which these can be developed, to address broad global themes. These sit under the overarching titles of Horizon 2020 (for missions to fly an operate up to this date) and Horizon 2020+, into the future.

- What are the conditions for planetary formation & the emergence of life?
- How does the Solar System work?
- What are the physical fundamental laws of the Universe?
- How did the Universe originate and what is it made of?

ESA currently has four mission types, Large (L), Medium (M), Small (S) and Opportunity (O). The large missions are 1 B€ European flagships, with a high innovation level, with a launch every 7-8 years. There may be non-enabling contributions from international partners. Medium missions use current cutting-edge technology and have a ~500M€ cost cap, with a flight every 3-4 years. They can have substantial international contributions, or can be a contribution to a mission of another agency. Small missions are a new, still experimental, element of the ESA programme. They are intended to be fast in execution with a 50M€ cost cap, with the aim of increasing flight opportunities for European scientists.

Missions of opportunity are contributions to the missions of other agencies. There are several missions currently approved for flight and others under consideration for the ESA Cosmic Vision programme:

- The approved missions in the Cosmic Vision programme, in order of launch date
  - Solar Orbiter (M1) – study of the SUN
  - CHEOPS (S1) – transits from known exo-planets
  - Euclid (M2) – search for dark energy
- The missions that have been selected, but have not yet received final approval are:
  - JUICE (L1) – Jupiter & its icy moons
  - PLATO (M3) – search for exo-planets down to Earth masses
  - Athena (L2) – X-ray observatory
  - SMILE (S2) - Earth's magnetosphere and Solar wind
- The Science theme has been selected for L3
  - Gravitational wave mission

There is potential for future involvement in UV missions through forthcoming calls. A European UV Observatory concept was submitted to the L theme selection process. We recognised there was a low likelihood of success but it was a useful placeholder to keep UV in the ESA consciousness. Strong interest in UV remains present in Europe. A series of UV missions have been proposed to ESA, although none have yet been successful.

### 7.2.3 Prospective UV Missions

In chronological order, the current European mission interests are outlined below:

The **WSO-UV mission (**Shustov et al. 2014; Sachkov et al. 2016**)** is a Russian led mission to build and operate a 170 cm primary space telescope that will be orbiting at HEO (geosynchronous) with intended launch in 2021. WSO-UV will work only in the UV (1150-3150 Å) and will be provided with instrumentation for high dispersion spectroscopy (55,000) in the full wavelength range, long-slit low dispersion (1,200) spectroscopy also in the full range and imaging capabilities: high angular resolution (<100 mas) imaging in the 1150-1750 Å range and moderate angular resolution (0.3 arcsec), wide field of view imaging in the 2000-6000 Å range. The instruments will be equipped with new technology CCD detectors develop by e2V for the project that will provide a unique sensitivity and dynamic range if lab expectations are fulfilled. Instrument integration will start in 2018.

Spain is a partner of ROSCOSMOS in the project, contributes to the development of the imaging instrument and the software for science operations.

WSO-UV science will be managed from two operations centers located in the Institute of Astronomy of the Russian Academy of Sciences, Moscow and in the campus of the Universidad Complutense de Madrid. WSO-UV will have a core science program and guaranteed time for the countries funding the project. Moreover, there will be open time for the world-wide science community.

The **Fresnel Interferometer Array (**Koechlin et al. 2009) is a French led proposal to carry out high angular resolution, high dynamic range imaging suitable to image exoplanets and resolve structures to scales of 0.01 mas in its optimal configuration. The optical design is being developed at the Toulouse University/Mid Pyrinees Observatory. An image is formed by diffraction (instead of the common reflection or refraction) in a two spacecraft, formation flying space telescope. One of the spacecraft deploys the screen diffracting the radiation, and another the optics. The proposal was submitted to an ESA call for medium size missions in 2009 but was rejected because of its low TRL. The concept has been successfully tested in the visible range on the ground using the Nice telescope, and in the near UV (air UV) in the lab. A prototype is under study to be installed on the International Space Station to reach high TRL in the far UV.

The **EUVO** proposal (Gómez de Castro et al. 2014) was submitted by the NUVA community in answer to the call issued in March 2013 by the European Space Agency (ESA) for white papers and science drivers for ESA Cosmic Vision program. EUVO was a proposal to build a large ultraviolet-visible observatory (EUVO) to address unexplored areas of the Cosmic Vision program. The consortium behind the EUVO proposal raised the concern about the feasibility of addressing key issues in the Cosmic Vision program such as the investigation of Planets and the search for Life, the chemical evolution of the Universe or the interaction between galaxies and intergalactic medium, without a large UV facility to observe diffuse matter in space. In a subsequent article, a list of requirements was produced for such a facility.

**ARAGO, SIRIUS** and **MESSIER** are proposals submitted by the European Astronomy Community to ESA for subsequent calls.

The **ARAGO** (Neiner et al. 2014) mission is intended to carry out spectropolarimetry in the 119–888nm with dispersion between 25,000 in the far UV to 35,000 in the red end to measure the magnetic fields of stars. The mission will be submitted to ESA call for M5; the diameter of the primary mirror is 130 cm.

**SIRIUS** is an S-class mission aimed at exploring stellar environments and the interactions between stellar sources in the nearby interstellar medium. It is a highly efficient normal incidence R~5000 EUV spectrograph where focusing and dispersion are deliver by a single optical element, a figure multilayer coated diffraction grating. It will observe the sky in two spectral bands covering 18-22nm and 19.5-25nm.

The **MESSIER** satellite is an S-class mission designed to explore the extremely low surface brightness universe at UV and optical wavelengths. The two driving science cases target the mildly and highly non-linear regimes of structure formation to test two key predictions of the ΛCDM scenario. The satellite is proposed to drift scan the entire sky in 6 bands covering the 200-1000 nm wavelength range to reach the unprecedented surface brightness levels of 34 mag/arcsec$^2$ in the optical and 37 mag/arcsec$^2$ in the UV.

Further details of these projects can be found in the NUVA series of publications. Europe's main instrumentation strengths are in detectors (CCDs and MCPs for UV astronomy) and optics (including optical elements and coatings) for UV instrumentation.

#### 7.2.4 Concluding remarks

There does seem to be renewed appetite within the ESA hierarchy for involvement in a flagship UVOIR facility to follow HST and JWST, along similar lines to past ESA contributions to these missions. We have set up an ad hoc European committee to promote this idea and future M calls may provide the capacity for ESA contributions.

## 7.3 Astrosat and CASTOR - John Hutchings

**Astrosat** is an Indian orbiting observatory launched in Sept 2015, observing from hard X-ray to blue-optical wavelengths. The nominal lifetime is 5 years. There are four co-aligned telescopes and one scanning sky monitor, and all instruments operate simultaneously, allowing for multi-wavelength science. There are two UV telescopes of ~40cm aperture (UVIT), with ~1/2 degree field of view and ~1" resolution. They provide simultaneous coverage of FUV, NUV, and VIS channels, each with a suite of filters. The UV channels also have gratings that provide R~100 spectra of all objects. Canada is a partner, having provided the UVIT detectors.

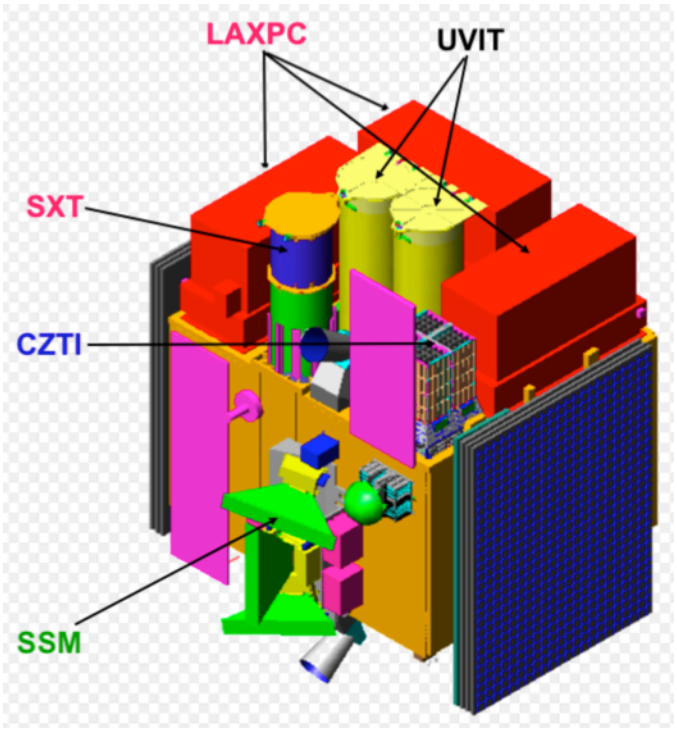

**Figure 7.3.1** - Astrosat orbiting observatory

Commissioning is completed as of March 2016, and guaranteed time science is under way until September 2016. Proposal time begins in October 2016, and that will include time open to all after year 2. All instruments are working well. The website is at http://astrosat.iucaa.in/.

**CASTOR** is a concept for a 1m-class UV-optical telescope with wide field of view and image quality ~0.15". A detailed concept study was carried out by the Canadian Space Agency (CSA) and contractors, and has been featured in a number of conferences in recent years.

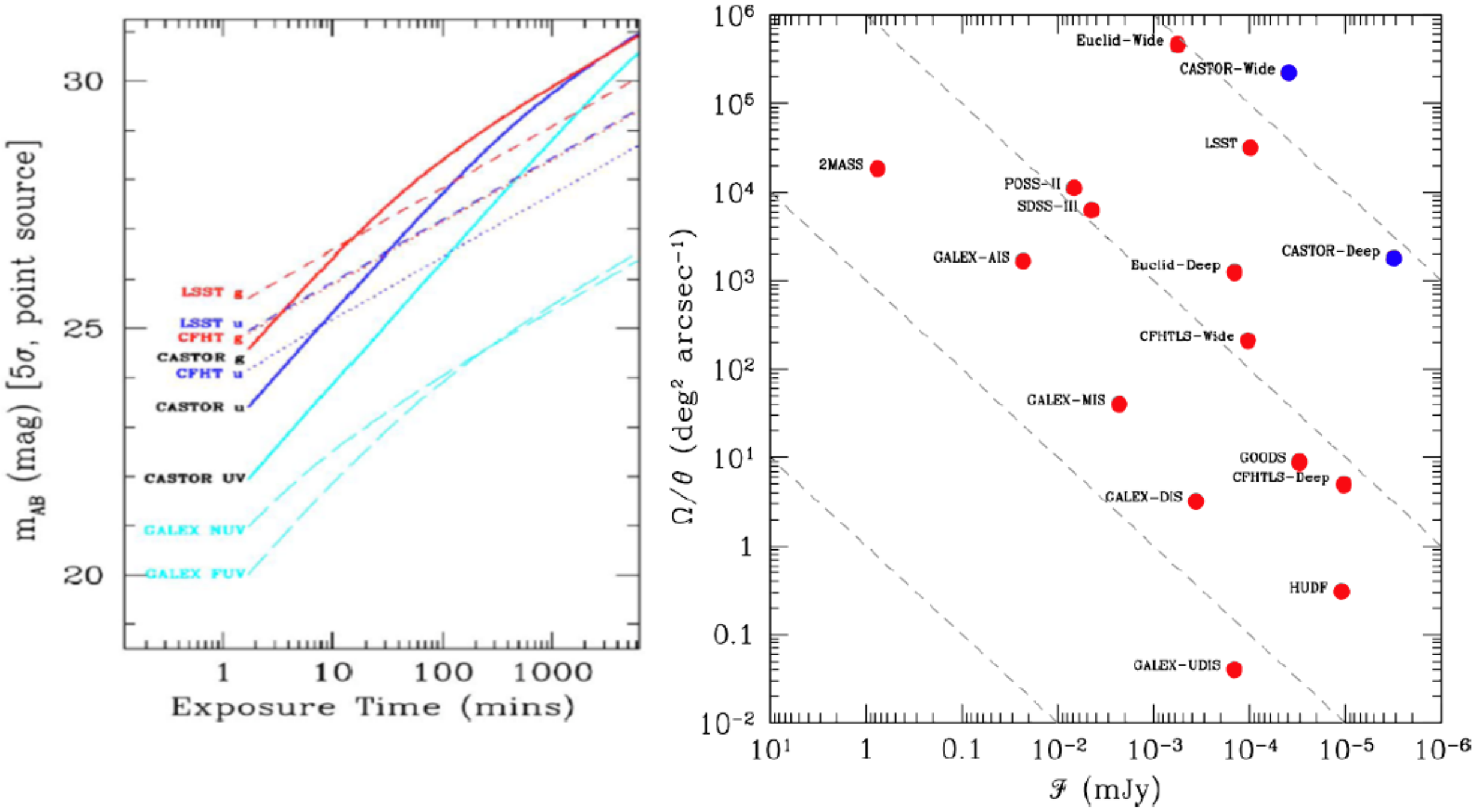

**Figure 7.3.2** - Some CASTOR performance metrics for surveys, compared with others.

The concept is a 1-m diameter unobscured Three Mirror Anastigmat telescope providing good image quality over a 1.16° x 0.58° field of view. A 725 Megapixel camera with wavelength coverage from 150-550 nm , in three wavelength channels, allows access to wavelengths not visible from the ground. The telescope can perform surveys of an area of 5,000 deg2 in <2 years, with planned efficient operations. The aperture provides u-band Wide Survey sensitivity of >27 AB magnitude. There is an option for dispersed imaging with the addition of one moving part.

Subsequent work has developed and prototyped a detector array, and further optimization of the opto-mechanical layout to provide improved coverage in wavelength bands, and provide the necessary red-leak elimination in the UV.  A CSA phase 0 and updated science reference study are in preparation.

CASTOR can provide valuable short-wavelength survey coverage for Euclid and WFIRST, as well as LSST.  CASTOR will also be a facility for a wide range of PI and legacy science.

Significant partnership opportunities are open for this proposed mission.

## 7.4 Suborbital Vehicles - Charles F. Lillie

### 7.4.1 Definition

NASA’s suborbital flight activities involve the use of sounding rockets, aircraft, high-altitude balloons, and suborbital reusable launch vehicles to advance science, and to train the next generation of scientists and engineers who will play a key role in future space missions.

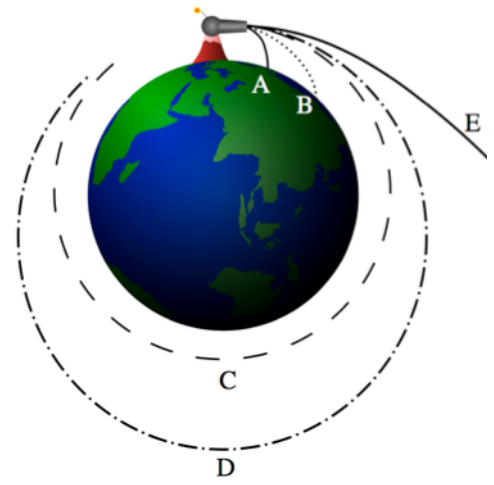

**Figure 7.4.1** - Trajectories versus burnout velocity

Sub-orbital space vehicles are vehicles that are able to reach space, but are unable to achieve the velocity required to orbit the earth.  Thus they have elliptical orbits with an apogee greater than 100 kilometers, a perigee less than 0 km. These vehicles use rocket motors that boost them to velocities between 1.1 and 7.7 km/sec, and then coast to their peak altitude after engine burnout. Their peak altitude and downrange capability increases with their burnout velocity.  Figure 7.4.1 illustrates the elliptical orbits for vehicles with a range of final velocities. Paths A and B depict sub-orbital trajectories; Paths C and D are orbital trajectories; and path E is an escape trajectory.

Although NASA’s aircraft and high altitude balloons cannot reach space, as defined by the “Von Karman Line” at an attitude of 100-km, they also provide an important opportunity to train personnel, develop new instruments, and obtain observations to address cutting-edge space and earth science research questions.

### 7.4.2 Purpose

Suborbital vehicles are an important tool for astronomers who wish to study the universe at wavelengths that are unobservable from the earth’s surface.  Figure 7.4.2 shows the altitudes to which electromagnetic radiation can penetrate the earth’s atmosphere. With the exception of the ~1-cm to 11-m “Radio Window” and the narrow 300-nm to 1100-nm “Optical Window”, sub-orbital vehicles and space vehicles such as NASA’s Great observatories (Figure 7.4.2) are required to open windows on the universe at wavelengths from the Gamma-ray to long wavelength radio region of the spectrum.

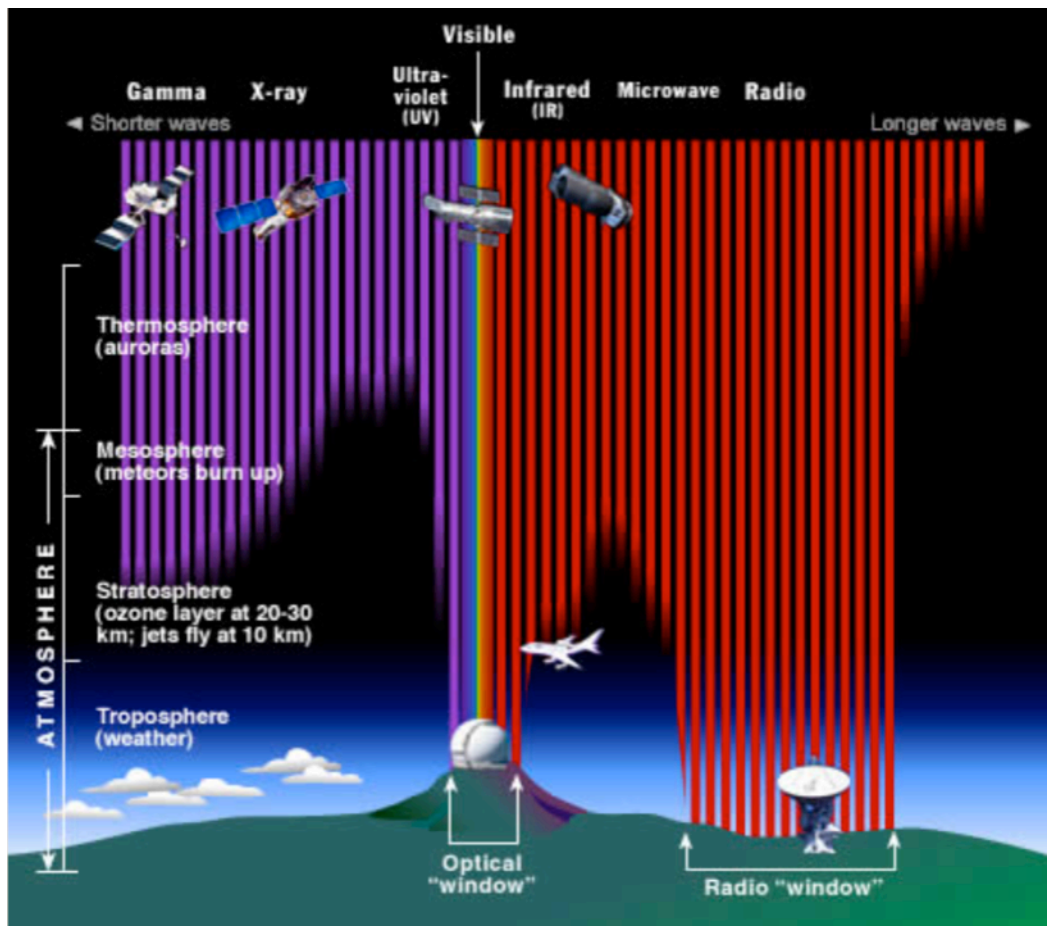

**Figure 7.4.2** – Altitude at which electromagnetic radiation is absorbed by the earth’s atmosphere

### 7.4.3 Available Vehicles

NASA's suborbital program utilizes a variety of vehicles for their Suborbital Research Program. The standard scientific Helium-filled balloons can carry payloads of 3,600 kg to an altitude of 42-km and remain there for as long as 2 weeks. And flights of up to 100 days are now possible with NASA's Ultra Long Duration Balloon (ULDB).

NASA also utilizes several aircraft to test new instruments and collect scientific data for Earth science, including calibration data for satellites in earth orbit. They also operate the SOFIA aircraft that can carry its 2.5-m telescope to an altitude of 41,000 ft. (well above most of the water vapor in the atmosphere) to observe celestial objects in the 0.3 to 655 micron spectral region.

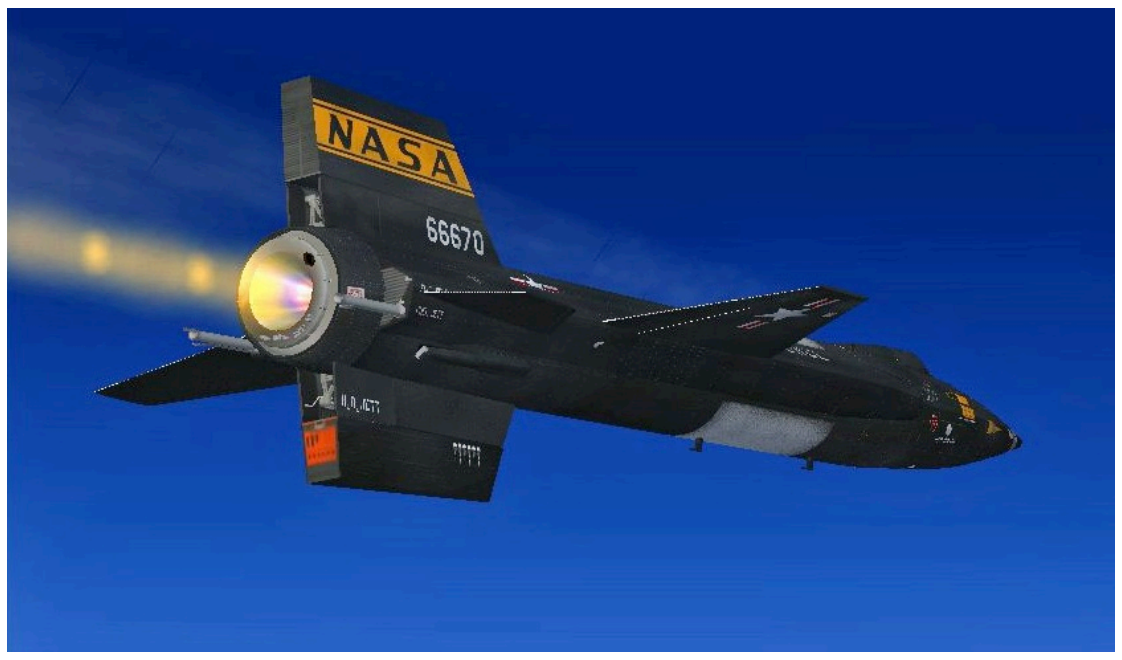

**Figure 7.4.3** - X-15-1 accelerating after launch from under the wing of a B-52 at 45,000 ft. and Mach 0.8

**Rocket-powered aircraft** have been used for scientific research in the past, and NASA has plans to do so again when the next generation of such aircraft become available. During the mid-1960s the X-15-1 was modified to carry scientific instruments to an altitude of 250,000 feet. This included the addition of a hatch over the instrument compartment that could be opened after engine burnout, allowing a pointing platform with four cameras to photograph stars in the 180 to 240 nm spectral region. X-15-1 flew 28 space science experiments, ranging from astronomical observations to micrometeorite collection. NASA is now planning to fly earth and space science experiments on reusable suborbital vehicles such as Virgin Galactic's Spaceship-2, and XCOR's Lynx when they become operational. These rocket-powered aircraft will be able to reach altitudes of 110-km, well above the edge of space. Unfortunately they will have only 4 or 5 minutes of time for observations above the atmosphere before falling back to earth and gliding to a landing.

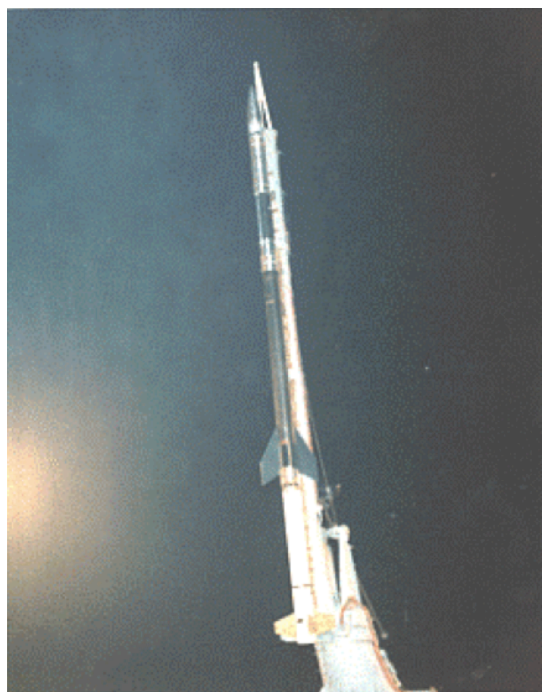

**Figure 7.4.4** - Black Brandt X

**Sounding rockets** have been the primary suborbital vehicle for scientific research. Since 1959 NASA has launched over 2900 sounding rockets to test instruments to be used on satellites and spacecraft and to provide information about the Sun, stars, galaxies and the Earth's atmosphere and radiation. Today there are 15 different types of sounding rockets ranging in size from the single stage, 2.1-m tall Super Arcas to the four-stage 20-m Black Brandt XII. These rockets carry scientific payloads to altitudes ranging from 48 to more than 1300 kilometers with observation times of 5 to 20 minutes. Figure 7.4.4 shows a 3-stage Black Brandt X shortly after launch. The 3$^{rd}$ stage motor is ignited once the vehicle system reaches exo-atmospheric conditions. The standard vehicle is 43.8-cm in diameter with a 3:1 ogive nose shape. The Black Brandt X can lift an 800-lb (362.9-kg) payload to 375-km or a 250-lb (113.4-kg) payload to 875-km.

### 7.4.4 Advantages of Suborbital Vehicles

Suborbital vehicles have several advantages for scientific research when compared with spacecraft in orbit: (1) missions are orders of magnitude less expensive; (2) The time from concept to launch (18 to 36 months) is far shorter, allowing graduate students to conduct a mission and obtain the data for a thesis; (3) they enable the maturation and flight demonstration of new instrument technologies, including instrument architectures, detectors, optical coatings and gratings, etc.; (4 they conduct focused cutting-edge science investigations that address current scientific questions; (5) they can obtain in-situ observations at sub-orbital altitudes; and (6) the payloads can be recovered, allowing them to be modified to correct for in-flight anomalies, updated with new enabling and enhancing technologies, and reused for additional missions to observe other objects.

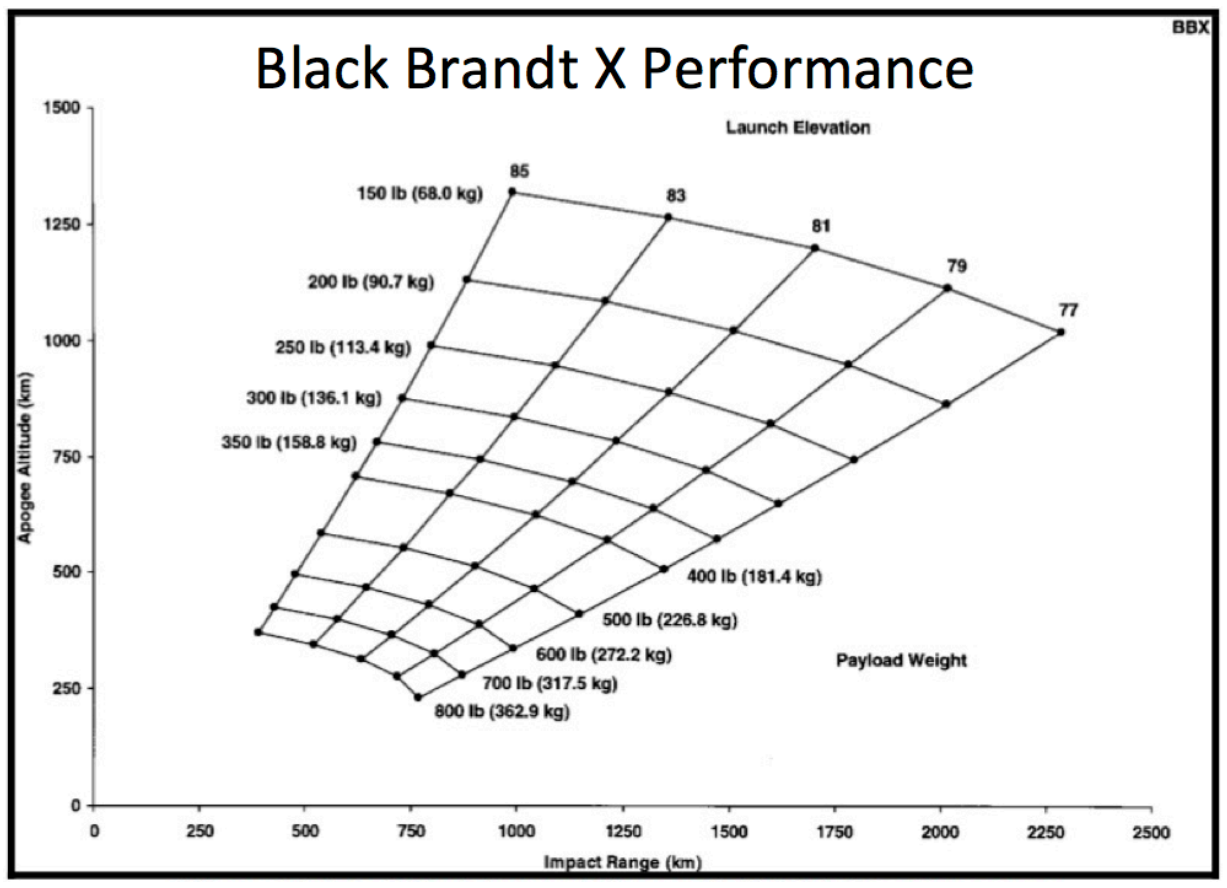

Figure 5. Black Brandt X altitude as a function of payload mass and elevation angle

The downside of course, is the limited observing time in space, i.e., only 5 to 20 minutes for sounding rockets and manned re-useable vehicles, and 5 to 50+ days for high altitude balloons. Nevertheless. Suborbital vehicles are an important, integral part of a balanced research program for NASA that includes small and medium explorers, probes, and large flagship missions.

# 8 Recommendations to the COPAG Executive Committee by the Workshop

## 8.1 Background

In January 2015, Dr. Paul Hertz asked the NASA's Astrophysics Program Analysis Groups (PAGs) to collect and analyze the community's thoughts on which future Flagship missions NASA should give input on for the 2020 NRC Decadal Survey on Astronomy and Astrophysics. This request to the PAGs was motivated by the need to ensure that any Flagship concept presented to the Decadal Survey was sufficiently developed both scientifically and technologically as to be considered plausible and feasible. Four possible Flagship concepts are currently under discussion: a Far-IR Surveyor, a Habitable-Exoplanet Imaging Mission, a UV/Optical/IR Surveyor, and an X-Ray Surveyor. NASA intends to invest limited funds in key technology development that would enable the science and mission concepts envisioned. To this end, the directed charge was:

- Each PAG should reach out to their communities to review the starting set of four Flagship concepts and to suggest additions, subtractions, and providing useful commentary.
- Each PAG will consider what mission studies should be performed to advance astrophysics as a whole.
- Each PAG should not consider that any one mission concept "belongs" to them.

Reports submitted by the PAGs will be used by NASA as the basis for the following subsequent steps to develop these mission concepts:

- Identify a small set of candidate large mission concepts
- Form a community-based Science and Technology Development Team (STDT) for each of these mission concepts
- Ask the teams to articulate the key science drivers and to identify critical technology studies needed in the interim.

- Fund studies of critical technology requirements and ask the teams to review these
- Prepare a case for delivery to the 2020 Decadal Survey committee.

In November 2014, the NASA Advisory Council (NAC) Astrophysics Subcommittee approved a request from the Cosmic Origins Program Analysis Group Chair, Dr. Ken Sembach, to establish a new Science Interest Group (SIG #2) on the future of UV-visible astronomy from space. This SIG met for the first time in January 2015 at the Seattle AAS meeting, and proceeded to solicit community input in direct support of the PAG charges with the specific focus on the UV-visible passband. In June 2015, the SIG held a 2-day workshop at NASA's Goddard Space Flight Center to consider not only the issue of Flagship science, but broader compelling science in the UV-visible, the technology needed to enable that science, and the spectrum of mission sizes needed to conduct that science. As part of that workshop, the assembled community took the opportunity to address the questions levied by Dr. Hertz in a spirited discussion. This summary outlines a set of recommendations to the COPAG Executive Committee motivated by the discussion and agreement arrived at during the workshop.

## 8.2 Recommendation Specifics – an Executive Summary

The main recommendations of this SIG Workshop to the COPAG Executive Committee (EC) are summarized here. More details on the presented science and technology germane to these recommendations are given in the following sections.

- The SIG does not suggest any additions or subtractions to the list of four concept studies.
- The SIG strongly recommends the endorsement and study of both a 10m+ class UVOIR Surveyor and the smaller UV-visible HabEx mission concepts.
- The SIG recognizes the potential of the 10+ m UVOIR Surveyor to make compelling discoveries in both cosmic origins and exoplanet science.
- Based on input from the ExoPAG, the SIG assumes that HabEx concept is smaller than the UVOIR Surveyor although the exact HabEx aperture has not yet been determined. At this time, the SIG did not explicitly explore the astrophysical science applications of a smaller aperture mission. Although a smaller aperture telescope may address many of the same astrophysical themes, it cannot achieve the sensitivity or resolution that a larger 10+ m telescope will deliver. Even more than its aperture, the suitability of HabEx for cosmic origins applications depends critically on two yet-to-be defined capabilities: its field of view in the UV/visible and its sensitivity into the far ultraviolet.
- Among the critical UV-visible technologies that need continued investment to be sufficiently mature for consideration by the 2020 Decadal survey are optical coatings, large format radiation-tolerant high efficiency photon-counting detectors, coronagraphs, and the accommodation of coronagraphic instruments in large-scale telescopes via technologies to address dynamics of the structure. It is particularly important to develop coatings and multiplexing detectors/instruments that maximize sensitivity into the far ultraviolet without compromising coronagraphic requirements.
- The SIG believes that a broad spectrum of precursor missions will be necessary to vet the new science, mature the required technologies and establish the credible workforce required to augment the scientific productivity and impact of large Flagships, while controlling risk and cost. These other missions include not only suborbital and explorer-class facilities, but also Probe-class spacecraft (cost <$1B), more ambitious than Explorers but more focused than Flagships.

- The SIG also believes there are several compelling opportunities to work with international partners towards Flagship-class missions that fulfill the science goals identified in this document, and we encourage NASA to explore these possibilities. In that regard, we see a benefit to including ESA, CSA and Asia-based scientists as observers in the STDTs and request that NASA pursue this possibility.

## 8.3 Compelling Science that Supports a Flagship Mission Implementation

The SIG and the COPAG solicited input from the astrophysics community in the form of short white papers and workshop presentations. We received a total of **35** of these relevant to the UV-visible and to Flagship-class implementation. While this document is purposely intended to be brief for easier digestion, we have asked all those contributors to summarize in Science Traceability Matrix (STM) form the goals and requirements of their science. We include this summary matrix below. This information has been collected in an attempt to represent the breadth of science that a Flagship-class UV-visible mission could address.

This STM details the range of science, the types of capability and how that science maps into fundamental mission properties such as aperture, spectral range, throughput, image scale, spectroscopic resolution, and other factors. On occasion our discussions included valuable relevant IR science goals and technologies. Many of the science problems presented here are motivated by the SIG's great interest in a 10m+ class telescope and require apertures of this size. We did not explore the extent to which 2030's era science goals could be partially addressed by smaller apertures. However, it is clear that wide-field imaging and high sensitivity into the Far-UV are essential requirements regardless of aperture size.

The scientific identity of each STM submission has been preserved to enable the STDT to interact with each submitter as appropriate. We could have further integrated the submissions to similar science questions and capabilities, but that would have represented additional work that NASA HQ has advised it did not want the PAGs to engage in to prepare these recommendations. We have merely presented work already completed as part of the SIG's deliberations running up to our workshop and provide it to the COPAG EC as information to be considered.

We have deliberately not tried to repeat the work of the AURA report "From Cosmic Birth to Living Earths", released July 6, 2015, which did an excellent job of summarizing many kinds of science that could be done with a Surveyor Flagship-class mission facility. However, we recognize that the AURA report was not complete due to their space limitations, and we wanted to make sure that no corner of the community was left unrepresented. That said, our summary is also not complete, but we believe it is more representative of the full range of potential UV-visible Flagship science.

This work should also be viewed in tandem with an earlier call from the COPAG about the future of UV-visible astrophysics in 2012 (Scowen et al 2014), where the focus was more centered on Hubble-class or slightly larger missions, and also in light of the findings of the Theia study (Kasdin 2009) which laid out the astrophysics goals for a 4m-class UVOIR mission.

## 8.4 Technologies that Need Investment

During the deliberations for the SIG workshop it became clear that technological advances for the UV-visible passband have been proceeding through a variety of development investments by both NASA's

Science Mission Directorate and Space Technology Mission Directorate. However, the low maturity level of some crucial technologies does give cause for concern and the SIG makes the strong recommendation that additional immediate investment be provided if those technologies are to be advanced enough for consideration by the Decadal Survey in 2020. The SIG expresses some concern that the timescale for the STDT process is not necessarily consistent with the pressing schedule needed to ensure that the required technologies will conform with SMD guidelines for readiness.

The state of those technologies, their impact, and what additional investments could yield, are summarized in the attached table. Specific conclusions and recommendations include:

- **Reflective Mirror Coatings**: development of a reflective coating that can be deployed in a relevant environment (i.e., mirrors for space missions) that improve upon $MgF_2$ over Al. In addition, the coating must be scalable to a large aperture. The goals would be a reflectivity of > 70% from 90 – 120 nm, a reflectivity of > 90% from 120nm to 1.7 microns, uniformity < 1% for wavelengths > 90 nm, and polarization < 1% over the bandpass
- **UV Detectors**: visible/NIR detectors are excellent devices with improvements mostly being incremental for cosmic origins science, unless exoplanet science is required. One key feature would be better radiation tolerance than is available to state of the art silicon detectors. The lifetime of an instrument using these visible/NIR detectors is limited by the detector, not the spacecraft. For the UV, improvements to DQE – greater than 70% at 90 – 120 nm, larger formats (> 4k × 4k resolution elements), and improvements in dynamic range would increase the science capabilities of a flagship mission. A stable (< 1 pixel) wavelength solution and the ability to observe to a very high signal to noise (>100) is critical to select scientific programs. UV detectors also need to be photon-counting in order to take advantage of the UV minimum in the natural sky background.
- **Opto-mechnical design and validation of large optical systems**: demonstrate fabrication of thermally-stable mirrors within a production schedule that have <7 nm RMS surface-figure. Demonstrate alignment and phasing of segments with gravity release and modeling to demonstrate on-orbit capability. Develop thermally- and dynamically-stable structures for mirror and instrument support. Demonstrate vibration isolation, metrology, and actuator performance to required levels. Validate structural-thermal-optical performance (STOP) models to the picometer level, and verify testing and on-orbit stability of the optical system. Additionally development of methods to reduce the areal cost of primary monolithic mirrors.
- **Polarization-preserving telescope coatings and configurations**: develop low polarization reflective coatings that can be deployed in a relevant environment (i.e., mirrors for space missions). In addition, the coating must be scalable to a large aperture. The goal would be a polarization uniformity of 0.01% to enable space-based precision polarimetry and coronagraph contrasts as high as 1E-11 necessary for terrestrial exoplanet characterization.

After good discussions with our exoplanet science colleagues, we recognize that any Flagship-class UV-visible mission will involve a partnership between the astrophysics and exoplanet communities and that it is essential that technology studies include the accommodation of both kinds of science.

The SIG recommends to the COPAG EC that additional investment be made in the listed critical technologies to enable the next generation science listed in the included Science Traceability Matrix.

## Science Traceability Matrix

| | | | Science Measurement Requirements | | Technical Requirements | | Instrument Performance | | Mission Requirements (Science Driven) | |
|---|---|---|---|---|---|---|---|---|---|---|
| **Science Goal & Author** | **Investigation Theme** | **Investigation Science Objectives** | **Observables** | **Physical Parameter** | **Type** | **Parameter** | **Baseline (ideal)** | **Threshold (minimum acceptable)** | **Parameter 1** | **Parameter 2** |
| To learn how the gas in circumstellar accretion disks is distributed and evolves (Patrick Hartigan - Rice U.) | Use high-resolution narrow-band images to observe UV emission line gas diagnostics in order to spatially resolve the structure is accretion disks and to observe the sites of planet formation. Use moderate-resolution longslit spectra to observe spectral line ratios to determine physical conditions in the gas. Spatially-resolved images of the gas disks will reveal gaps, image accretion streams onto the star and forming planets, show sites of gas accretion onto newly formed planets, and reveal processes of disk photoevaporation. Time-resolved observations will reveal orbital and pattern motion, a critical factor in understanding the physics of accretion disks. | Spatially resolve gas disks in order to understand disk evolution and the formation of planetary systems | Emission Lines ($H_2$, CO) from cool circumstellar gas | Well measured line ratios of many lines will constrain the excitation (X-ray, thermal) and determine the temperature of the gas | Spectral | Spectral range | 912A-2000A | 1210-1700A | Longslit spectrograph, Pixel-size chosen as $\lambda/D$ for UV, Various gratings and slit widths | Spectral capability shortward of Ly $\alpha$, High-spectral resolution to resolve blends and detect Doppler broadening |
| | | | | | | Resolution | 1000-10000 | 5000 | | |
| | | | | | | Spatial coverage | 30 arcsec long slit | 10 arcsec long slit | | |
| | | | Emission Lines (atomic, high ionization states) from gas accreting onto young planets or gas flowing out in a photoevaporative wind | UV and optical spectra of plasma between $10^4$K and $10^5$K, including kinematic information | | Spectral range | 905A-7000A | 1000A-7000A | | |
| | | | | | | Resolution | 1000-10000 | 5000 | | |
| | | | | | | Spatial coverage | 30 arcsec long slit | 10 arcsec long slit | | |
| | | | | | | | | | | |
| | | | Emission Lines ($H_2$, CO) from cool circumstellar gas | Narrow-band high-resolution imaging of emission lines and local continuum in the UV | Imaging | Pixel size | 2mas | 5mas | Imager with full suite of molecular and atomic filters | Optical and UV pixel size, wavelength range and filter specs |
| | | | | | | Filters | Several $H_2$ and CO lines spread across the 912 - 1700 Ang region. | 16 filter minimum | | |
| | | | | | | Field of View | 5 arcminute | 3 arcminute | | |
| | | | Emission Lines (atomic, high ionization states) from gas accreting onto young planets or gas flowing out in a photoevaporative wind | Narrow-band high-resolution imaging of emission lines and local continuum in optical and UV | | Pixel size | 2mas | 5mas | | |
| | | | | | | Filters | Lines: O VI, N V, CII, CIII, CIV, Si IV, He II, H-alpha, H-beta, [OI], [SII], [NII] | 16 filter minimum | | |
| | | | | | | Field of view | 5 arcminute | 3 arcminute | | |
| | | | | | | | | | | |
| Understand when the first stars in the universe formed and how they influenced the environments around them (Dennis Ebbets - Ball Aerospace) | Confirm the identity of and begin to characterize the stellar astrophysics of the first stars (Pop III objects). | 1. Distinguish stellar objects from accretion disks of early black holes. Determine the redshift at which the earliest stars are observed. | 1. Redshifted wavelengths of Hydrogen and Helium lines from stellar atmosphers and surrounding gaseous environment. | 1. For redshift z=10 Hydrogen Lyman series will be observed between 1 to 1.4 microns. He II will be near 1.8 microns. | IFS and/or MOS spectroscopy | Wavelength range | 1000Å to > 2μm to cover stellar signatures from z = 0 to z =10 | | Field of Regard | Any place on celestial sphere over the course of one year |
| | | 2. Determine the end of the eopch of the first stars as the redshift at which the products of stellar nucleosynthesis first appear. | 2. Redshifted wavelengths of C IV, N V and O VI lines in stellar atmospheres. Redshifted wavelengths of nebular emission lines of ejecta from stellar winds and supernovae. | 2. Stellar atmospheric lines will be observable shortward of 1.8 microns. | | Spectral resolving power | R = $\lambda/D\lambda$ = 100 to 200 to measure redshifts with precision of D$\lambda$ = 0.05 | | Pointing stability, jitter control | 1/10 pixel of high resolution imager |
| | | 3. Investigate clustering characteristics, size of star-formation regions, number of objects per region. | 3. Spatial extent of star-forming regions. Number and distribution of individual objects resolvable. | 3. High resolution imaging at wavelengths near peaks of SED of Pop III objects at the observed redshift. | High-resolution imaging | Wavelength range | 1000Å to > 2μm to cover stellar signatures from z = 0 to z =10 | | Sky background | Minimize foregrounds to allow imaging of faintest diffuse objects |
| | | 4. Measure Spectral Energy Distribution, luminosity and effective temperatures. | 4. Flux of light in many spectral bands from observers frame ultraviolet through near infrared. | 4. Spectral Energy Distribution from photometry and/or low-resolution spectroscopy. | | Spectral bandpasses | R = $\lambda/D\lambda$ = 5 to 50 with selectable central wavelength and selectable width. | | Exposure times | Exposures of 10 days or longer for spectroscopy of faint objects |

| | | | | | | | | | |
|---|---|---|---|---|---|---|---|---|---|
| | | 5. Detect and characterize supernova explosions of Pop III objects | 5. Rise time, peak magnitude and decay rate. | 5. Multi-band light curves with a cadence sufficient to sample rise-time, isolate peak brightness and characterize decay. Low resolution spectroscopy to detect signatures of products of explosive nucleosynthesis. | | Spatial Resolution | 20 milli-arc seconds FWHM at λ = 1.0 μm. | Mission duration | > 5 years to allow complete temporal coverage of light curves of 10 or more Pop III supernovae |
| Understand how the universe came to be (mostly) ionized (Stephan McCandliss - JHU) | Determine galaxy and stellar cluster luminosity functions at energies above the hydrogen ionization edge from 0< z<3, covering 11 Gyrs of evolution | Detect and measure the flux above and below the 1 Rydberg in the rest frame of at least 25 galaxies and stellar clusters per redshift bin per luminosity bin to yield a confidence level of 20% per bin. | Mesure Spectral Energy Distribution above and below 912(1+z) with SNR>5 at 900(1+z) from stellar clusters and galaxies. | Intensity | Sensitivity (Expected limits for escape fraction: 100% at low, 1% escape at high z) | $f_{900(1+z)} = 10^{-15}$ erg $cm^{-2}$ $s^{-1}$ $Å^{-1}$ at z = 0.02 $f_{900(1+z)} = 10^{-20}$ erg $cm^{-2}$ $s^{-1}$ $Å^{-1}$ at z = 3 | | Aperture Driver (likely requires 12 meter Gregorian 2 bounce to meet low end) | |
| | | | | | Background | Interplanetary zodi and Lyman alpha limited | | Orbit Driver (L2) | |
| | | | | | Dynamic Range (whole sample) | 100000 | | Detector Driver | |
| | | | | Wavelength (Energy) | Bandwidth | 900 to 3650 Å | | Handle data rate and volume (high rates for bright multiplexed targets) | |
| | | | | | Spectral Resolution | 10 Å | | | |
| | | | | | Redshift resolution | ~ 1 Gyr bins over 11 Gyrs (0<z<3) (11 bins) | | | |
| | | | | | Luminosity resolution | 0.25 magnitude bins over 5 magnitudes of apparent magnitude (20 bins) | | | |
| | | | Hot Stellar Cluster 30 to 100 pc in diameter, Evolution from 0.02<z<3 | Angular | Resolution | 0.075 to 0.250 arcseconds at z= 0.02, 0.004 to 0.013 arcseconds at z=3 | | Attitude (pointing) hold to 0.02 arcseconds per several hour observation | |
| | | | Galaxies 1 kpc to 100 kpc in extent, Evolution from 0.02<z<3 | | | 2.50 to 250 arcseconds at z=0.02 , 0.13 to 13 arcseconds at z=3 | | | |
| | | | Total angular coverage for galactic luminosity functions > 1 degree to reduce cosmic variance | | Instantaneous FOV | Slits ~ 1.5 x 3 $arcseconds^2$ | | Detector and Microshutter Array (MSA) Driver | |
| | | | | | Pointing, multiplexing | Multiobject spectroscopy over ~ 6 x 6 $arcminutes^2$ | | | |
| | | | 25X20=500 Lyman continuum leaking objects per luminosity function. 11 Luminosity functions. Total galactic targets ~ 5500. Total cluster targets ~ 5500. | Temporal samples | Integration time | Depends on Aperture, Detector and MSA | | With multiplexing and high efficiency optical design program could be carrried out in 5Msecond. | |
| | | | | | Single observation duration | Several (5) hours for faintest, few seconds for brightest. Multiplexing is required. | | | |
| | | | | | Cosmic Time and resolution | 0 to 11 billion yrs, ~1Gyr resolution | | | |

| | | | | | | | | | | |
|---|---|---|---|---|---|---|---|---|---|---|
| Understand how the first stars influenced their environments, how the chemical elements were dispersed through the CGM, and how galaxies formed and evolved (Ian Roederer - U. Michigan) | (1) What were the properties (e.g., masses, rotation rates, binary fractions) of the first stars, and what were their supernova explosions like? (2) Better understand stellar nucleosynthesis by studying its products. Identify the nature and site or sites of the r-process. Characterize the physical parameters of the s-process. | Characterize the detailed abundance patterns of metals (Be, B, Si, P, S, Sc, Ti, V, Cr, Mn, Fe, Co, Ni, Zn, etc.) detected in absorption in long-lived, low-mass second generation stars. | absorption spectrum with S/N = 100/1 | Intensity | Sensitivity | S/N >= 100/1 per exposure (for brightest targets) | | | above earth's atmosphere | |
| | | | | | Dynamic Range | N/A | | | | |
| | | | hundreds of absorption lines of 10-20 species | Wavelength (Energy) | Bandwidth | 1700 to 3100 Angstroms | | | Echelle spectrograph should cover this wavelength range (or more) in one or (at most) two exposures | |
| | | | resolve the stellar line widths, or come close | | Resolution | 60,000 sufficient (30,000 minimum; 100,000 ideal) | | | | |
| | | | N/A | Angular | Resolution | N/A | | | | |
| | | | | | Instantaneous FOV | N/A | | | | |
| | | | single-object point-source mode is sufficient | | Pointing, scanning, etc | N/A | | | | |
| | | | no time domain requirements (can co-add multiple exposures taken at different epochs) | Temporal | Integration time | tens of minutes | | | | |
| | | | | | Single observation duration | whatever maximizes time on target to overhead | | | | |
| Conduct spectral imaging observations over the 1200-6600 A range, to advance our understanding of the formation, structure, and evolution of stars and stellar systems. (Ken Carpenter - GSFC) | Use ultra-high spatial resolution spectral-imaging to study the evolution of structure and transport of matter within, from, and between stars and to study stellar magnetic activity by resolving stellar surfaces. | Resolve stellar disks and the surface manifestations of magnetic activity in their atmospheres and the mass flows to, from, around, and between stars to understanding magnetic activity and its impact on the formation, structure, and evolution of stars and stellar systems. | Cover a range of wavelengths from 120 to 660 nm, from the UV into the mid-optical. | Find and determine the structure of physical manifestations of magnetic activity on and between stars. | Spectral | Spectral range | 120-660nm | 120-500nm | 20 filters OR energy resolving detectors | |
| | | | | Determine the physical conditions inside those magnetic structures. | | Number of filters | 20 | 8 | | |
| | | | Take time-resolved, high-quality spectral-images in UV of the surfaces of sun-like stars to 4pc and larger stars to further distances. | Find and follow the dynamic evolution of stellar magnetic structures; determine drivers of stellar magnetic activity with goal of understanding the | Spectral | Minimum SNR | 50 (all targets all filters) | | instrument throughput requirements | telescope aperture (sparse array diameter and individual mirror size) |
| | | | | | | Exposure times | 1 -60 min | 10 min | | |
| | | | Take time-resolved, high quality spectral images in UV of accretion, convection, shocks, pulsations, winds, and jets. | Find and determine the structure of such features. | | filter widths | 1 nm | 1 nm | 20 filters with well-characterized bandpasses OR energy-resolving | |
| | | | | Determine the physical conditions inside these features and measure their evolution in time. Improve theoretical models. Cadence from few hours to years. | | Central wavelengths | 120-660nm | 120-500nm | | |
| | | | | | | Repeat Obs. with cadence of few hours to a year or more. | 1 hr to 10 yr | 1 hr to 5 yr | stable instrument response | calibration program requirements |
| | | | Take high-quality images in UV at multiple wavelengths of stellar surfaces and intra-system flows. | Detect and measure manifestations of magnetic activity (e.g., plages, spots) | Angular | Resolution (defined at 150 nm) | sub-milliarcsec | sub-milliarcsec | Fizeau interferometer beam combiner that can handle ~30 separate beams. Energy-resolving detectors preferred. | Sparse array of ~30 spacecraft, each containing 1 m mirror, with array baselines adjustable from 100 to 1000 m maximum diameter. Plus beam combiner s/c at 5 km distance. |
| | | | | Detect and measure intrasystem mass flows. | | Instantaneous FOV | 4x4 milli-arcsec | 4x4 milli-arcsec | | |
| | | | Optical intensity (10 nm wide filters) variations as function of location on stellar surface with cadence of 1 min. | Perform astereoseismology to measure internal stellar structure. | | Acquire and readout time (optical) | 1 min | 1 min | readout time < 1 min | |
| | | | | Changes in internal structure as function of magnetic activity cycle. | | | | | | |
| | | | Ultra-high angular resolution | Stellar surface and intra-system mass flows. | Precision Formation Flying | orbital location | L2 | | | Mission must be at Sun-Earth L2 to permit precision formation-flying of array. |
| | | | | | | Flight duration and timing | 10 year | 5 year | | Observe stars over significant fractions of magnetic activity cycles (5 yr min, 10 yr desired). |
| | | | | | | | | | | Other intra-system mass flows require 10 year mission. |
| | Systematic survey of circumstellar environments to determine distribution of matter. | Detect spatial distribution of dust, rings and protoplanets. | Extended circumstellar emission. | Intensity | Sensitivity | =Limiting Flux/SNR | | | 1:1e9 contrast ratio w/ CID | |
| | | | | | Dynamic Range | =Max Flux/ sensitivty | | | Roll control +/- 15 degrees within a visit. | |
| | | | | Wavelength (Energy) | Bandwidth | 2.0 microns+ | | | LEO, GEO or L2 for stable PSF | |
| | | | | | Resolution | Broadband | | | | |
| | | | | Angular | Resolution | 0.1" | | | All sky survey | |
| | | | | | Instantaneous FOV | 60" | | | 1/10 Pointing better than pixel size. | |
| | | | | | Pointing, scanning, etc | 0.01" | | | | |
| | | | | Temporal | Integration time | Background limited | | | Operate for 5+ years. | |
| | | | | | Single observation duration | Jitter limited | | | | |
| | Detect clouds and surface features of exoplanets. | Time resolved reflectivity of exoplanets. | Exoplanet light curves. | Intensity | Sensitivity | =Limiting Flux/SNR | | | Maintain fine guidance lock between orientations with muliple GS acquisitions. | |
| | | | | | Dynamic Range | =Max Flux/ sensitivty | | | | |
| | | | | Wavelength (Energy) | Bandwidth | 550 nm | | | | |
| | | | | | Resolution | Broadband | | | | |
| | | | | Angular | Resolution | 0.01" | | | | |
| | | | | | Instantaneous FOV | 10" | | | | |
| | | | | | Pointing, scanning, etc | 0.001" | | | | |
| | | | | Temporal | Integration time | Background limited | | | | |
| | | | | | Single observation duration | Jitter limited | | | | |
| Conduct observations over UVOIR wavelengths, that contribute to the understanding of exoplanets and the circumstellar environment, the low mass end of the stellar IMF, the background | Constrain low end of the stellar | Deep imaging of star | Faint cool stars | Intensity | Sensitivity | =Limiting Flux/SNR | | | | |
| | | | | | Dynamic Range | =Max Flux/ sensitivty | | | | |
| | | | | Wavelength (Energy) | Bandwidth | 2.0 mcrons+ | | | | |
| | | | | | Resolution | Broadband | | | | |
| | | | | Angular | Resolution | 1" | | | | |
| | | | | | Instantaneous FOV | 300" | | | | |

| | | | | | | | |
|---|---|---|---|---|---|---|---|
| background fields of bright stars, and the nature of QSO host galaxies. (Dan Batchelador - FIT) | IMF. | clusters and the field. | Faint cool stars. | Angular | Pointing, scanning, etc | 0.1" | 5-8m monolithic Cassegrain in oversize faring. |
| | | | | Temporal | Integration time | Background limited | |
| | | | | | Single observation duration | Jitter limited | |
| | Deep survey of QSO hosts to determine distribution of matter. | Deep imaging of quasars. | Circumnuclear and extended field around quasars. | Intensity | Sensitivity | =Limiting Flux/SNR | |
| | | | | | Dynamic Range | =Max Flux/ sensitivty | |
| | | | | Wavelength (Energy) | Bandwidth | 500 nm - NIR | |
| | | | | | Resolution | Broadband | |
| | | | | Angular | Resolution | 0.1" | |
| | | | | | Instantaneous FOV | 60" | |
| | | | | | Pointing, scanning, etc | 0.01" | |
| | | | | Temporal | Integration time | Background limited | |
| | | | | | Single observation duration | Jitter limited | |
| | Unknown discovery space around bright stars. | Deep imaging of bright star fields. | Bright stars. | Intensity | Sensitivity | =Limiting Flux/SNR | |
| | | | | | Dynamic Range | =Max Flux/ sensitivty | |
| | | | | Wavelength (Energy) | Bandwidth | UVOIR | |
| | | | | | Resolution | Broadband | |
| | | | | Angular | Resolution | 1" | |
| | | | | | Instantaneous FOV | 300" | |
| | | | | | Pointing, scanning, etc | 0.1" | |
| | | | | Temporal | Integration time | Background limited | |
| | | | | | Single observation duration | Jitter limited | |

| | | | | | | | | | | |
|---|---|---|---|---|---|---|---|---|---|---|
| Tracing the galaxy evolution and rejuvenation processes in nearby (< 40 Mpc) early ty-pe galaxies (ETGs) in low density environments (LDEs). Combined observations with space (e.g. X-ray) and ground based radio and sub-millimiter new generation observatories are required | Derive mechanisms of evolution investigating ETGs, from giant to dwarfs, members of associations of different galaxy richness. Separate secular vs. external evolutionary mechanisms. | Reveal and map sub-structures in galaxies, e.g. stellar and gas streams, rings, shells, tidal tails, external UV disks. Investigate their link with the IGM. Determine the kinematics of the sub-structure, derive aboundances tracing stellar evolution and metallicity enrichment. | Multi-Object-Spectroscopy (R>=3000) large FOV (>=4') | Measure stellar and gas aboundances from absorption/emission features | Spectral | Spectral range | 350-1600nm | 350-900nm | Obtain 50-100 spectra per galaxies in Near UV, Optical, NIR | |
| | | | | Measure kinematics of substructures stars vs. gas from absorption/emission features | | Number of slits | 100 | 50 | | |
| | | | Far UV-Optical Integral Field Spectroscopy (possibility of Intermediate to high R<=100000) large FOV (>=4') | Measure stellar and gas aboundances from absorption/emission features | Spectral | Spectral range | 90-900nm | | Obtain 50-100 spectra per galaxy in Far UV-Optical | |
| | | | | Measure kinematics of Far UV bright sub-structures | | Number of slices | 100 | 50 | | |
| | | | UV-optical Imaging | Detect sub-structures. Build HR diagramsfor nearby ETGs. Measure physical properties and distribution of the gas e.g. | Photometry | Minimum SNR | 300 (all targets all filters) | 100 (all targets all filters) | an ample set of filters with well-characterized standard bandpasses including narrow band filters. | telescope aperture |
| | | | | | | Wavelengths range UV-Optical | 90-900nm | | | |
| | | | imaging NIR+MIR. | Detect sub-structures. Build HR diagrams for nearby ETGs. Measure physical properties and distribution of the gas (atomic, molecular) e.g. with | | Exposure times | 1–200 sec | 1–200 sec | | stable instrument response |
| | | | | | | Narrow band filter widths | <=10 nm | <=10 nm | | |
| | | | | | | Wavelengths range NIR+MIR | 1000-10000nm | 1000-5000nm | | calibration program required |
| | | | Cover a FOV as large as possible | Spatial resolution and FOV for | | Resolution (defined at 400 nm) | 0.05" | 0.1" | plate scale of 0.02" to 0.05" per pixel (for 2k | |
| are required. (Roberto Rampazzo - INAF, Padova) | | | (>6') to map sub-structures at high resolution. | Spatial resolution and FOV for imager. | Angular | Pointing, scanning, etc | 0.1" error over FOV | 0.1" error over FOV | x 2k array); stable focal plane and telescope assembly | |
| | | | | | | Instantaneous FOV | >=6' | 4' | | |
| | | | | | Temporal | Total Integration Time | >10 hours | | | |
| | | | | | | Single Observation duration | 1-200 sec | 1-200 sec | | |
| To learn how accretion disks in young stars collimate and accelerate supersonic jets (Patrick Hartigan - Rice U.) | Use high-resolution narrow-band images to observe jets as they become collimated. Use spectral line ratios to define temperature, ionization fraction and density. Spatially-resolved higher-spectral resolution observations define dynamics within flows as they are launched. Time-resolved observations follow knots as they are ejected. UV spectra connect the flows to accretion events. | Derive physical conditions in jet collimation regions and observe time-evolution in order to understand MHD disk wind collimation and acceleration | Emission Lines from radiative shock waves | Spatially-resolved emission line ratios from the optical through the UV | Spectral | Spectral range | 912A-9000A | 1216-7000A | Longslit spectrograph, Pixel-size chosen as λ/D for UV, Various gratings and slit widths | Spectral capability shortward of Ly α, High-spectral resolution to resolve blends and detect Doppler broadening |
| | | | | | | Resolution | 1000-30000 | 1000-20000 | | |
| | | | | | | Spatial coverage | 5 arcminute slit | 3 arcminute slit | | |
| | | | Stellar accretion shock diagnostics | UV spectra of plasma between 10^4K and 10^6K | | Spectral range | 912A-3000A | 1000A-3000A | | |
| | | | | | | Resolution | 1000-30000 | 1000-20000 | | |
| | | | | | | Spatial coverage | N/A | N/A | | |
| | | | Emission Lines from reconnection point | Spatially-resolved UV spectra of plasma between 10^4K and 10^6K | | Spectral range | 912A-7000A | 912A-3000A | | |
| | | | | | | Resolution | 1000-10000 | 1000-3000 | | |
| | | | | | | Spatial coverage | 30 arcsec slit | 20 arcsecond slit | | |
| | | | Emission Lines from radiative shock waves | Narrow-band high-resolution imaging of emission lines in optical and UV | Imaging | Pixel size | 3mas | 5mas | Imager with full suite of nebular filters | Optical and UV pixel size, wavelength range and filter specs |
| | | | | | | Filters | SII 6716, SII 6731, NII 6583, Hα, OI 6300, NI 5200, OII 3727, OIII 5007, MgII 2800, CIII 1909, HeII 1640, CIV 1550, OVI 1036, etc. | 16 filter minimum | | |
| | | | | | | Field of View | 5 arcminute | 3 arcminute | | |
| | | | Emission Lines from reconnection point | Narrow-band high-resolution imaging of emission lines in optical and UV | | Pixel size | 3mas | 5mas | | |
| | | | | | | Filters | Similar to above | 16 filter minimum | | |
| | | | | | | Field of view | 20" | 10" | | |
| Obtain high-resolution ultraviolet spectroscopy of ~200 white dwarfs that are polluted by the debris of planetary debris. (Boris Gaensicke - U. Warwick) | Derive the bulk abundances of the planetary debris in these systems using model atmosphere and diffusion analyses. | Obtain a deep statistical understanding of the abundances of exo-planetary systems, that is comparable to what we achieved (primarily via meteorite studies) in the solar system. These data will guide our understanding of, and the theoretical models of planet formation | Ultraviolet spectroscopy, Photospheric abundances | Cover a wide range of atomic transitions, including C, N, O, Si, Mg, Ca, Al, Ti, S, P, Ti | Spectral | Spectral range | 92-360nm | 100-320nm | Detectors, optical elements, coatings | |
| | | | | Resolve line blends, separate photospheric and ISM lines | | Spectral resolution | 40000 | 20000 | Optical elements, detector size | |
| | | | | Increase sample volume accessible to detailed abundance studies | | Sensitivity | S/N~50 at 1e-15 continuum flux in 1h | S/N~50 at 5e-15 continuum flux in 1h | Coatings, no. of reflections, detector efficiency | Aperture size |
| Survey the extent and feeding of the CGM about galaxies. (Chris Howk - U. Notre | *Use multi-object spectroscopy (MOS) to map H I in CGM using multi-object spectroscopy of background galaxies/QSOs, examining external source of CGM flows (i.e., the connection to the IGM). *Use MOS to map individual launch points of feedback-driven outflows about | *Column densities and kinematics of CGM gas to R_vir. *Kinematics of outflows being fed by individual OB associations | H I Ly-alpha + Lyman break. O VI, C III, C IV. | Intensity | SNR | 25 | 10 | | Large aperture required to allow spectroscopy of faint background galaxies. | |
| | | | H I Ly-alpha + Lyman break. O VI, C III, C IV. | Wavelength | Range | 900-3200 Å | 1000-2000 Å | | Resolution+SNR push large apertures. Need good resolution for probing outflows / doing physics on CGM absorption. | |
| | | | Resolve Ly-series absorption from interloping IGM. Detect galactic outflows in O VI absorption toward OB assocations. | | Resolution | 5,000 | 2,000 | | | |
| | | | | | Resolution | | | | Sufficient to separate individual OB associations. | |
| Howk - U. Notre Dame) | feedback-driven outflows about galaxies. Target OB associations across face of galaxies. *Map CGM in emission with long "stares" in an MOS mode. | individual OB associations. | | Angular | Instantaneous FOV | 6' – 10' | 2' | | FoV matched to virial diameter of massive galaxies at z~0.1 in the ideal case. Larger field provides more | |
| | | | | | Pointing, scanning, etc | | | | | |
| | | | | Temporal | Integration time | <50 ksec | | | Exposure times short enough to allow surveys of 10s of galaxies. | |
| | | | | | Single observation duration | 5 ksec | | | | |
| Conduct a survey of the baryonic and metal content in the circumgalactic medium (CGM) of galaxies as a function of gas "phase." Trace the exchange of matter between galaxies and the intergalactic medium. (Chris Howk - U. Notre Dame) | *Map the circumgalactic medium absorption about individual galaxies toward several background sources at high resolution and signal-to-noise. *Measure velocities of circumgalactic gas with respect to galaxy host. *Assess metallicity from inner circumgalactic medium through the virial radius and into the intergalactic medium. *Use multiple ions to probe the distribution of baryons, metals with temperature. Probe the hottest 10^6 K gas at nearly the virial temperature with FUV/EUV transitions (O VI, Ne VIII, Mg X, Si XII) ; probe the cooler gas with H I, Si II, Si III, Si IV, C II, C III, C IV. | High resolution spectroscopy of UV wavelengths giving access to redshifted EUV/UV absorption/emission diagnostics | Detect weak EUV transitions tracing the hottest gas, low metal column density tracing IGM inflow. | Intensity | SNR | 50 | 25 | | High SNR needed for weak metal lines. Pushes to large aperture, good reflectivity. | |
| | | | Observe H I features incl. 1 Ryd break throughout Universe; access critical EUV transitions. | Wavelength | Range | 900-3200 Å | 1000-3200 | | Resolution+SNR push large apertures. Need for multiple background objects (notably QSOs) also | |
| | | | Resolve internal motions in CGM/diff. between phases. | | Resolution | 45,000-60,000 | 20,000 | | | |
| | | | | Angular | Resolution | | | | Resolution sufficient to provide necessary spectral resolution. | |
| | | | | | Instantaneous FOV | | | | Key for CGM science is stable, well-characterized spectral LSF, preferrably very | |
| | | | | | Pointing, scanning, etc | | | | | |
| | | | | Temporal | Integration time | <50 ksec | | | Exposure times short enough to do many QSOs behind individual galaxies | |
| | | | | | Single observation duration | 5 ksec | | | | |
| Conduct ultraviolet to optical observations, that contribute to the understanding of the evolution and fates of massive stars. (Aida Wofford - Institut d'Astrophysique de Paris) | Collection of statistically significant samples over the entire HRD and in environments that are extreme in terms of SFR, metallicity, reddening. | Derive statistics on the physical properties of massive stars, their evolved descendants, and the precursors of supernovae (temperature, luminosity, mass loss rate, rotation speed, binary masses and periods, chemical composition of atmospheres and ejecta, spatial location as a function of type, variability) | Cover a range of wavelengths in UV and visible | Measure stellar temperature, luminosity, mass, mass loss rate, terminal velocity, rotation speed | Spectral | Spectral range | 90-1000nm | 120-1000nm | ??+ filter imaging in UV/vis range | |
| | | | | Measure chemical composition of atmospheres and ejecta | | Spectral resolution | - | 20,000 | | |
| | | | Take high-quality images of individual massive stars in | Study spatial location of massive stars and their | Photometry | Minimum SNR | - | 25 | instrument throughput | telescope aperture |
| | | | | | | Exposure times | - | - | | |
| | | | Cover a range of wavelengths in UV and visible | Measure extinction due to dust towards many sightlines | | Central wavelengths | 90-1000nm | 120-1000nm | ??+ filters with well-characterized bandpasses | |
| | | | | | | Accuracy and systematic error | - | - | | |
| | | | | | | Resolution (defined at 150 nm) | 0.05" | 0.1" | stable instrument response | calibration program requirements |
| | | | | | Angular | Stable distortion solution | 0.1" error over FOV | 0.1" error over FOV | plate scale of 0.02" to 0.05" per pixel (for 2k x 2k array); stable focal plane and telescope assembly | - |
| | | | | | | Instantaneous FOV | 60" | > 50" | | |
| | | | Sample at cadence tied to binary orbital motions | Obtain masses and periods of binaries | Target temporal sampling and sky location | Acquire and readout time | ?? min | ?? min | readout time < ?? min | - |
| | | | Sample at multiple epochs to study history of mass loss | Study eruptions of LBVs and other massive star descendants | | | | | | |
| | | | | Study conditons prior to supernova explosions | | | | | | |

| | | | | | | | |
|---|---|---|---|---|---|---|---|
| Conduct observations from the UV to the radio to understand star formation within galaxies and how it drives galaxy growth througout space and time. (Daniela Calzetti - U. Massachusetts) | Observe nearby galaxies within 100 Mpc to obtain statistics within a representative volume of the Universe of star formation histories with lookback time from ~1 Myr to 13 Gyr. | Determine the Upper End (slope, maximum star mass) of the Stellar Initial Mass Function | Isolate massive stars within star clusters out to 10 Mpc and individual star clusters out to 100 Mpc | Intensity | Sensitivity | 2 M_sun star at 10 Mpc in 0.1 micron band | |
| | | | | | Dynamic Range | 2-300 M_sun stars at 10 Mpc | |
| | | | by getting their UV-to-nearIR spectra plus | Wavelength (Energy) | Bandwidth | 0.09-1.2 micron | |
| | | | | | Resolution | 3000-5000 | |
| | | | with spatial resolution of 0.05 pc (per pixel) at 5 Mpc and 1 pc at 100 Mpc | Angular | Resolution | 0.004" (2 pixels) | |
| | | | and spatial coverage of 3 arcmin x 3 arcmin (rough galaxy size) | | Instantaneous FOV | 3'x3' | |
| | | | | | Pointing, scanning, etc | | |
| | | | Simultaneous coverage of multiple stars and star clusters (multi-object spectroscopy) | Angular | Aperture size (spectroscopy) | 0.006" | |
| | | | | | number of apertures | >10000 each pointing | |
| | | Reconcile the Local and Cosmic Star Formation Histories | Recent-past star formation histories (past 1 Gyr with 10 Myr time resolution) of local | Intensity | Sensitivity | m_UV~31 mag | |
| | | | | | Dynamic Range | 8-10 mag | |
| | | | by getting their UV-to-nearIR broad/medium/narrow band | Wavelength (Energy) | Bandwidth | 0.15-1.5 micron | |
| | | | | | Resolution | 10-300 | |
| | | | with spatial resolution of 0.05 pc (per pixel) at 5 Mpc | Angular | Resolution | 0.004" (2 pixels) | |
| | | | and spatial coverage of 3 arcmin x 3 arcmin (rough galaxy size) | | Instantaneous FOV | 3'x3' | |
| | | | | | Pointing, scanning, etc | | |
| | | Determine the growth of galaxies by studying the stellar population content, star formation modes, and star formation histories of their outskirts | Color-magnitude diagrams of stars down to 0.5-1 mag below the Main Sequence turn off at | Intensity | Sensitivity | m_V=35 mag | |
| | | | | | Dynamic Range | 8-10 mag | |
| | | | UV-to-nearIR medium/broad band photometry | Wavelength (Energy) | Bandwidth | 0.15-1.5 micron | |
| | | | | | Resolution | 10 -- 50 | |
| | | | with spatial resolution of 0.1 pc (per pixel) at 10 Mpc | Angular | Resolution | 0.004" (2 pixels) | |
| | | | and spatial coverage of 3 arcmin x 3 arcmin (rough galaxy size) | | Instantaneous FOV | 3'x3' | |
| | | | | | Pointing, scanning, etc | | |
| Understand how protoplanetary disks evolve and form planetary systems (Kevin France - U.Colorado) | Determine the accretion luminosity of protostars, the composition of the inner 10 AU of planet-forming disks, and the lifetime of gas disks in young planetary systems | Mass accretion rates of protostars; abundances, physical conditions, and lifetimes of molecular gas in the inner regions of protoplanetary disks | 1) Flux calibrated and spectrally resolved ultraviolet/optical spectra of > 30 protostellar systems in each of 10 star-forming regions with a variety | Intensity | Sensitivity | S/N = 10 @ 1E -16 [erg /cm2/s] per 60s | In space, high QE detectors |
| | | | | | Dynamic Range | 1E-11 - 1E-19 [erg/cm2/s/A] | high count rate detectors |
| | | | 2) Spectrally resolved abosrption lines in high-inclination disks: H2, CO, H2O, atomic species | Wavelength (Energy) | Bandwidth | 2a) 91 - 170 nm, 2b) 100 - 400 nm | optical coatings down to 91 nm. high-resolutuion, low-scatter gratings. one high-resolution spectrograph and one multi-object spectrograph. flux standards for 2% absolute spectrophotometry |
| | | | 2b) H2 and CO fluorescent emission in disks of any orientation | | Resolution | 2a) 3 km/s for lines absorption lines, 2b) 100 km/s for emission lines | |
| | | | | | Resolution | 2" | N/a |
| | | | | Angular | Instantaneous FOV | ~20' for MOS | FUV-MOS, e.g., microshutter device |
| | | | 3) Observe temporal variability of mass accretion, reverberation mapping of molecular emission lines | Temporal | resolution | 10 seconds | photon-counting detectors, L2 or elliptical orbit |
| | | | | | Single object observation duration | 5 hrs | |
| Characterize nearby habitable exoplanets (Kevin France - U. Colorado) | Determine the absolute level, the spectral energy distrubution, and the temporal variability of the energetic radiation environment around exoplanets to determine atmospheric photochemistry on habitable exoplanets and control for biosignature false positives | Chromospheric, transition region, and coronal luminosity and activity level of low-mass stars (G, K, and M) | 1) Broadband spectrally resolved ultraviolet irradiance spectra of all habitable planet candidates, N = TBD | Intensity | Sensitivity | S/N = 10 @ 1E -16 [erg /cm2/s] per 60s | In space, high QE detectors |
| | | | | | Dynamic Range | 1E-11 - 1E-19 [erg /cm2 /s /A] | high count rate detectors |
| | | | 2) Absolute fluxes of spectrally and temporally resolved upper atmosphere emission lines: | Wavelength (Energy) | Bandwidth | 95 - 400 nm | optical coatings down to 91 nm. gratings. Flux standards for 2% absolute spectrophotometry |
| | | | C III, O VI, LyA, O I, C II, Si IV, C IV, He II, Fe II, Mg II, Ca II | | Resolution | 15 km/s for lines, 100 km/s broadband | |
| | | | 3) ang resolved stellar LyA from background (geo, interplanetary) | Angular | Resolution | 0.2 arcsec at LyA | N/a |
| | | | | | Instantaneous FOV | > 2 arcsec (no hard requirement) | N/a |
| | | | 4) Temporal variability of high-energy emission lines on typical timescales of UV flares | Temporal | resolution | 1 sec | photon-counting detectors, L2 or elliptical orbit |
| | | | | | Single object observation duration | 8 hrs | |
| Understand the processes that determine the structure and evolution of planetary atmospheres (Kevin France - U. Colorado) | Determine the heating rates, mass-loss rates, compositions, and thermodynamic structures of the atmospheres of extrasolar planets | Atmospheric mass-loss rates from short-period planets of multiple atmospheric constituents. The incident stellar high energy radiation spectrum. | 1) Spectrally resolved far-UV transit observations of > 30 Jupiter-mass planets, > 20 Neptune-mass planets, and > 10 rocky planets. >= 3 transits | Intensity | Sensitivity | S/N = 50 @ 1E -15 [erg /cm2/s] per 60s | In space, high QE detectors |
| | | | | | Dynamic Range | 1E-11 - 1E-18 [erg /cm2 /s /A] | high count rate detectors |
| | | | 2) Transit depth as a function of wavelength and orbital phase for key atmospheric tracers: | Wavelength (Energy) | Bandwidth | 100 - 300 nm | optical coatings down to 100 nm. high-resolutuion, low-scatter gratings |
| | | | LyA, O I, C II, Mg II, H2 (superposed on O VI, C II, N II, and C III profiles) | | Resolution | 3 km/s | |
| | | | 3) ang resolved stellar LyA from background (geo, interplanetary) | Angular | Resolution | 0.2 arcsec at LyA | N/a |
| | | | | | Instantaneous FOV | > 2 arcsec (no hard requirement) | N/a |
| | | | 4) Observe pre-ingress, transt, and post-egress stellar flux | Temporal | resolution | 1 min | photon-counting detectors, L2 or elliptical orbit |
| | | | | | Single object observation duration | 8 hrs | |
| Understand the nature of stellar winds, magnetic fields, and circumstellar material in massive evolved stars and their influences on single and | Use UVV time-domain broadband polarimetric and spectropolarimetric observations to characterize changes with time (and orbital phase, for binaries) of CIRs, magnetic field lines, disks, and other stellar wind and | Shapes, sizes, extents, temperatures, densities, and compositions of electron- and resonance-line scattering regions in the atmospheres, winds, | Orientation of primary system axis (e.g., binary orbital plane, elongation of SN ejecta) | Broadband linear polarimetry (Stokes I, Q, U) | Sensitivity | S/N in total light = 100 in 30 min | Instrumental polarization < 3%. Polarized and unpolarized standard |
| | | | | | Dynamic Range | P = 0-10% with $s_P/P$ < 0.1 | |
| | | | Gas distribution, clumpiness, composition, temperature, density, ionization state; magnetic field strength and geometry via Hanle effect | Linear spectropolarimetry (Stokes I, Q, U) | Bandwidth | 100-900 nm for key diagnostic lines (Lya, Ha, UV wind lines) | Instrumental polarization < 3%. Polarized and unpolarized standard stars. |
| | | | | | Resolution | R = 25,000 (UV) to 35,000 (visible) | |
| | | | Magnetic field strength and | Circular spectropolarimetry | Bandwidth | 100-900 nm for key diagnostic lines (Lya, Ha, UV wind lines) | Good pointing stability (~0.1 km/s between spectra in a sequence) for precise |

| | | | | | | | | | | |
|---|---|---|---|---|---|---|---|---|---|---|
| single and binary stellar evolution and SN/GRB progenitor pathways. (Jennifer Hoffman - U. Denver) | disks, and other stellar wind and CSM structures; as well as illuminating the characteristics of SN ejecta and surroundings that trace the progenitor's mass-loss history. | and CSM of massive single and binary stars and supernovae. 3-D magnetic field geometries in single and binary evolved stars. | magnetic field strength and geometry via Zeeman effect | circular spectropolarimetry (Stokes V) | Resolution | R = 25,000 (UV) to 35,000 (visible) | | | line combination. Simultaneous wavelength calibration. | |
| | | | | Angular | Resolution | | | | | |
| | | | | | Instantaneous FOV | | | | | |
| | | | Periodicities including rotation, orbital motion, and CIR modulations; time evolution of SN ejecta and CSM | Temporal | Resolution | 30 min | | | Monitoring capability to characterize orbital periods up to timescales of months | |
| | | | | | Single object observation duration | few days for entire rotation period | | | | |
| Investigate the dispersal machanism of proto-planetary disks through observations of gas forbidden lines (Patrick Hartigan - Rice U.) | Use narrow-band imaging and high resolution spectroscopy to map the morphology and kinematics of photoevaporating winds and invesigate the time scale for the disk dispersal in systems that are in the process of forming planets | Derive statistics of the disk dispersal time scale as a function of the stellar and environmental properties, including stars with different spectral types and ages, and in stellar clusters with different star formation histories. Compare the dispersal of proto-planetary disks in low-mass star forming regions such as Taurus and Chamaleon, with that in intermediate mass star forming regions such as Orion, and in high mass star forming regions such as Carina. | A range of emission lines in near UV, visible, and near IR, including [O I] 6300A, Hα, and Mg II 2800A | Gas temperature in the photoevaporating wind | Spectral | Spectral range | 1200A - 9000A | 2700A - 7000A | Spectrograph with long slit capability. An IFU system would be great if no spatial resolution is lost | Possible synergy with needs of extragalactic community to spectro-image multiple sources in a field |
| | | | | Gas velocity structure in the photoevaporating wind | | Spectral resolution | R up to 30,000 | R up to 10,000 | | |
| | | | | Gas density in the photoevaporating wind | | Spatial resolution | 20 arcsecond slits @ λ/D pixel scale | 5 arcsecond slits @ λ/D pixel scale | | |
| | | | | Morphology of the gas emission as a function of the distance from the star | Imaging | Spatial Resolution (defined at 400 nm) | λ/D pixel scale | λ/D pixel scale | Imager with excellent filter options and a modest FOV | 12 m primary to give spatial resolution, flux sensitivity, and Strehl ratio needed to image faint extended structures reliably around bright point sources |
| | | | | | | FOV | 5 arcminutes | 3 arcminutes | | |
| | | | | | | Filters | 16 narrowband nebular filters, among them CI 9850, NII 6583, NI 5200, OI 6300, OII 3727, SII 6716, SII 6731, Mg II 2800, Hα | 12 filters minimum | | |
| What controls the mass-energy-chemical cycles within galaxies? (Ben Williams - U. Washington) | Observationally constrain models of the formation and evolution of the massive stars that drive metal production and distribution through libraries of resolved massive stars in a wide range of formation environments. | Count and measure the physical parameters to 10% accuracy of individual stars down to 5 solar masses in star clusters out to 5 Mpc (e.g., NGC253,M82), covering metallicities from 0.1-2.0 solar and galaxy masses ranging from 10^5-10^12 solar masses | Individual stars in young clusters down to 5 solar masses at 5 Mpc. | Angular Resolution | | | 0.008" | | | |
| | | | | Field of View | | | 15 arcmin | | | |
| | | | Brightnesses of such individual stars in at least 6 independent bands covering 0.2-1.0 micron. | Number of Filters | | | 6 | | | |
| | | | | Wavelength Coverage | | | 0.2-1.0 microns | | | |
| Measure magnetic field and dust characteristics in the interstellar medium (BG Andersson - NASA Ames) | Use spectropolarimetry - over the UV (and optical) range - to determine the amount and orientation of the dichroic extinction polarization of background stars due to aligned dust. | 1) Measure the magnetic field strength in diffuse gas and 2) characterize the dust properties by analyzingth the polarization spectra in the context of modern grain alignment theory | Medium to high-spectral resolution spectropolarimetry in the UV | Size distribution alignment fraction and mineralogy of the small dust | Spectral | Spectral range | 160-1600nm | 300-500nm | Instrumental polarization, stability and variations over the FOV | Sensitivity to assemble statistically significant samples of polarimetry of background stars |
| | | | | Particle heights, cloud/haze thickness, gas abundances | | Resolving power | 1000 | 100 | | |
| | | | Spectropolarimetry of the 2175Å extinction feature | Statistics of polarization of feature | Spectral | Spectral range | 210-130nm | | Sensitivity, instrumental polarization | |
| | | | | Establish carrier and improve it's use in extinction curves etc. | | Resolving power | 1000 | 200 | | |
| | | | Tracing of the polarization curve to FUV wavelengths | Test ans utilize the theoretical prediction that paramagnetic alignment dominates for the very smallest grains, in which | Photometry | Sensitivity and systematic error | 0.10% | | Sensitivity, instrumental polarization | |
| | | | | | | number of fileters | 4 for 120-250nm | 2 for 120-250nm | | |
| | | | Line polarimetry of fine structure lines | Measure magnetic field strengths through the Hanle | Spectral | Spectral range | 120-1600nm | | instrument throughput | telescope aperture |
| | | | | | | Spectral resolution | 50,000 | 5,000 | | |